\documentclass[acmlarge,nonacm]{acmart}
\usepackage{caption}
\usepackage[T1]{fontenc}
\usepackage{subcaption}
\usepackage{graphicx}
\usepackage{xcolor}
\usepackage{soul}
\usepackage{url}
\usepackage{tabularx}
\usepackage{multirow}
\usepackage{multicol}
\usepackage{adjustbox}
\usepackage{hyperref}
\usepackage{array}
\usepackage{algorithm}
\usepackage{algpseudocode}
\usepackage{textcomp}

\usepackage{amsmath,amssymb,amsfonts,amsthm}
\usepackage{balance}

\newcommand{\nsf}[1]{\href{https://www.nsf.gov/awardsearch/showAward?AWD_ID=#1}{#1}}

\author{Anthony Etim}
\affiliation{%
  \institution{Yale University}
  \city{New Haven}
  \state{CT}
  \country{USA}
}
\email{anthony.etim@yale.edu}

\author{Kenan Erol}
\affiliation{%
  \institution{Yale University}
  \city{New Haven}
  \state{CT}
  \country{USA}
}
\email{kenan.erol@yale.edu}

\author{Jakub Szefer}
\affiliation{%
  \institution{Northwestern University}
  \city{Evanston}
  \state{IL}
  \country{USA}
}
\email{jakub.szefer@northwestern.edu}
\renewcommand{\shortauthors}{Anthony Etim, Kenan Erol and Jakub Szefer}

\title{ML-Assisted Power Side-Channel Attacks on ML Accelerators in Shared FPGAs}

\begin{document}

\begin{abstract}
Cloud providers offer on-demand access to Field-Programmable Gate Arrays (FPGAs), and multi-tenant FPGAs let multiple users share one device to improve utilization. This sharing introduces security risks: an adversary co-located with a victim on the same FPGA can mount power side-channel attacks against the victim's Machine Learning (ML) accelerator. Prior on-chip attacks on such accelerators based on Time-to-Digital Converter (TDC) sensors have recovered only small, MNIST-scale grayscale inputs, including with ML-assisted reconstruction. This work is the first on-chip, TDC-based attack to produce full-scale \emph{color} reconstructions of ImageNet-scale inputs, by combining ML-based upscaling and recoloring with TDC measurements of a remote victim's ML accelerator. We frame this as a proof-of-concept leakage study: the victim implements the first convolution layer of a CNN processing 224$\times$224 ImageNet inputs on a shared FPGA, which we isolate to characterize the input-dependent leakage, and we quantify how much of each reconstruction reflects genuinely recovered information versus ML-generated post-processing. As a preliminary countermeasure, we study a pixel-substitution defense that replaces each pixel value, within a fixed intensity budget, with one of different Hamming Weight, producing a different TDC signature. We find that this transformation costs the victim's own model more accuracy than it costs the attacker and, because it applies the same fixed, publicly reconstructable mapping to every image, is largely defeated by an attacker retrained on the transformed data; robust defenses will require randomized or input-dependent transformations.

\end{abstract}

\maketitle

\section{Introduction}
\label{sec_introduction}

Field-Programmable Gate Arrays (FPGAs) have become a key component of modern computing infrastructures. In cloud environments, users can readily access on-demand FPGA resources to accelerate computation using custom hardware logic. Major cloud providers, such as Amazon Web Services~\cite{amazonf1}, now offer a range of FPGA-enabled services. Amazon's current, second-generation FPGA offering, the EC2 F2 instance family, has been generally available since December 2024 and has since expanded to additional regions~\cite{aws_f2_2024}. Users can deploy standalone FPGAs or, more recently, FPGA-integrated SmartSSDs~\cite{smartssd} provided by vendors such as VMAccel~\cite{vmaccel}.

In order to help improve the efficiency of these FPGAs, there is active interest in multi-tenancy. The research on multi-tenant FPGAs has made it possible to increase the efficiency and utilization of the FPGAs~\cite{10.1145/3474058,istvan2018providing,multi_defense_attack}. However, multi-tenancy opens up the FPGAs to different security threats. In particular, a malicious user or adversary who is co-located next to a victim can attack the victim or steal information from the victim. Already, a range of attacks have been demonstrated: from power-side channels~\cite{tian_remote_2021,islam_exploiting_2017,boutros_neighbors_2020,provelengios_power_2021} to long-wire crosstalk~\cite{8457631,giechaskiel2018leaky}.

While prior work, such as~\cite{giechaskiel2018leaky}, has primarily focused on attacks targeting cryptographic keys, this work investigates threats against the increasingly important machine learning applications deployed on FPGAs. There is growing interest in leveraging FPGAs to accelerate machine learning models for image and video recognition tasks. In these scenarios, input images and videos are typically unencrypted when supplied to the models. Consequently, during inference, the data being processed remains vulnerable to side-channel attacks, even if it was previously protected by encryption during storage or transmission. Due to the significant computational overheads of fully homomorphic encryption, this work does not consider machine learning on encrypted data.

In the realm of attacks on machine learning algorithms running on FPGAs, Moini et al. demonstrated a remote power-based side-channel attack on a binarized convolutional neural network (BNN) in an FPGA to recover MNIST images being processed by the BNN. A BNN is a type of convolutional neural network (CNN), and they were able to recover the input black and white MNIST images used with the BNN with high accuracy~\cite{moini_remote_2021}. Further, a number of other works have also explored attacks on neural networks~\cite{wei2018know,maji2021leaky,huegle2023power2picture} processing MNIST images, in order to show that the attacker can recover the input~images. A comparison to the most related work is shown in Table~\ref{tab:work_comparison}. While prior ML-assisted attacks such as Power2Picture~\cite{huegle2023power2picture} recover small grayscale MNIST-scale inputs, in this work we demonstrate a remote power side-channel attack on a CNN that, for the first time, produces full-scale \emph{color} reconstructions of complex ImageNet inputs from an on-chip TDC-based power side-channel. We achieve this by introducing a new class of ML-assisted attacks where the attacker leverages data from remote TDC measurements and processes them through an ML algorithm to recover information about the user's input images. We emphasize at the outset that this is a proof-of-concept leakage study: our victim implements only the first convolution layer of the CNN, which directly processes the raw input pixels and which we deliberately isolate to characterize the input-dependent leakage under favorable measurement conditions (see Section~\ref{section:discussion} for a candid discussion of the resulting assumptions and limitations). This work presents two distinct ML-assisted attack methodologies, differentiated by the type of information they extract from the power side-channel to overcome the limitations of prior work.

\begin{table*}[t]
    \centering
    \small
    \caption{\small Comparison with prior work, all works aim to use side channels to recover sensitive inputs to a neural network.}
    \label{tab:work_comparison}
    \small
    \begin{adjustbox}{width=\textwidth}
    \begin{tabular}{|p{1.1cm}|p{1.5cm}|p{0.7cm}|p{1.2cm}|p{1.3cm}|p{1.0cm}|p{1.5cm}|p{1.8cm}|p{1.4cm}|}
        \hline
        \textbf{Paper} &
        \textbf{Image Size (pixels)} &
        \textbf{Color} &
        \textbf{Image Type} &
        \textbf{Victim Neural Network} &
        \textbf{Side-channel Type} &
        \textbf{Sensor} &
        \textbf{Attack Cost / Complexity} &
        \textbf{Counter-measure Proposed} \\
        \hline \hline
         Wei et al.~\cite{wei2018know} & 28x28 & $\times$ & MNIST & BNN &  Power & Oscilloscope & High - requires oscilloscope & No  \\ \hline
         Maji et al.~\cite{maji2021leaky} & 28x28 \newline 32x32 \newline (unspecified) & \checkmark \newline \checkmark \newline \checkmark & MNIST \newline CIFAR10 \newline ImageNet  & CNN \newline CNN \newline CNN & Timing \newline Timing \newline Timing & Oscilloscope \newline Oscilloscope \newline Oscilloscope & High - requires oscilloscope & Yes\\ \hline
          Moini et~al.~\cite{moini_remote_2021} & 28x28 \newline 28x28 & $\times$ \newline $\times$ & MNIST \newline Fashion- MNIST & BNN \newline BNN &  Power \newline Power &  TDC \newline TDC & Low - on-chip sensor & No \\ \hline
         Huegle et~al.~\cite{huegle2023power2picture} & 28x28 \newline 28x28 & $\times$ \newline $\times$ & MNIST \newline Fashion- MNIST & CNN \newline CNN &  Power \newline Power &   TDC \newline TDC & Low - on-chip sensor & No\\ \hline \hline
         {\bf This Work} & 224x224 & \checkmark & ImageNet & CNN  &  Power &  TDC & Low - on-chip sensor & Yes  \\ \hline
    \end{tabular}
   \end{adjustbox}
\end{table*}

Our first methodology aims to produce a high-fidelity visual reconstruction of the victim’s image. It assumes the side channel leaks information correlated with actual pixel values and addresses two limitations of prior work. First, inputs to neural networks are often downsampled versions of original images, which degrades reconstruction fidelity. Second, prior attacks that recover only aggregate power traces, rather than per-channel measurements for the red, green, and blue components, can produce only grayscale approximations. To overcome these issues, we introduce an attacker pipeline that, unlike prior generative recovery such as Power2Picture~\cite{huegle2023power2picture}, couples interpolation-based upscaling with learned recoloring: recovered grayscale reconstructions are upscaled using interpolation and recolored using learned models to generate full-scale color reconstructions of the victim’s inputs. Because the recoloring step is generative, we are careful throughout to distinguish the information genuinely recovered from the side channel from the plausible detail synthesized by the ML post-processing, and we quantify this distinction in Section~\ref{sec:reconstruction_quality}.

Our second, complementary attack, assumes the side channel leaks information correlated with Hamming weight of the pixel values. The method exploits the fact that power side channels reveal the Hamming weight of data processed by the device. To the best of our knowledge, this is the first work to demonstrate reconstruction of \emph{color} inputs, as opposed to the grayscale inputs recovered by prior on-chip TDC-based attacks, from a remote, on-chip power-based side channel. The attack leverages voltage fluctuations caused by variation in the power consumption of the convolution unit within a CNN. Because the attacker recovers Hamming weights rather than exact pixel values, we introduce a scaled Hamming-weight representation of images. This representation is used to train a suite of machine learning models for classification, and we evaluate the models’ accuracy on scaled Hamming-weight inputs. Once models are trained, the attacker can classify victim inputs directly from observed power traces.

Our attacks can be executed remotely, requiring no physical hardware access by the attacker. We illustrate the details of our attack on the convolution unit of a CNN-based circuit designed for image recognition from the ImageNet database~\cite{imagenet}. Successful implementation of our attack and the corresponding image recovery is demonstrated on the Xilinx ZYNQ System-on-Chip (SoC) based PYNQ-Z2~FPGA~\cite{pynq-z2}.

In order to present a defense, we analyze root-causes of the attacks. During evaluation of the attacks, we found that visually highly-contrasting images are most easily detected and attacked. As a result, we first define a metric for determination of how contrasting images are and confirm that indeed contrasting images, according to our metric, are the ones more easily attacked using our method. Then, second, we develop a pixel-substitution method that replaces each pixel value, within a fixed intensity budget, with a value of different Hamming Weight, which lowers the Hamming Weight contrast of the image the attacker observes while the image still looks much like the original. We then, third, evaluate this substitution against both the victim's and the attacker's models. On the selected test images it prevents the attacker's original models from classifying several inputs correctly, but on the full ImageNet validation set it costs the victim's accuracy far more than the attacker's, and an attacker that retrains on substituted images regains accuracy at the largest budget for every architecture; we therefore present it as a preliminary obfuscation rather than a robust~defense.

\subsection{Contributions}

The contributions of this work are as follows:

\begin{enumerate}

  \item We present a new side-channel attack targeting a machine learning accelerator deployed on remote FPGAs. The attack utilizes Time-to-Digital Converters (TDCs) to collect power traces, then uses custom-trained ML to classify the data and learn the type of the input image used by the victim with their victim convolutional neural~network.

  \item This work is the first to leverage ML-based recoloring combined with software and hardware interpolation-based upscaling in an attack pipeline for recovering color images from TDC-based power side channels. This allows an attacker to obtain full-scale color reconstructions from the downscaled, and potentially grayscale, data recovered from power traces. We analyze cases for recovering both per-channel (color) and aggregate (grayscale) power signatures.

  \item We develop the idea of scaled Hamming Weight representation of colored images. Scaled Hamming Weight images are ones where each pixel is represented by its Hamming Weight. We build a database of scaled Hamming Weight images and train a set of neural networks to categorize the scaled Hamming Weight images.

  \item We demonstrate how TDC power measurements can be used to recover the scaled Hamming Weight version of the victim user's input image, which is then fed into our machine learning models to correctly classify the image only using the scaled Hamming Weight data but not full color pixels; the approach is validated on the ImageNet images.

  \item We study a pixel-substitution countermeasure that replaces each pixel value with one of different Hamming Weight within a fixed intensity budget, reducing the Hamming Weight contrast the attacker observes. We evaluate it on the full ImageNet validation set against both the victim's own models and adaptive attackers retrained on substituted images, and find that it is not a robust defense: it costs the victim considerably more accuracy than the attacker, and adaptive retraining regains accuracy at the largest budget for every attacker~architecture.

\end{enumerate}

\section{Background}
\label{background}

In this section, we provide a brief overview of convolutional
neural networks, and in particular of VGG-19 and VGG-16 which are used in this work as the example victim machine learning models.

\subsection{Convolutional Neural Network}

Deep neural networks (DNNs) represent a class of artificial neural networks characterized by multiple layers~\cite{dnn_survey}. In a DNN, each layer is responsible for extracting relevant features, with the output from each layer serving as the input for the subsequent layer. By combining feature extraction with the classification capabilities of traditional neural networks, DNNs effectively map input data to a set of predictions, making them suitable for tasks like image classification.

\begin{figure*}[t]
    \centering
    \includegraphics[width=0.5\textwidth]{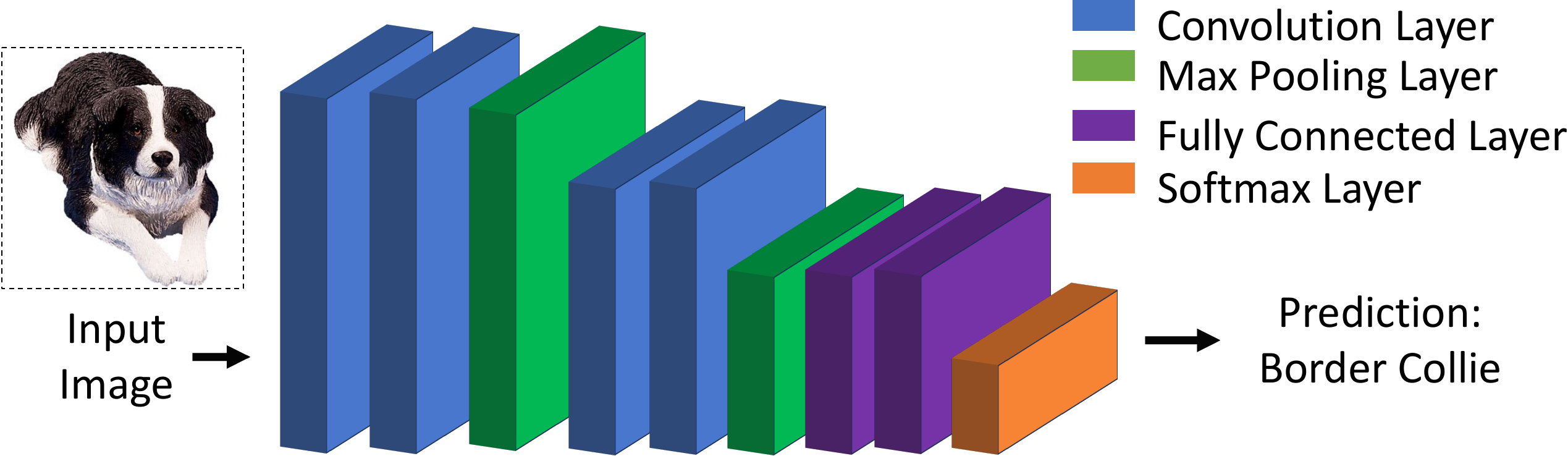}
    \caption{\small Simplified overview of a typical Convolutional Neural Networks (CNNs), modeled after the VGG-19 network.}
    \label{fig:dnn}
\end{figure*}

One prevalent form of DNNs is Convolutional Neural Networks (CNNs), which consist of multiple convolutional layers, as depicted in Figure~\ref{fig:dnn}.
CNNs are widely used in deep learning for image processing and computer vision. Comprising three main types of layers (convolutional layers, subsampling layers, and fully-connected layers), CNNs seamlessly integrate feature extraction and classification.
The convolutional layer conducts a two-dimensional (2D) convolution between input data and a kernel, followed by the application of an activation function to generate a feature map. Typically, the kernel size of 3x3 elements is used in the convolution. The ReLU (rectified linear unit) activation function is commonly used in CNNs after the convolution is done. Multiple kernels in each convolutional layer are used to produce diverse feature maps to extract various features from the input data.
The subsampling layer reduces the spatial size of feature maps from the preceding convolutional layer, aiding in the extraction of dominant features that are rotationally and positionally invariant. This process ensures the efficacy of the model's training and reduces computational complexity. Two types of subsampling operations, max-pooling and average-pooling, exist, with max-pooling generally preferred for its superior performance. The common subsampling operation involves 2x2 max-pooling.
After feature extraction, the output of the subsampling layer is flattened into a single vector and fed into the fully-connected layer. The fully-connected layer performs the classification task, producing a label indicating the correct category for the input image. The output of the last fully-connected layer then goes into a softmax layer. The softmax layer is typically used in the last layer of a neural network to predict the class of an input image. \emph{Among all network layers, the convolutional layer is the most computationally intensive. In this study, we demonstrate that this layer leaks important information to an attacker.}

\subsection{VGG-19 and VGG-16}

\begin{table*}[t]
\centering
\caption{\small Architecture of the VGG-16 and VGG-19 model which is used as the victim model in our attacks.}
\small
\label{table_vgg_19}
\small
\begin{adjustbox}{width=0.7\textwidth}
\begin{tabular}{|c|p{4.0cm}|p{2cm}|p{2cm}| p{1cm}|}
\hline
\textbf{Layer No} & \textbf{Layer Type } & \textbf{Input Size} & \textbf{Kernel Size} & \textbf{Stride}
\\ \hline \hline
1  & 2 X Convolution &  224x224x3 & 3x3x64 & 1\\ \hline
2  & Max Pooling     &  224x224x64 & 2x2 & 2\\  \hline
3  &  2 X Convolution &  112x112x128 & 3x3x128 & 1\\  \hline
4  &  Max Pooling &  112x112x128 & 2x2 & 2\\ \hline
5  &  3 X Convolution (VGG-16) \newline 4 X Convolution (VGG-19)  &  56x56x256 & 3x3x256 &
1\\ \hline
6  & Max Pooling     &  56x56x256 & 2x2 & 2\\ \hline
7  & 3 X Convolution (VGG-16) \newline 4 X Convolution (VGG-19) &  28x28x512 & 3x3x512 &
1\\ \hline
8  & Max Pooling     &  28x28x512 & 2x2 & 2\\ \hline
9  & 3 X Convolution (VGG-16) \newline 4 X Convolution (VGG-19) &  14x14x512 & 3x3x512 &
1\\ \hline
10  & Max Pooling     &  14x14x512 & 2x2 & 2\\ \hline
11  & Fully Connected    &  7x7x512 & - & -\\ \hline
12  & Fully Connected    &  4096 & - & -\\ \hline
13  & Fully Connected    &  4096 & - & -\\ \hline
14  & Softmax    &  1000 & - & -\\
\hline
\end{tabular}
\end{adjustbox}
\end{table*}

One example of a CNN is the VGG-19 model~\cite{simonyan2015deep}, which serves the purpose of image classification and comprises a total of 19 weight layers. The model employs multiple 3x3 kernels in each layer. This architecture consists of three fully connected layers, five pooling layers, sixteen convolution layers, and a softmax layer. In VGG-19, two fully connected layers with 4096 channels each are utilized, followed by another fully connected layer with 1000 channels for predicting 1000 labels. The final layer employs a softmax activation function for classification. The configuration of VGG-19 architecture is shown in Table~\ref{table_vgg_19}.

The VGG architecture encompasses three fundamental types of layers: Convolution layer, Pooling layer, and Fully Connected layer. The convolution layer performs the essential operation of convolution. In this process, a kernel or filter is applied to an input image of size 224x224, generating a feature map that aids in understanding specific aspects of the input image in detail. CNNs enforce local connectivity patterns, enhancing edges and embossing.
The pooling layer is a crucial step in convolutional-based systems, reducing the dimensions of feature maps. This layer addresses the overfitting problem by summarizing features in patches of the feature map based on their presence. Common strategies include average pooling or maximum pooling. In the VGG-19 model, maximum pooling is predominantly employed.
In the VGG-19 model, the last three layers constitute fully connected layers. Input to these layers originates from the previous layers (convolution or pooling), and the input image is flattened before feeding into this layer. Matrix multiplication and bias offset are then applied to compute their activations. %

The VGG-16 model~\cite{simonyan2015deep}, a closely related predecessor to VGG-19, features a total of 16 weight layers, including 13 convolutional layers and three fully connected layers. Like VGG-19, it employs 3x3 convolutional filters, which allow for capturing intricate image details while maintaining computational efficiency. VGG-16 also utilizes ReLU activation functions and maximum pooling layers to reduce spatial dimensions and enhance feature extraction. The primary difference lies in the number of convolutional layers, as VGG-16 has fewer than VGG-19, which leads to a slightly reduced computational complexity. Despite this, VGG-16 maintains robust performance across various image classification tasks, making it a popular choice in scenarios where computational resources are limited.  The configuration of VGG-16 architecture is shown in Table~\ref{table_vgg_19}.

\section{Threat Model}

This paper centers on a multi-tenant FPGA scenario where the targeted user is conducting a machine learning inference algorithm on a hardware module co-located with the modules of a malicious user on the same FPGA. The adversary utilizes the FPGA platform concurrently with the victim, but is logically separated from the victim. This necessitates the use of side-channels for any data extraction from the victim.

Existing research in~\cite{multi_tenant_2025_survey} has explored and promoted multi-tenancy in cloud FPGAs to increase utilization. While cloud providers implement security measures, the co-location of mutually untrusting tenants on a shared hardware fabric inherently creates side-channel risks.
For example, consider a cloud user deploying a proprietary medical image analysis model. A malicious co-tenant could use this attack to infer the types of medical conditions being screened (e.g., classifying images as ``malignant'' or ``benign'') without seeing the model's output, which is highly sensitive information. Other plausible scenarios include: a financial services firm using an FPGA-accelerated model to classify check images or identity documents, where the attacker could infer the document type or content; an autonomous vehicle company testing perception models in the cloud, where a competitor could infer the types of road scenarios being processed; or a surveillance system vendor processing video frames, where an attacker could learn what types of scenes are being analyzed. In each case, even learning the \emph{class} of input image (rather than exact pixel values) constitutes a significant privacy breach.

Regarding co-location, we assume the attacker can provision their circuit on the same FPGA as the victim. In current cloud FPGA deployments, including both Amazon EC2 F1 and the newer, second-generation EC2 F2 family that became generally available in December 2024~\cite{aws_fpga_2021, aws_f2_2024}, each FPGA is still allocated to a single tenant at a time, so the shared-fabric threat model we study is not yet directly realizable on these commercial offerings. However, academic research has actively developed multi-tenant FPGA frameworks~\cite{multi_tenant_2025_survey, 10.1145/3474058, istvan2018providing, multi_defense_attack} and cloud providers are exploring shared models to improve utilization. Our work is motivated by this trajectory: as multi-tenancy becomes more prevalent, understanding the associated security risks becomes critical. We note that even without explicit multi-tenancy, recent work~\cite{schellenberg_remote_2018} has shown that FPGA side-channel attacks can be conducted across supposedly isolated partitions.

We assume the victim's circuit inputs, such as input images for a machine learning accelerator, are securely transmitted to the FPGA. While the data may be encrypted during transmission and storage, we assume it is decrypted when the machine learning operations are applied, which is when the attacker steals the information. During the neural network inference operation, we assume the attacker, co-located on the same FPGA, can collect power traces (from which they recover the input image class they want to steal).

We also assume the architecture of the victim's CNN is known to the attacker. This assumption is grounded in several practical scenarios. Many users build ML accelerators based on well-known, open-source architectures like VGG, making their structure public knowledge. Furthermore, for proprietary or black-box designs, an attacker could first perform a preliminary reverse-engineering attack. Prior work has demonstrated that even encrypted DNN accelerator Intellectual Property (IP) cores can be reverse-engineered using side-channel analysis to reveal data flow, scheduling, and other crucial architectural parameters~\cite{probst2025remote, sp_side}. Therefore, our work focuses on the data-extraction phase of an attack, proceeding on the basis that the victim's architecture is either public or has been previously compromised.

Beyond the algorithmic architecture (e.g., that the victim runs VGG-19), our attack also benefits from knowledge of the victim's circuit-level \emph{microarchitecture}, specifically, how the CNN algorithm is mapped to FPGA hardware, including the degree of parallelism in the convolution engine, the data scheduling policy, and the memory access patterns. This microarchitecture directly determines the cycle-by-cycle switching activity and thus shapes the power traces captured by the TDC. We note that the same algorithmic architecture can yield many different hardware implementations with different parallelism and switching behavior. In practice, this microarchitectural knowledge may be obtained because: (1)~the victim uses a well-known open-source hardware accelerator framework (e.g., High-Level Synthesis (HLS)-generated designs from tools like Vitis~AI, or published Register Transfer Level (RTL) implementations), where the microarchitecture is effectively public; (2)~prior reverse-engineering attacks~\cite{sp_side} have demonstrated that data flow, scheduling, and parallelism of even encrypted DNN accelerator IPs can be extracted through side-channel analysis; or (3)~the attacker can profile the victim's design by observing its power signature across multiple inference runs to learn its scheduling behavior, a technique whose feasibility has been demonstrated in related TDC-based works~\cite{moini_journal_2021, tuneable_tdc}. We discuss the impact of imprecise microarchitectural knowledge on attack effectiveness further in Section~\ref{section:discussion}.

Additionally, it is presumed that the outputs, securely transmitted back to the user, remain undisclosed to the adversary. The focus of this work lies in employing a Time-to-Digital Converter (TDC) to measure voltage changes of the victim. The TDC data aids the adversary in estimating the voltage drop across the FPGA power distribution network (PDN) during the execution of the convolution layer within the neural network's accelerator's image classification process. These voltage estimates serve as a side channel, offering a means to extract the victim's input image. We assume the attacker can leverage ML techniques to help in effectively amplifying the information they can recover from the side channel by training ML on various types of images that may be possibly used by the~victim.

Our threat model also presumes the attacker can synchronize their TDC measurements with the victim's inference process to isolate the power trace of the first convolution layer. While this is a non-trivial engineering challenge, it is considered a plausible step in side-channel attacks. An attacker could achieve this by monitoring shared resources for activity indicative of the victim's operation. For instance, monitoring Peripheral Component Interconnect Express (PCIe) bus traffic for large data transfers characteristic of a new image being loaded into the victim's on-chip memory could serve as a trigger to begin capturing TDC data.

Finally, our experimental analysis is conducted in a controlled environment with only the victim and attacker circuits co-located on the FPGA. This allows us to establish a clean baseline and quantify the potential information leakage under ideal conditions. We acknowledge that a real-world, noisy multi-tenant environment would present additional challenges, a limitation we discuss further in Section~\ref{section:discussion}.

\section{Attack Implementation and Steps}

This section presents the attack implementation and attack steps taken by the adversary to learn the classification of the victim's input images just from TDC measurement data, which is then processed by attacker's ML algorithm.

\subsection{Basic Power Side-Channel Attack Setup}
\label{sec_basic_power_side_setup}

\begin{figure*}[t]
    \centering
    \includegraphics[width=0.5\textwidth]{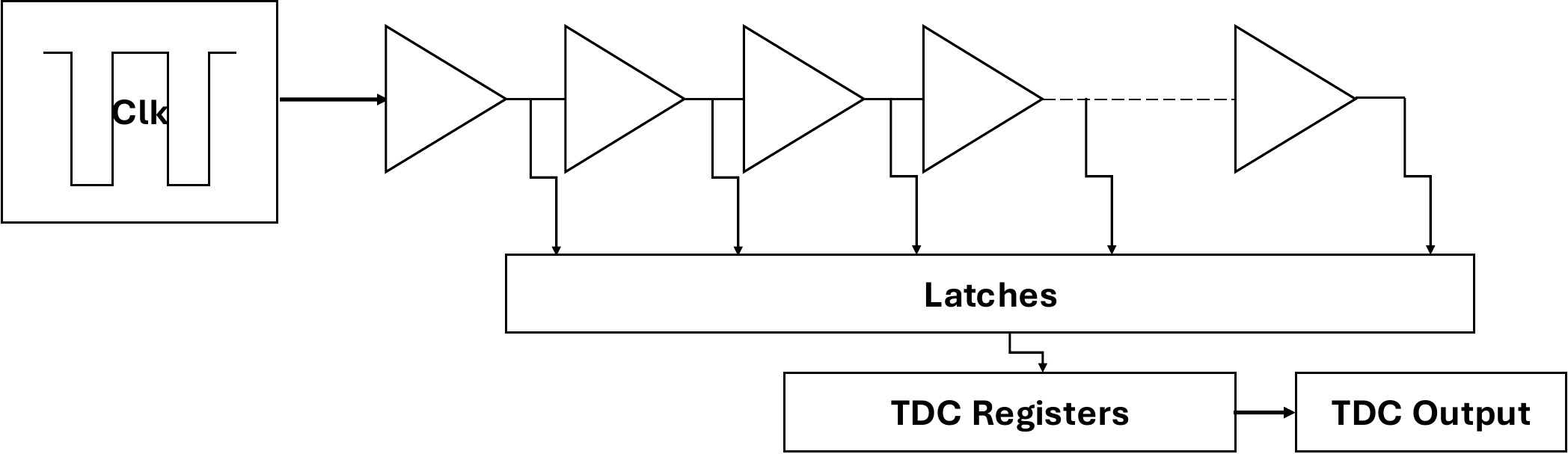}
    \caption{\small Architecture of the TDC Sensor.}
    \label{fig:TDC_diagram}
\end{figure*}

Following our assumptions, both the victim circuit and the attacker circuit share the same FPGA. The adversary's objective is to extract the input image from the neural network accelerator, with a particular focus on the initial convolution layer responsible for direct processing of the input image. Utilizing a Time-to-Digital Converter (TDC), voltage change estimates are collected by the attacker when the neural network accelerator is engaged in processing the first convolution layer.  Within the first convolution layer, the input image undergoes convolution with multiple distinct kernels, resulting in the creation of various output feature maps. To be precise about what the victim implements: our hardware realizes a single output filter of the first convolution layer, i.e., one $3\times3$ kernel per input channel (R, G, B) producing one output feature map, rather than all output filters of the layer. We refer to this throughout as the ``first convolution layer'' for brevity, but the attacker observes the leakage of this one filter's multiply-accumulate datapath. For our attack, we employ a TDC to get the voltage change estimate trace during the execution of this filter for a given input image. The TDC architecture can be seen in Figure~\ref{fig:TDC_diagram}.

The TDC operates on a fundamental principle of Complementary Metal-Oxide-Semiconductor (CMOS) circuits: the propagation delay through logic gates is sensitive to the supply voltage. As shown in Figure~\ref{fig:TDC_diagram}, the TDC consists of a chain of delay elements (e.g., carry-chain elements or inverters) through which a signal propagates. A reference clock periodically samples the state of this delay chain via a bank of latches, capturing how far the signal has traveled. The resulting thermometer-coded output is converted to a digital count that is directly proportional to the propagation delay and, inversely, to the instantaneous supply voltage.

The connection between TDC measurements and the Hamming Weight of the victim's data operates through the following physical mechanism. When the victim's convolution engine processes a pixel, it performs multiply-and-accumulate operations on the pixel value and the kernel weights. In CMOS logic, dynamic power consumption is dominated by switching activity: each logic transition (0$\rightarrow$1 or 1$\rightarrow$0) charges or discharges internal capacitances, drawing current from the shared power supply. The number of such transitions during a multiplication is directly correlated with the Hamming Weight of the operands: a pixel value with more bits set to 1 produces more switching activity during the multiplication, drawing more current. This increased current draw causes a larger transient voltage drop on the shared Power Distribution Network (PDN). The TDC, connected to the same PDN, measures this voltage drop as a change in propagation delay. Thus, the TDC output for each pixel is correlated with the Hamming Weight of that pixel's value.

The TDC is sensing voltage fluctuations on the shared power rail that supplies the FPGA fabric. These fluctuations are an aggregate effect of all activity, but the attack isolates the power consumption of the
first convolution layer's computation by synchronizing the measurements with the start of the victim's inference process. Specifically, we focus on the power consumption as the first convolution layer is processing the input~image.

\subsection{Recovering Input Image Information}

Each ImageNet input is first resized to $224\times224$ before being processed by the victim accelerator (the input size expected by VGG-19 and similar networks). The resized image is then split into three individual color channels. Each channel is processed through a convolution layer with a known convolution kernel. For each channel, TDC counts are taken as the convolution is performed on that channel. The input image is then processed by the first convolution layer. We analyze two distinct attack scenarios based on how the victim accelerator processes the image's color channels.

\paragraph{Scenario 1 (Per-Channel Attack)} In a resource-constrained accelerator design, the Red, Green, and Blue channels may be processed sequentially by a shared convolution engine to conserve hardware resources. This allows an attacker to synchronize TDC measurements with each channel's computation, thereby isolating distinct power traces for each color. This more powerful attack scenario enables direct reconstruction of a color image, as explored in our Attacks E and F of Section~\ref{section:six_attack_scenarios}.

\paragraph{Scenario 2 (Aggregate Attack)} In a more common high-throughput design, the three color channels are processed in parallel. In this case, the TDC captures an aggregate power trace that combines the consumption of all three channels. This leakage allows the attacker to recover a grayscale representation of the image. Our attack gracefully degrades to this scenario, which we analyze in Attacks A through D of Section~\ref{section:six_attack_scenarios}. By analyzing both sequential and parallel processing, we provide a comprehensive evaluation of the information leakage.

Based on the leakage model, these power traces can be interpreted in two ways: as being proportional to the actual pixel values or, more commonly, to the Hamming Weight of the pixel data. We stress that these two interpretations are not on equal physical footing. As explained above, the dominant source of dynamic power in CMOS is switching activity, so the physically faithful model is the one in which the trace is correlated with the Hamming Weight (and, more generally, the Hamming distance between successive operands) rather than with the raw integer intensity. The pixel-value interpretation is therefore an \emph{idealized, optimistic} assumption: it holds only to the extent that, for the specific victim datapath and input statistics, pixel intensity is approximately monotonically related to switching activity. We include the pixel-value attack because it characterizes an upper bound on visual-reconstruction quality when this favorable relationship holds, but we regard the Hamming-Weight attack as the physically grounded result. Our work explores both scenarios. The following subsections detail the two distinct attack methodologies we developed: one for visual reconstruction based on pixel-value leakage, and another for image reconstruction and then classification based on Hamming-Weight leakage.

We take a total of $2^{20}$ TDC samples for each image measurement.
This large number was used to average out noise in our specific setup. A better sensor or a less noisy environment would require fewer samples. We leave experiments on the impact of varying the number of measurements for future work.
Since the FPGA board used is not optimized for power side-channel analysis, we amplify the signal from the victim neural network accelerator to increase the power consumption based on the Hamming Weight of the pixel value so that TDC measurements are amplified. Second, we slow down the convolution to a frequency of 25~MHz so that more TDC measurements are collected per pixel. With a specialized FPGA, or better TDC, these techniques may not be needed, and our attack does not depend on them. Our attack works in any setting where attacker can obtain Hamming Weight of each pixel through some side-channel means.
Slowing the victim's clock frequency relative to the sensor's sampling rate is a standard practice in the side-channel literature~\cite{moini_remote_2021, huegle2023power2picture}. It serves to increase the number of measurement samples per computational operation, thereby improving the signal-to-noise ratio (SNR) of the captured trace and enabling a clearer analysis of the leakage. We discuss the implications of this experimental parameter in Section~\ref{section:discussion}.

\subsection{Attack Steps --- Pixel Value Leakage}
The first methodology aims to produce a high-fidelity visual reconstruction of the victim's input image. This attack assumes that the power leakage is proportional to the actual pixel values being processed; as discussed above, this is an idealized assumption that overstates what the physical side channel provides, and the results of this methodology should be read as a best case for visual reconstruction rather than as a claim about what is recoverable in general.
Each pixel of the power side-channel trace is mapped one-to-one to its original intensity \(v\in\{0,\dots,255\}\). Unlike the Hamming Weight image, which abstracts bit-level detail, the pixel-value image retains full amplitude nuance.

\begin{figure*}[!htbp]
    \centering
    \includegraphics[width=0.8\textwidth]{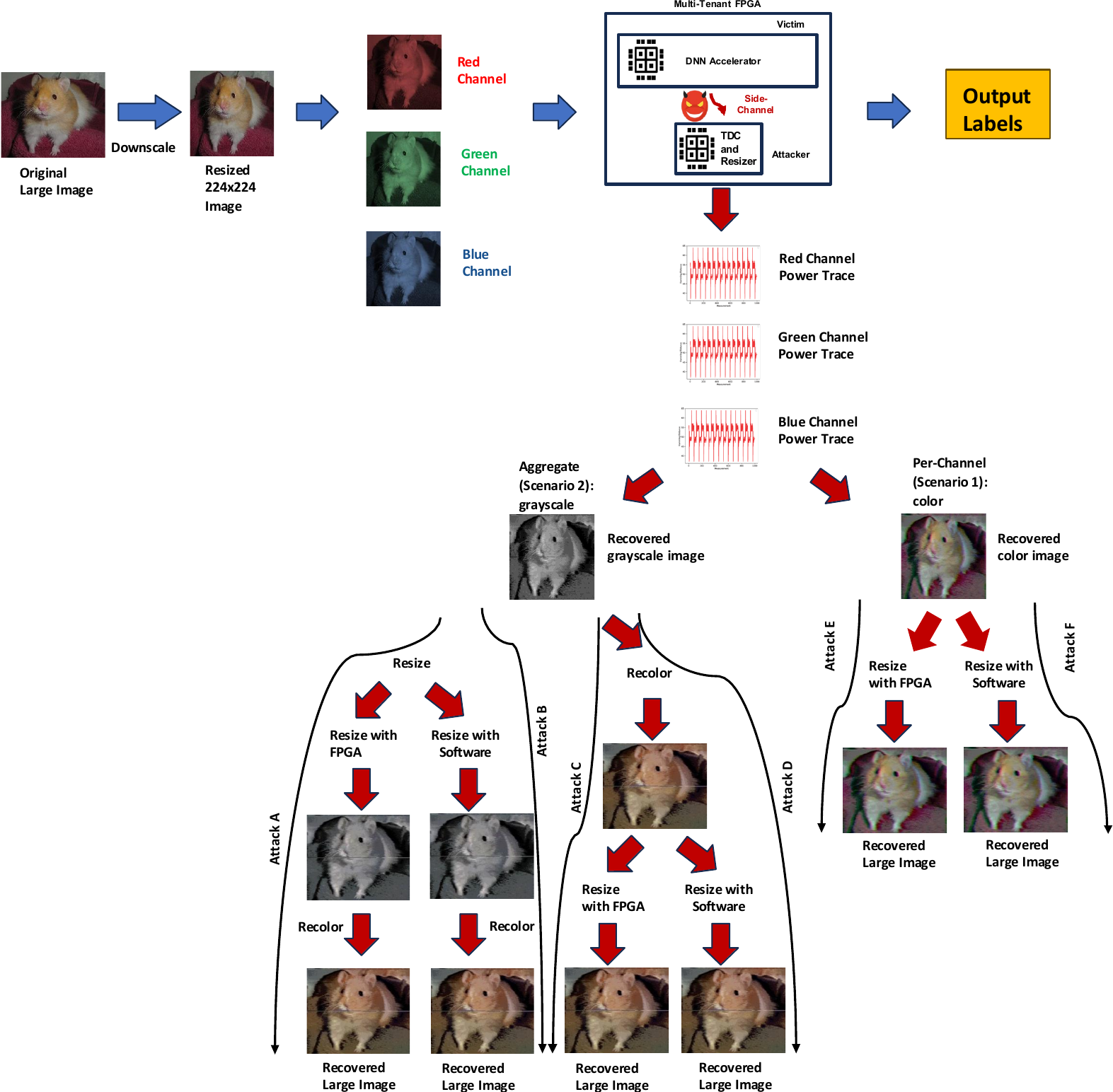}
     \caption{\small End-to-end overview of the attack process for the pixel value leakage.
    The top portion shows the capture phase: the original image is downscaled to $224\times224$, split into red/green/blue channels, processed by the victim's convolution layer accelerator on the multi-tenant FPGA, and the attacker's TDC sensor records per-channel power traces.
    Depending on whether the attacker can isolate per-channel traces, the result is either a recovered grayscale image (Aggregate / Scenario~2, left) or a recovered color image (Per-Channel / Scenario~1, right).
    The bottom portion shows the six attack strategies derived from these two outputs: Attacks A--D operate on the grayscale image (resize-then-recolor for A, B; recolor-then-resize for C, D), while Attacks E--F upscale the recovered color image directly using FPGA-based or software-based interpolation, respectively.}
    \label{fig:end_to_end_implementation_pixel}
\end{figure*}

Figure~\ref{fig:end_to_end_implementation_pixel} shows the full end-to-end pixel-value attack flow.
The top portion illustrates how, from the TDC power traces, the attacker can either recover a grayscale image (Aggregate / Scenario 2) or directly recover the colored image (Per-Channel / Scenario 1).
The bottom portion then demonstrates the six attack strategies proposed in this paper:  Attacks A--D start from the recovered grayscale image and combine resize-then-recolor (A, B) or recolor-then-resize (C, D); Attacks E and F start from the recovered color image and need only a final resize step (no recoloring required).

While raw, downscaled grayscale images might be sufficient for some automated analysis, visual reconstruction of full-scale, color images is crucial for attacks where the goal is human interpretation or public leakage of sensitive visual data (e.g., leaking a still from a protected video stream or a copyrighted photograph). This frames the upscaling/recoloring not as a distraction, but as the final step in achieving a specific adversarial goal. The goal is to reconstruct an image that is visually intelligible to a human, which is a different adversarial goal than simply classifying the image.

\subsubsection{Six Attack Scenarios}
\label{section:six_attack_scenarios}

The bottom part of Figure~\ref{fig:end_to_end_implementation_pixel} illustrates six new attack scenarios proposed in our work. As the neural network processes the input image, the attacker collects the TDC power traces. A less powerful attacker can only recover a grayscale image (bottom left), whereas a more powerful attacker, who can isolate traces for the red, green, and blue channels, can reconstruct a full-color image (bottom~right).

\paragraph{Attack A}
In Attack A, the less powerful attack scenario, the victim runs the convolution operations in parallel on all the red, green, blue channels, while attacker collects TDC measurements; in this case the attacker effectively learns only how dark or light the pixels are, and they can recover only grayscale image. Given the attacker is able to learn only a $222\times222$ grayscale image, the attacker then resizes the recovered grayscale image using the simple resizer module inside the FPGA to recover the full-scale image size. After that, a machine learning model is used to recolor the image to get the final, full-scale, colored image.

\paragraph{Attack B}
Similar to Attack A, Attack B also leverages the TDC to recover a grayscale image. The attacker then resizes the recovered grayscale image using software-based bicubic interpolation~\cite{gao2011bilinear} to recover the full-sized image. After that, the recoloring takes place using a machine learning model and produces the final full-scale, colored image.

\paragraph{Attack C}
In Attack C, similar to Attack A and Attack B, the attacker leverages the TDC to recover the $222\times222$ grayscale image. In this scenario, however, the attacker first applies a machine learning algorithm to recolor the image to get a full-color $222\times222$ image. After that, attacker uses the resizer inside the FPGA to upscale the image to get the final full-scale, colored image.

\paragraph{Attack D}
Similar to Attack C, in Attack D, the attacker leverages the TDC to recover only a $222\times222$ grayscale image.
The attacker first recolors the recovered $222\times222$ image, similar to Attack C. After that, the attacker uses software-based bicubic interpolation to upscale the image, resulting in the final full-scale, colored image.

\paragraph{Attack E}
In Attack E, we consider a more powerful scenario where the attacker can collect information about voltage changes for each of the three color channels.
 A plausible mechanism is if the victim's accelerator processes the Red, Green, and Blue channels sequentially. The attacker could synchronize their TDC capture with the start of each channel's processing block, thus isolating the power traces for each color.
The attacker thus directly recovers a colored $222\times222$ image, one recovered value per output pixel of the first convolution layer. In this scenario, the attacker then upscales the image to the original size using the resizer inside the FPGA, resulting in the final full-scale, colored image.

\paragraph{Attack F}
In Attack F, similar to Attack E, the attacker leverages the TDC to recover all 3 channels of the power traces and gets a $222\times222$ colored image.
After that, the attacker uses software-based bicubic interpolation to upscale the image, resulting in the final full-scale, colored image.

\subsection{Training and Attack Phases --- Hamming Weight Leakage}

In this subsection, first, the training phase by the attacker is described; second, the inference, or the attack, phase is described.

\subsubsection{Training}

\begin{figure*}[t]
    \centering
    \includegraphics[width=0.9\textwidth]{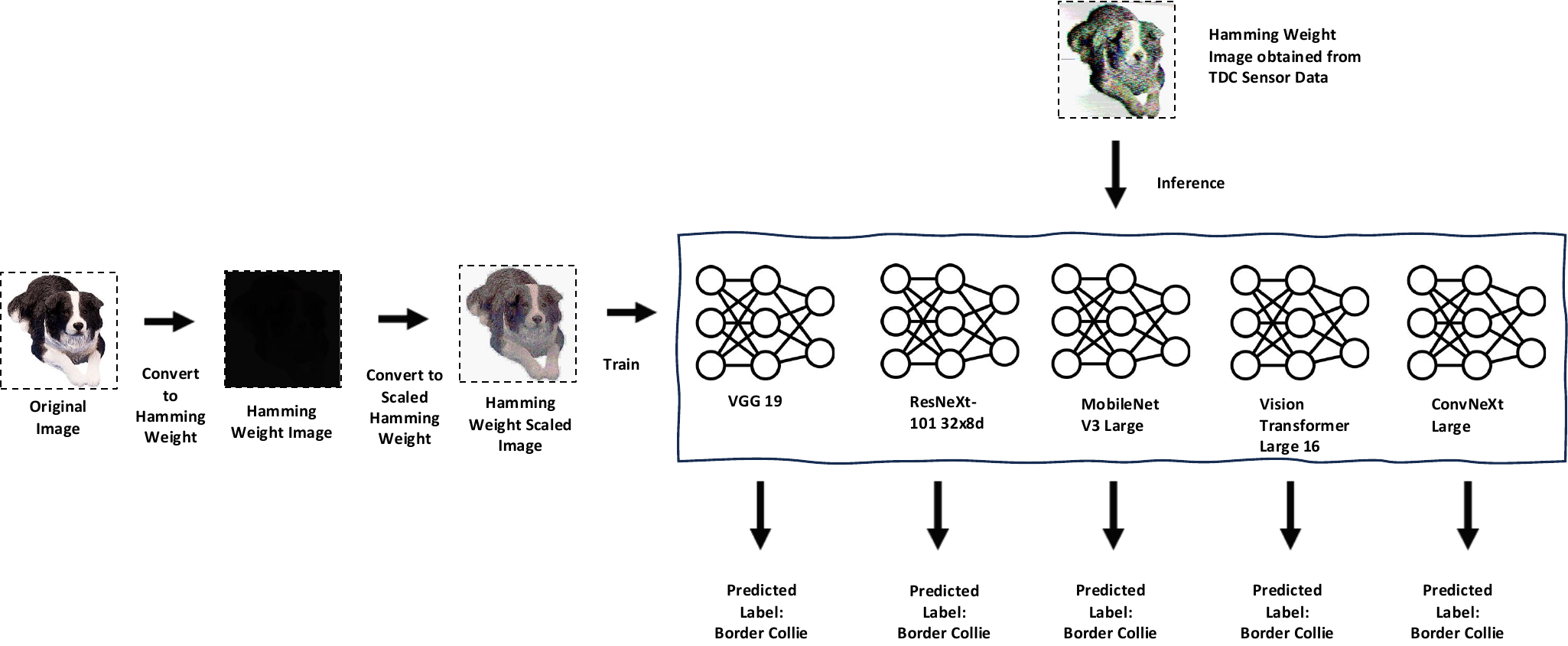}
    \caption{\small Recovering image class from only Hamming Weight data about pixels of the victim's image.}
    \label{fig:end_to_end_implementation}
\end{figure*}

Figure~\ref{fig:end_to_end_implementation} shows how the data obtained from power side channel attack in Figure~\ref{fig:end_to_end_implementation_pixel} is used to classify the input image. After the input victim image is processed by the victim neural network, the attacker collects the TDC power traces and learns the Hamming Weight version of the input victim image. In order to learn what that image is, the attacker needs to use a special neural network as shown in Figure~\ref{fig:end_to_end_implementation}.

In the left and bottom part of Figure~\ref{fig:end_to_end_implementation} we show the attacker's steps of converting ImageNet images into their Hamming Weight versions.
Since Hamming Weight of pixels with integer values 0 to 255 ranges from only 0 to 8, we have experimentally found that training ML models on images where each pixel is between 0 and 8 is very difficult. To make the images more recognizable to machine learning models, we develop a new scaled Hamming Weight version of the images, in which the Hamming Weight of each pixel is multiplied by 31. Each pixel of the resulting scaled Hamming Weight image still takes only one of 9 possible values (the Hamming weights $0$ through $8$), but their magnitudes are increased by a factor of~31. Given a pixel value \(v\in\{0,\dots,255\}\) with binary representation \(v_7 v_6 \dots v_0\), its Hamming weight and scaled Hamming Weight are defined as:
  \[
    HW(v) = \sum_{i=0}^7 v_i,
    \quad
    sHW(v) = 31 \times HW(v).
  \]

We experimented with scaling factors in the range [1, 31] and found that a factor of 31 yielded the highest classification accuracy for our models on a validation set, as it stretched the 0-8 range of Hamming weights to a 0-248 range, better utilizing the dynamic range expected by standard CNNs.
The scaling factor of 31 was chosen empirically, as it provided a significant boost in classification accuracy compared to unscaled weights, while avoiding excessively large pixel values that offered no further benefit. This scaling expands the dynamic range of the image, making it more suitable for training standard CNN architectures.
The scaled Hamming Weight images are fed into $5$ different neural networks for training. This is done for all images from the ImageNet database, resulting in a new set of $5$ neural networks that can classify scaled Hamming Weight images.

\subsubsection{Inference}

With the trained attacker neural networks, the attacker can take the Hamming Weight image obtained from the TDC sensors, and input it to the attacker neural networks. The networks then classify the image and the attacker learns what was the input to the victim's neural network accelerator co-located on the same FPGA as the attacker's (TDC) circuit.

\subsection{Experimental Setup}
This section details the hardware, software, and procedures used to execute and evaluate the ML-assisted side-channel attacks.

\subsubsection{Hardware Configuration}

\paragraph{FPGA Platform}In our work, we use the PYNQ-Z2 FPGA board, which is built around a Xilinx ZYNQ SoC device, featuring a dual-core ARM Cortex-A9 processor coupled with FPGA fabric. Within this configuration, the dual-core ARM Cortex-A9 processor is designated as the Processing System (PS), while the FPGA fabric is identified as Programmable Logic~(PL).

\paragraph{FPGA Resource Utilization}
The Xilinx ZYNQ SoC on the PYNQ-Z2 contains 53,200 Look-Up Tables (LUTs), 106,400 Flip-Flops (FFs), and 140 Block RAMs (BRAMs).
The victim's convolution layer accelerator is a lightweight implementation, occupying approximately 16\% of the FPGA's LUTs, 0.68\% of its FFs and 46\% of its BRAMs.
The attacker's TDC sensor is designed to be minimal, consuming less than 7\% of the total available logic resources. While this low footprint shows that the sensor is inexpensive in area and easily fits alongside a victim circuit, we caution that low resource usage alone does not establish that a useful attack is possible: whether the attacker actually obtains exploitable leakage also depends on placement, routing, coupling through the power distribution network (PDN), and floorplanning constraints, all of which govern how strongly the sensor observes the victim's switching activity.

\subsubsection{Victim Circuit}
The victim hardware module comprises a convolution module developed by the authors using Verilog. This module is designed to perform convolution operations on input images, specifically handling images of size 224x224 pixels. The output of the network is an image of size 222x222 pixels, which is fed into next layer; note the size change, reflecting the reduction in dimensions typically associated with convolutional layers due to the absence of padding.

In our implementation, only the first layer of the convolutional network is realized, focusing exclusively on the convolution operation applied to the input image. This choice allows us to isolate and analyze the effects of convolution on the input data without the complexities introduced by additional network layers.

The input image is stored in BRAM within the FPGA, providing fast and efficient access to the data during the convolution process. The use of BRAM is crucial for maintaining high performance and low latency, which are essential for real-time image processing tasks.

The Verilog code for the convolution module includes mechanisms for loading the input image into BRAM, performing the convolution operation using a specified kernel, and storing the resulting output image. The kernel applied in the convolution operation is designed to extract features from the input image, such as edges and textures, which are fundamental to subsequent image processing~tasks.

\paragraph{Convolution datapath and data representation.} To make the leakage analysis reproducible, we specify the datapath of the implemented first convolution layer. Each input pixel is an $8$-bit unsigned integer in $\{0,\dots,255\}$, and the three color channels (R, G, B) are each convolved with a $3\times3$ kernel. Depending on the scenario analyzed in Section~\ref{sec_basic_power_side_setup},the channels are convolved either sequentially (yielding per-channel leakage) or in parallel (yielding aggregate leakage). The layer uses a stride of $1$ and no zero-padding, which is why a $224\times224$ input produces a $222\times222$ output. Each output pixel is the sum of nine multiply-accumulate (MAC) operations between a $3\times3$ neighborhood of input pixels and the kernel weights; representative learned weights for the three channels are listed in Tables~\ref{table_vgg_19_red}--\ref{table_vgg_19_blue} and lie in the approximate range $[-0.24,\,0.23]$, so the coefficients are stored as \emph{signed fixed-point} values rather than integers. Following standard VGG-style preprocessing, the per-image mean is subtracted from each pixel before convolution. It is the switching activity of these MAC operations, dominated by the Hamming Weight of the $8$-bit pixel operands, that the co-located TDC observes through the shared PDN.

The kernel coefficients are stored as 32-bit signed fixed-point numbers with a 16-bit fractional part (scale factor $2^{16}$): each floating-point weight $w$ is quantized as $\mathrm{round}(w\cdot 2^{16})$, i.e., a signed Q15.16 format. Each 8-bit pixel operand multiplies a coefficient in a full $32\times32\rightarrow64$-bit signed multiplier; the 64-bit product is right-shifted by 16 bits (dropping the fractional part by truncation) to recover an integer-valued partial product, and the nine partial products are summed in a 32-bit signed accumulator to form the layer output.

\begin{table}[t!]
    \centering
    \small
    \caption{\small VGG-19 first convolution layer $3\times3$ kernel values for the red, green, and blue channels.}
    \label{table_vgg_19_kernels}
    \begin{subtable}[b]{0.31\linewidth}
        \centering
        \caption{Red channel}
        \label{table_vgg_19_red}
        \begin{tabular}{|*{3}{c|}}
            \hline
            $-0.053$ & $-0.049$ & $-0.068$\\ \hline
            $0.015$  & $0.045$  & $0.002$ \\ \hline
            $0.036$  & $0.02$   & $0.02$  \\ \hline
        \end{tabular}
    \end{subtable}
    \hfill
    \begin{subtable}[b]{0.31\linewidth}
        \centering
        \caption{Green channel}
        \label{table_vgg_19_green}
        \begin{tabular}{|*{3}{c|}}
            \hline
            $0.017$ & $0.055$ & $-0.006$\\ \hline
            $0.142$ & $0.227$ & $0.138$ \\ \hline
            $0.12$  & $0.2$   & $0.092$ \\ \hline
        \end{tabular}
    \end{subtable}
    \hfill
    \begin{subtable}[b]{0.31\linewidth}
        \centering
        \caption{Blue channel}
        \label{table_vgg_19_blue}
        \begin{tabular}{|*{3}{c|}}
            \hline
            $-0.045$ & $0.013$ & $-0.014$\\ \hline
            $0.06$   & $0.14$  & $0.054$ \\ \hline
            $-0.001$ & $0.058$ & $-0.03$ \\ \hline
        \end{tabular}
    \end{subtable}
\end{table}

\subsubsection{Attacker Circuit}
The attacker's hardware module is a Time-to-Digital Converter (TDC), an open-source Tunable Dual Polarity sensor designed specifically for performing side-channel attacks~\cite{tuneable_tdc}. The TDC's operating principle and its connection to the victim's power consumption via the shared PDN are described in detail in Section~\ref{sec_basic_power_side_setup}. This particular TDC design is based on the tunable TDC architecture from Drewes et al.~\cite{tuneable_tdc}, which has been proven to be highly effective in side-channel attack scenarios. The tunability allows the sensor to be calibrated for optimal sensitivity to voltage fluctuations in a specific FPGA placement, enhancing its ability to capture subtle variations in the victim's power consumption.

The TDC module measures time intervals between digital events with high precision. The dual polarity feature of the sensor enables it to detect both positive and negative voltage transitions on the PDN, providing a more comprehensive view of the victim's switching activity. In the context of our attack, each TDC sample captures a snapshot of the instantaneous voltage on the shared PDN, which is correlated with the Hamming Weight of the pixel being processed by the victim's convolution engine at that moment, as explained in Section~\ref{sec_basic_power_side_setup}.

\subsubsection{Image Preparation and Splicing}
\label{sec:image_prep_splicing}
A key limitation of our current experimental setup is the data-acquisition bandwidth from the on-chip TDC sensor to the host system. Each full TDC capture sequence is constrained by the TDC buffer to \(2^{20}\) samples \((1{,}048{,}576)\). At a \(50\,\mathrm{MHz}\) TDC clock \((20\,\mathrm{ns/sample})\), this corresponds to a \(\approx 21\,\mathrm{ms}\) window of signal. This single window can only record the power side-channel information corresponding to approximately \emph{half} of a \(224\times 224\) input image, but not the full frame, since the convolution requires approximately twice that duration. To reconstruct a complete image, we therefore employ a two-step process. First, the \emph{top half} of the victim’s input \((112\times 224)\) is processed and the corresponding TDC trace is captured. Second, the process is repeated for the \emph{bottom half}. The two resulting power-trace captures are transferred to a host computer, where they are manually stitched to form the data for a complete \(224\times 224\) image. This manual, two-part acquisition and stitching procedure is time-intensive; its duration is dominated by streaming each \(2^{20}\)-sample, 32-bit trace \((4\,\mathrm{MiB})\) over a limited-bandwidth host interface rather than by on-chip computation.

In our setup the two halves are obtained by having the victim process the image and re-running the (deterministic) inference a second time, with the attacker's fixed-length capture window aligned to the first half on one run and the second half on the next; exact stitching therefore requires the victim to process the \emph{identical} input more than once. A video or streaming pipeline offers repeated inferences, but successive frames are only approximately similar, so using two different frames would introduce inter-frame reconstruction error that we do not characterize here; we therefore present repeated-identical-input as the setting in which the two-half stitching is exact, and treat the video case as an approximation whose error would need separate evaluation. This requirement disappears with a larger on-chip buffer: writing the TDC samples to on-chip BRAM/UltraRAM and reading them back after the inference completes (rather than streaming them live over the host link) would allow a single-pass capture of a full-frame inference. Such an on-chip-buffered, streaming acquisition---which also removes the I/O bottleneck discussed below---is the most direct way to eliminate both the two-half splicing and the long per-image acquisition time, and we identify it as the primary engineering improvement for a more realistic implementation.

We note, however, that on-chip buffering is capacity-limited on our specific board. A single full-length trace of $2^{20}$ samples at $32$~bits is $4$~MiB, which exceeds the roughly $4.9$~Mb ($\approx0.6$~MiB) of block RAM available on the PYNQ-Z2's Zynq-7020. On this board the realistic single-pass path is therefore not pure on-chip BRAM but streaming the samples into PS-attached DDR via a DMA engine, using the block RAM only as a small FIFO; a genuinely on-chip, UltraRAM-buffered capture would require a larger device (e.g., a UltraScale+ FPGA with UltraRAM). Either way, the acquisition rate is bounded by the chosen off-chip link rather than by sensing.

\subsubsection{Experimental Procedure and Runtimes}
\label{sec_run_times}
The end-to-end process for recovering a single victim image is lengthy due to the data acquisition and manual stitching steps described above.
The end-to-end runtime is dominated by off-chip I/O rather than on-chip computation.

 \paragraph{Data Acquisition}
We use \(2^{20}\) samples per capture because this is the maximum contiguous window supported by the TDC buffer, not because the computation demands that length. At \(50\,\mathrm{MHz}\) sampling, a capture spans \(\approx 21\,\mathrm{ms}\), easily covering the convolution burst for a half image while also recording surrounding memory and control activity. With 32-bit samples, each capture is \(4\,\mathrm{MiB}\), and the effective throughput of the host interface yields \(\approx 5.5\,\mathrm{min}\) to transfer one trace \emph{per channel per half image}. Aggregated over three channels and two halves, this results in \(\approx 33\,\mathrm{min}\) per full image. Compute and on-chip capture contribute negligibly to the wall-clock time relative to I/O.

We want to be clear that this slow acquisition is a property of our specific academic prototype, not of the attack itself. The bottleneck is the low-bandwidth link we use to move raw samples from the FPGA to the host; the sensing and on-chip capture are effectively instantaneous by comparison. Several standard interfaces would remove this cost by orders of magnitude: an FPGA debug/JTAG bridge, a PCIe-attached FPGA (as in commercial cloud cards), or---most simply---buffering the samples in on-chip BRAM/UltraRAM and bursting them off-chip only after the inference completes, as described in Section~\ref{sec:image_prep_splicing}. We did not have access to a PCIe-connected platform for this study, but we expect any of these paths to reduce per-image acquisition from tens of minutes to well under a second, which is what would make the larger-scale, end-to-end hardware evaluation practical.

\paragraph{Image Reconstruction and Classification}
Once the data is acquired, the software-based processing begins. The first step, reconstructing a coherent image from the raw TDC traces, requires a careful data alignment process to correct for timing jitter and systematic noise, which can take several minutes per image. Following this reconstruction, the subsequent steps of upscaling, ML-based recoloring, and final classification are computationally faster, typically completing in under one minute combined. While the software processing requires time, it is the time-intensive nature of the manual data acquisition that remains the most significant bottleneck in our experimental workflow. For this reason, our evaluation was focused on a carefully selected set of 9 test images that represent high- and low-contrast scenarios, as detailed in Section~\ref{sec:test_image_selection}.

\subsubsection{Test Images Used}

\begin{figure*}[t]
    \centering
    \begin{subfigure}[b]{0.31\textwidth}
        \centering
        \includegraphics[width=0.55\linewidth]{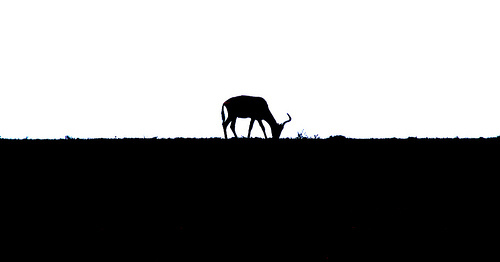}
        \caption{\footnotesize Original Image, \\Top Image 1}
        \label{fig:original_top_image_1}
    \end{subfigure}
    \hfill
    \begin{subfigure}[b]{0.31\textwidth}
        \centering
        \includegraphics[width=0.55\linewidth]{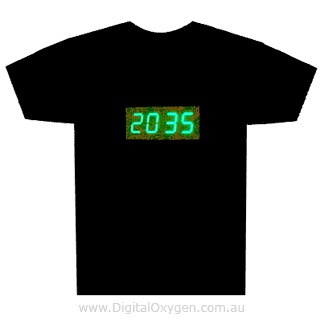}
        \caption{\footnotesize Original Image, \\Top Image 2}
        \label{fig:original_top_image_2}
    \end{subfigure}
    \hfill
    \begin{subfigure}[b]{0.31\textwidth}
        \centering
        \includegraphics[width=0.55\linewidth]{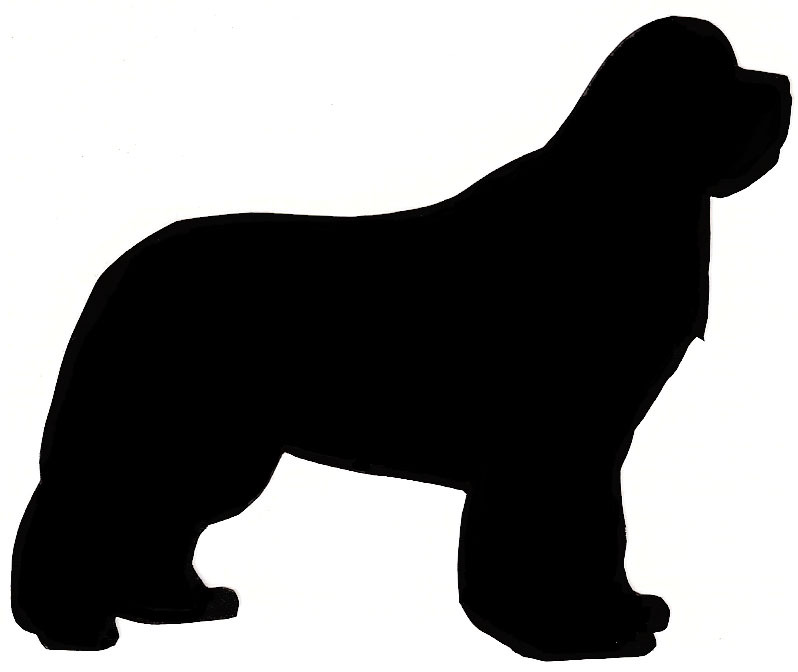}
        \caption{\footnotesize Original Image, \\Top Image 3}
        \label{fig:original_top_image_3}
    \end{subfigure}

    \vspace{1em} %

    \begin{subfigure}[b]{0.31\textwidth}
        \centering
        \includegraphics[width=0.55\linewidth]{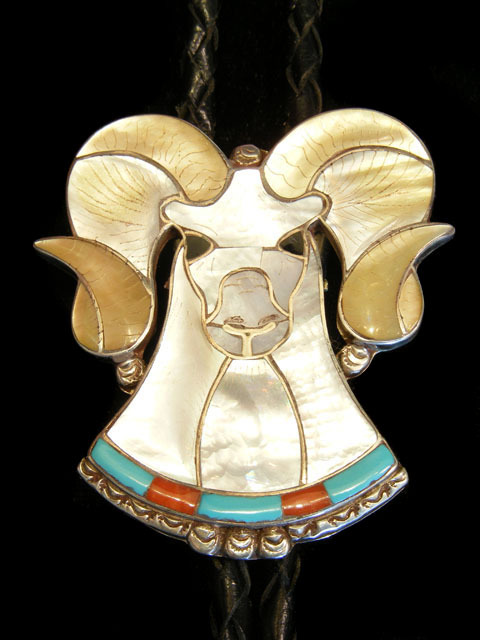}
        \caption{\footnotesize Original Image, \\Top Image Colored 1}
        \label{fig:original_top_image_color_1}
    \end{subfigure}
    \hfill
    \begin{subfigure}[b]{0.31\textwidth}
        \centering
        \includegraphics[width=0.55\linewidth]{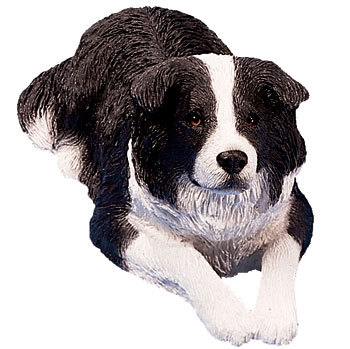}
        \caption{\footnotesize Original Image, \\Top Image Colored 2}
        \label{fig:original_top_image_color_2}
    \end{subfigure}
    \hfill
    \begin{subfigure}[b]{0.31\textwidth}
        \centering
        \includegraphics[width=0.55\linewidth]{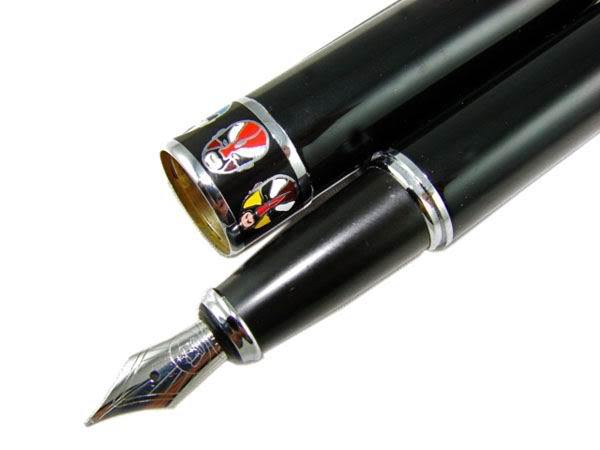}
        \caption{\footnotesize Original Image, \\Top Image Colored 3}
        \label{fig:original_top_image_color_3}
    \end{subfigure}

    \vspace{1em} %

    \begin{subfigure}[b]{0.31\textwidth}
        \centering
        \includegraphics[width=0.55\linewidth]{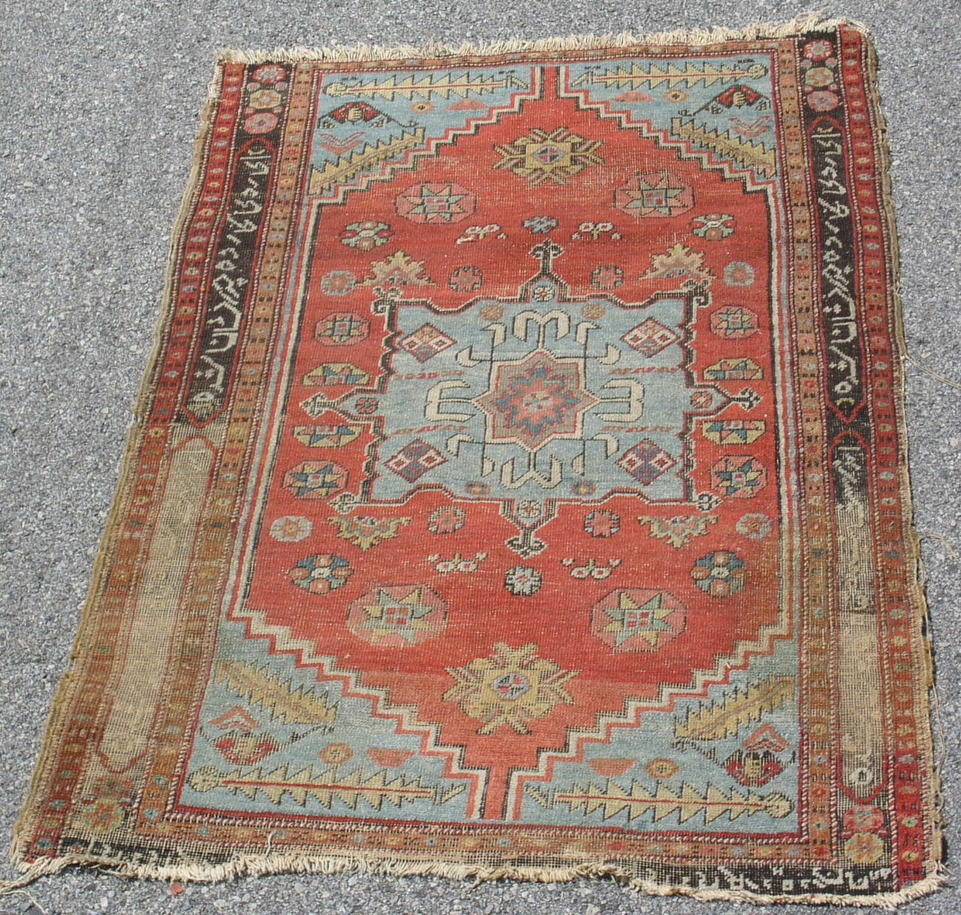}
        \caption{\footnotesize Original Image, \\Bottom Image Colored 1}
\label{fig:original_bottom_image_color_1}
    \end{subfigure}
    \hfill
    \begin{subfigure}[b]{0.31\textwidth}
        \centering
        \includegraphics[width=0.55\linewidth]{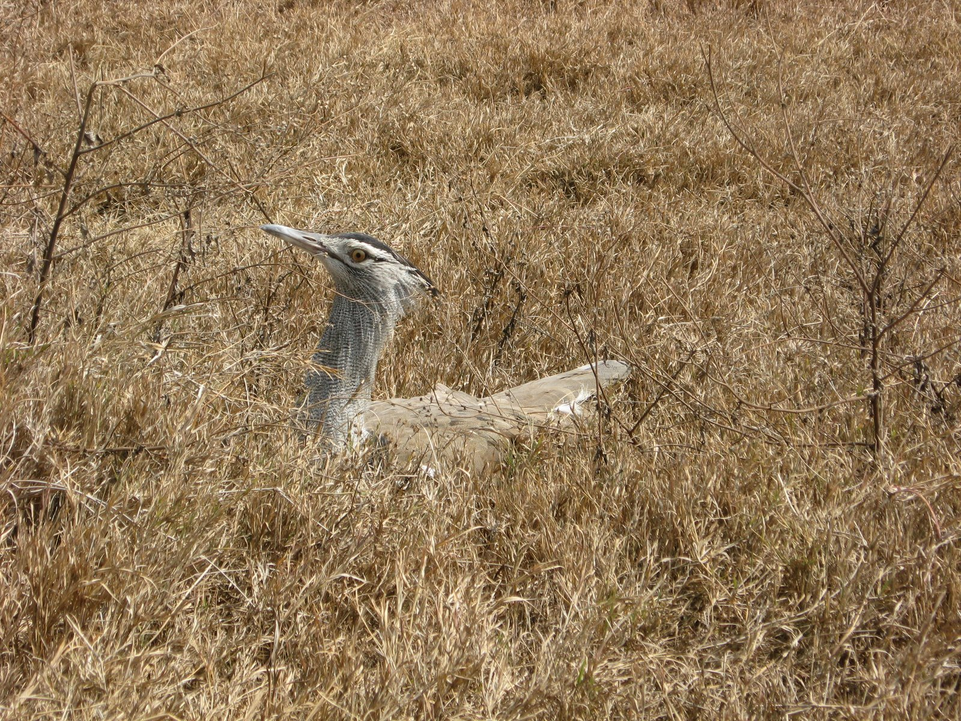}
        \caption{\footnotesize Original Image, \\Bottom Image Colored 2}
        \label{fig:original_bottom_image_color_2}
    \end{subfigure}
    \hfill
    \begin{subfigure}[b]{0.31\textwidth}
        \centering
        \includegraphics[width=0.55\linewidth]{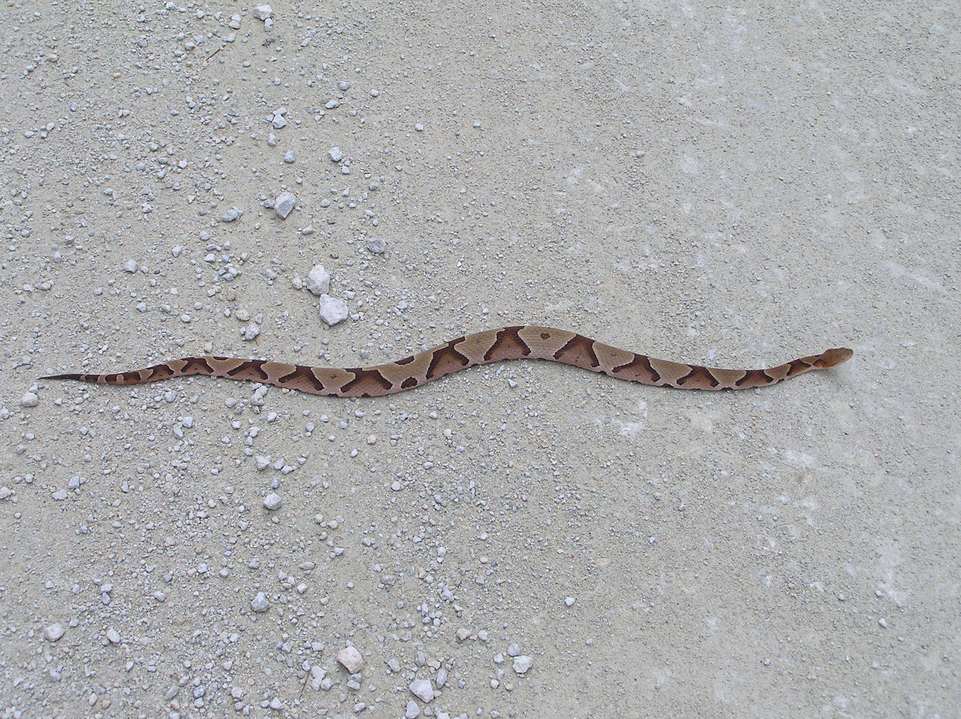}
        \caption{\footnotesize Original Image, \\Bottom Image Colored 3}
\label{fig:original_bottom_image_color_3}
    \end{subfigure}
    \caption{\small Original ImageNet images used in experimentation and testing, images were selected based on their contrast values. The labels ``Top Image'', ``Top Image Colored'', and ``Bottom Image Colored'' are explained in Section~\ref{sec:test_image_selection}.}
    \label{fig:original_image_net_images_first}
\end{figure*}

\begin{figure*}[!htbp]
    \centering
    \begin{subfigure}[b]{0.15\textwidth}
        \centering
        \includegraphics[width=0.75
        \linewidth]{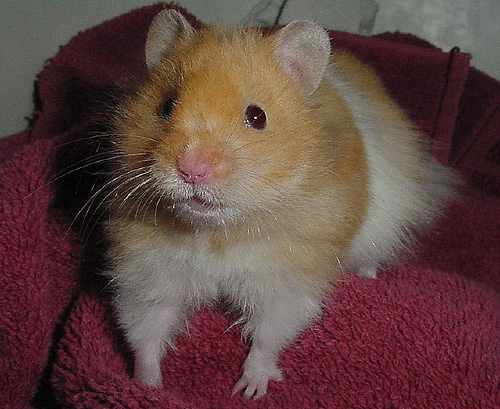}
        \caption{Image 1}
        \label{fig:image_net_original_1}
    \end{subfigure}
    \hfill
    \begin{subfigure}[b]{0.15\textwidth}
        \centering
        \includegraphics[width=\linewidth]{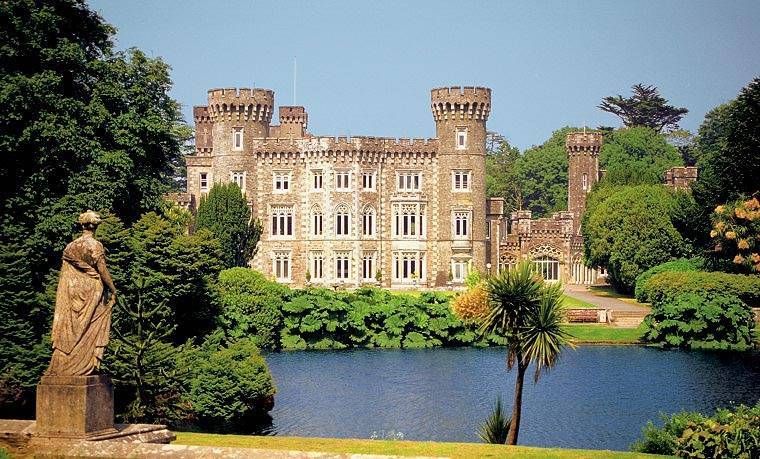}
         \caption{Image 2}
         \label{fig:image_net_original_2}
    \end{subfigure}
    \hfill
    \begin{subfigure}[b]{0.15\textwidth}
        \centering
    \includegraphics[width=0.8\linewidth]{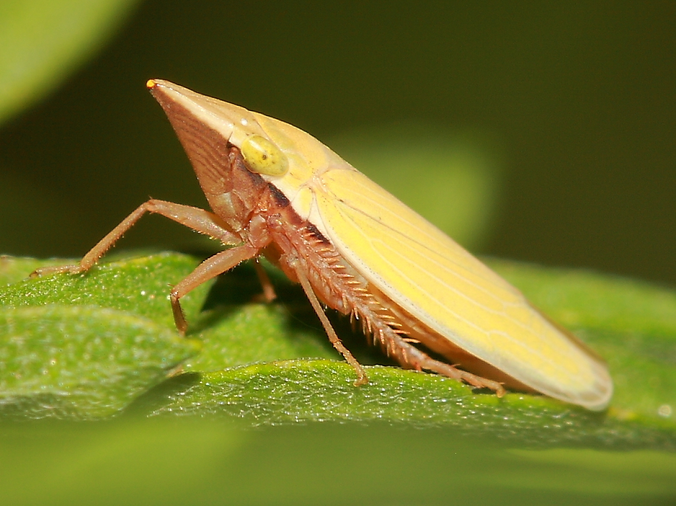}
        \caption{Image 3}
        \label{fig:image_net_original_3}
    \end{subfigure}
    \hfill
    \begin{subfigure}[b]{0.2\textwidth}
        \centering
        \includegraphics[width=0.65\linewidth]{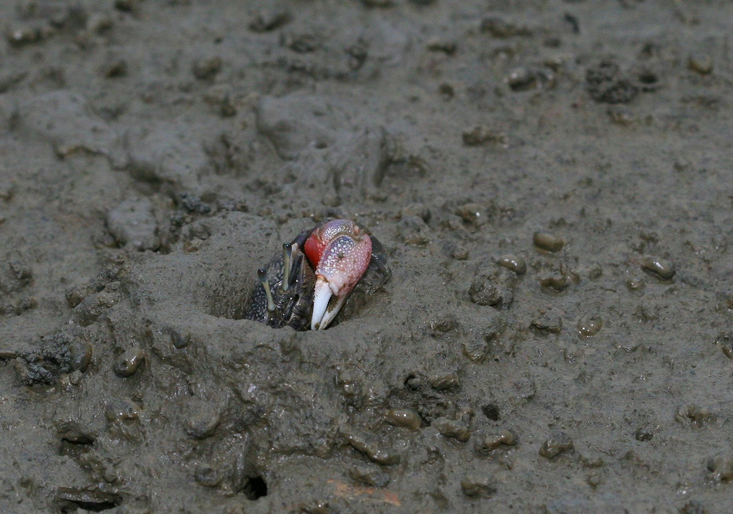}
        \caption{Image 4}
        \label{fig:image_net_original_4}
    \end{subfigure}
    \hfill
    \begin{subfigure}[b]{0.2\textwidth}
        \centering
        \includegraphics[width=0.65\linewidth]{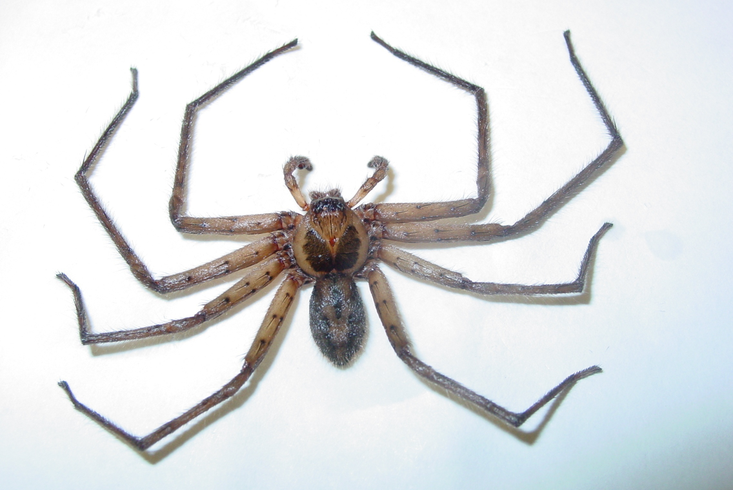}
        \caption{Image 5}
        \label{fig:image_net_original_5}
    \end{subfigure}
    \caption{\small Test images used in this work for pixel value leakage.}
    \label{fig:original_image_net_images}
\end{figure*}

For the Hamming Weight leakage experiments, we use 9 test images from the ImageNet dataset. The images are shown in Figure~\ref{fig:original_image_net_images_first}. The labels ``Top Image'', ``Top Image Colored'', and ``Bottom Image Colored'' are explained in Section~\ref{sec:test_image_selection}.
For the pixel value leakage experiments, we use 5 test images from the ImageNet dataset. The images are
shown in Figure~\ref{fig:original_image_net_images}. We use short names Image 1 to Image 5.

\subsubsection{Resizer}
A simple bilinear interpolation~\cite{kirkland2010bilinear} resizer was implemented in the FPGA fabric for the attacker's use in some scenarios.  Resizing is necessary only if the attacker's goal is to restore the image to its original dimensions for full visual fidelity. If the goal were just to get a small, recognizable thumbnail, it would not be necessary. Further, note that the resizer could be done in software as well by the attacker.
The software-based alternative used in Attacks B, D, and F employs bicubic interpolation via the Python Imaging Library (PIL)~\cite{pillow_lib}.

\section{Test Image Selection and Attacker's ML Algorithms}

\begin{figure*}[t]
    \centering
    \begin{subfigure}[b]{0.31\textwidth}
        \centering
        \includegraphics[width=0.55\linewidth]{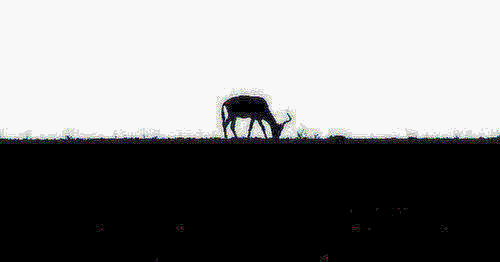}
        \caption{\footnotesize Hamming Weight Scaled Image, Top Image 1}
        \label{fig:original_top_image_1_hw_scaled}
    \end{subfigure}
    \hfill
    \begin{subfigure}[b]{0.31\textwidth}
        \centering
        \includegraphics[width=0.55\linewidth]{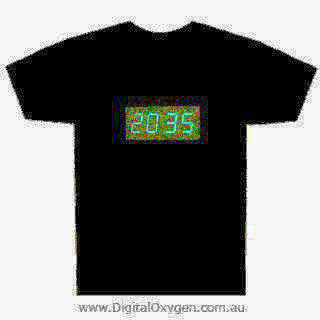}
        \caption{\footnotesize Hamming Weight Scaled Image, Top Image 2}
        \label{fig:original_top_image_2_hw_scaled}
    \end{subfigure}
    \hfill
    \begin{subfigure}[b]{0.31\textwidth}
        \centering
        \includegraphics[width=0.55\linewidth]{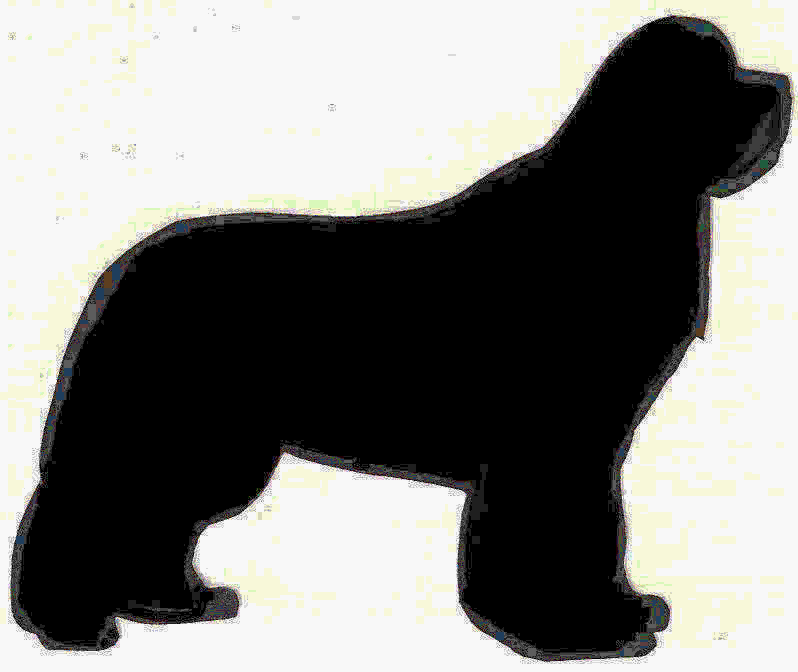}
        \caption{\footnotesize Hamming Weight Scaled Image, Top Image 3}
        \label{fig:original_top_image_3_hw_scaled}
    \end{subfigure}

    \vspace{1em} %

    \begin{subfigure}[b]{0.31\textwidth}
        \centering
        \includegraphics[width=0.55\linewidth]{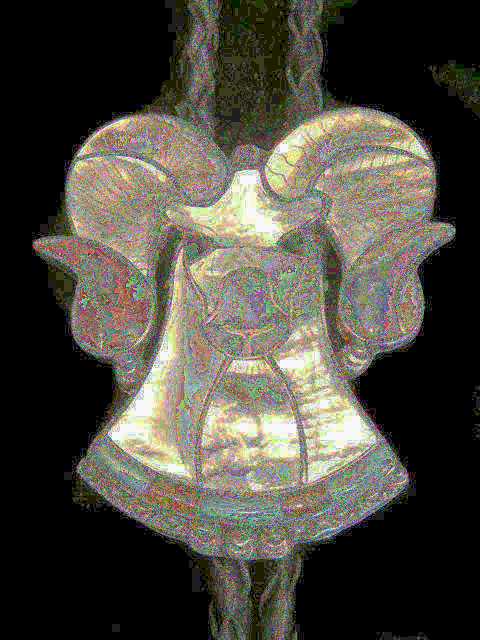}
        \caption{\footnotesize Hamming Weight Scaled Image, Top Image Colored 1}
        \label{fig:original_top_image_color_1_hw_scaled}
    \end{subfigure}
    \hfill
    \begin{subfigure}[b]{0.31\textwidth}
        \centering
        \includegraphics[width=0.55\linewidth]{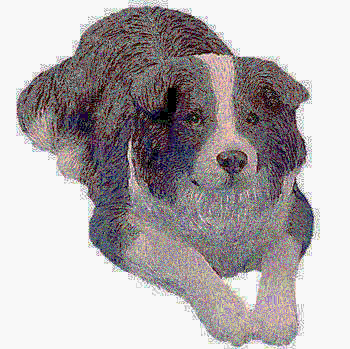}
        \caption{\footnotesize Hamming Weight Scaled Image, Top Image Colored 2}
        \label{fig:original_top_image_color_2_hw_scaled}
    \end{subfigure}
    \hfill
    \begin{subfigure}[b]{0.31\textwidth}
        \centering
        \includegraphics[width=0.55\linewidth]{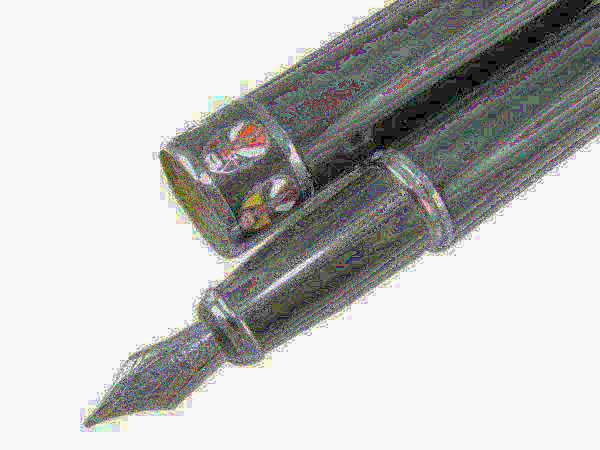}
        \caption{\footnotesize Hamming Weight Scaled Image, Top Image Colored 3}
        \label{fig:original_top_image_color_3_hw_scaled}
    \end{subfigure}

    \vspace{1em} %

    \begin{subfigure}[b]{0.31\textwidth}
        \centering
        \includegraphics[width=0.55\linewidth]{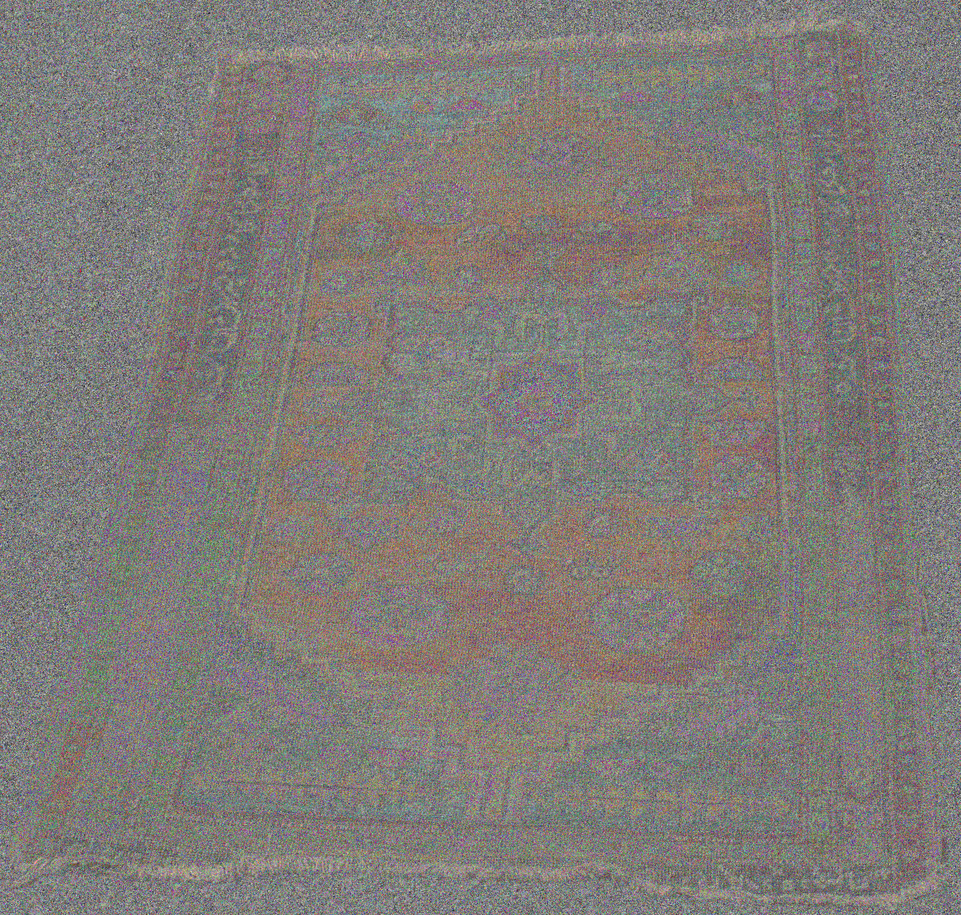}
        \caption{\footnotesize Hamming Weight Scaled Image, Bottom Image Colored 1}
        \label{fig:original_bottom_image_color_1_hw_scaled}
    \end{subfigure}
    \hfill
    \begin{subfigure}[b]{0.31\textwidth}
        \centering
        \includegraphics[width=0.55\linewidth]{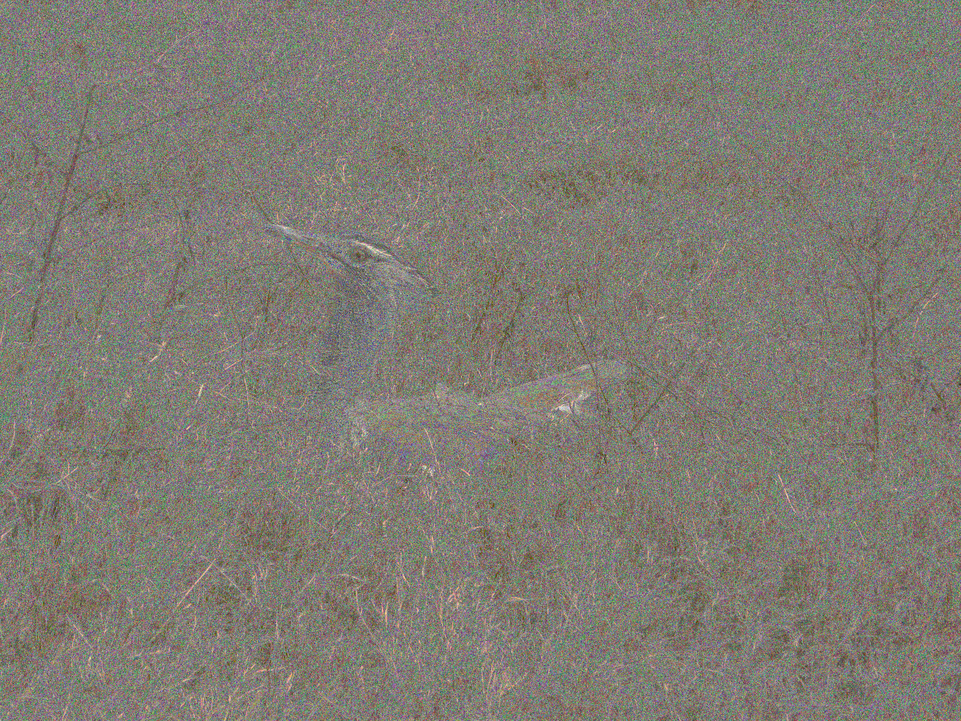}
        \caption{\footnotesize Hamming Weight Scaled Image, Bottom Image Colored 2}
        \label{fig:original_bottom_image_color_2_hw_scaled}
    \end{subfigure}
    \hfill
    \begin{subfigure}[b]{0.31\textwidth}
        \centering
        \includegraphics[width=0.55\linewidth]{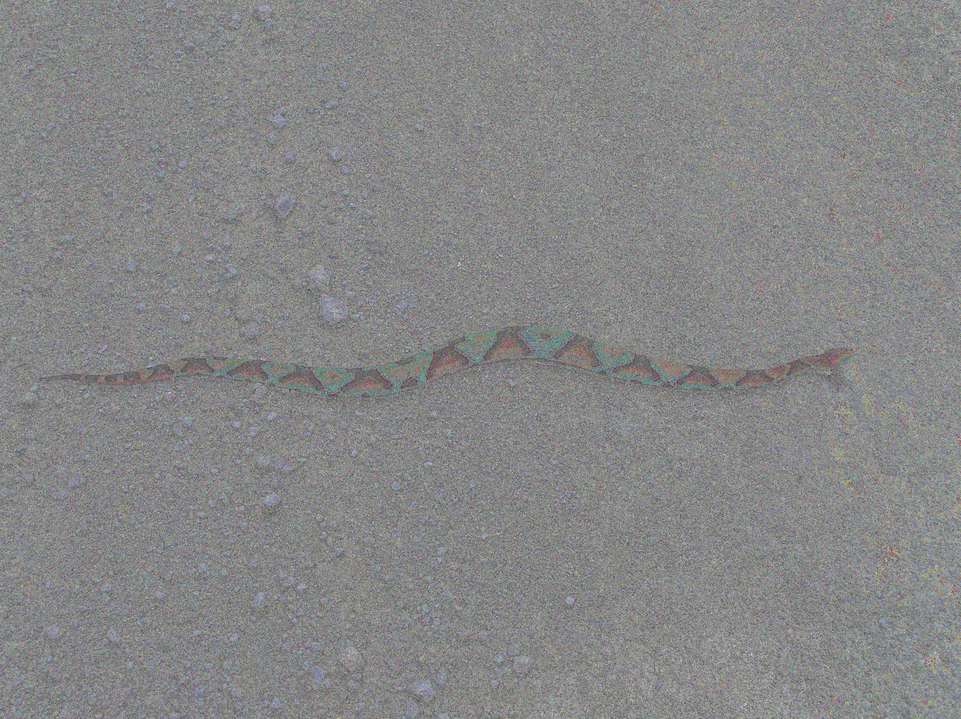}
        \caption{\footnotesize Hamming Weight Scaled Image, Bottom Image Colored 3}
        \label{fig:original_bottom_image_color_3_hw_scaled}
    \end{subfigure}

    \caption{\small Scaled Hamming Weight image versions of images shown in Figure~\ref{fig:original_image_net_images_first}.}
    \label{fig:original_image_net_images_hw_scaled}
\end{figure*}

For our attack scenario, the TDC is used to capture the Hamming Weight of each pixel, but not the actual integer value of the pixels of the victim's image. As a result, the attacker needs a machine learning model to classify the Hamming Weight images. There is today no such model; existing machine learning models are all trained on black and white images, or on colored images. We are the first to train and test image inference on Hamming Weight versions of the images.

\subsection{Test Image Selection and Analysis of Image Contrast}
\label{sec:test_image_selection}

\begin{algorithm}[t]
\caption{Image Standard Deviation Calculation}
\footnotesize
\label{alg:image_std}
\begin{algorithmic}
\Require Resized image $I$ of size $(224, 224)$
\Ensure $overall\_std$
\If{Grayscale image}
    \State $overall\_std \gets \text{std}(I)$
\Else
    \State $std\_devs \gets [\text{std}(I_c) \ \text{for each channel} \ I_c]$
    \State $overall\_std \gets \text{mean}(std\_devs)$
\EndIf
\State \Return $overall\_std$
\end{algorithmic}
\end{algorithm}

\begin{algorithm}[t]
\caption{Std Dev, Histogram, and Frequency Threshold}\label{alg:std_hist}
\begin{algorithmic}
\Require Resized image $I$ of size $(224, 224)$
\Ensure $overall\_std$, $bins\_above\_threshold$
\If{Grayscale}
    \State $std\_devs \gets \text{std}(I)$
    \State $hist \gets \text{histogram}(I)$
\Else
    \State $std\_devs \gets [\text{std}(I_c) \ \text{for each channel} \ I_c]$
    \State $hist \gets \text{sum of RGB histograms}$
\EndIf
\State $overall\_std \gets \text{mean}(std\_devs)$
\State $bins\_above\_threshold \gets \text{sum}(hist > 1000)$
\State \Return $overall\_std$, $bins\_above\_threshold$
\end{algorithmic}
\end{algorithm}

To select a subset of ImageNet images for testing, we analyzed the contrast of various images. We calculated the contrast for each image in the ImageNet database by computing the standard deviation of the pixel values. For each image, we determined the standard deviation for each color channel and averaged them to obtain a comprehensive contrast metric. Algorithm~\ref{alg:image_std} implements this process.

Based on this analysis, we divided the images into three categories: top 3  contrast images, top 3 color contrast images, and bottom 3 contrast images. The top 3  contrast images, which had the highest contrast values, were primarily black and white, although they still had three channels of red, green, and blue data. The bottom 3 contrast images had low contrast and were predominantly~colored.

To identify the top 3 color contrast images, we used a combination of standard deviation and histogram analysis. We computed histogram values for each image by summing the histograms of each channel, with 256 bins representing pixel values from 0 to 255. We then compared this combined histogram value against a frequency threshold of 1000 to determine the number of bins exceeding this~threshold.

If the combined bin count was 0, it indicated that the image was mostly black and white. However, when the combined bin count was 1 or greater, it identified the top color contrast images we required for testing. Similarly, Algorithm~\ref{alg:std_hist} implements this process.
Figure~\ref{fig:original_image_net_images_first} shows the images with the top 3 contrasting, top 3 colored contrasting, and bottom 3 contrasting images. Their corresponding Hamming Weight scaled version can be seen in Figure~\ref{fig:original_image_net_images_hw_scaled}.

\subsection{Attacker's Machine Learning Algorithms and Training}

\begin{table*}[t]
    \centering
    \caption{\small Model training parameters for the attacker's ML models used to try to classify victim's images only from the recovered Hamming Weight data. Convolutional models (VGG19, ResNeXt-101, MobileNet~V3) use SGD with a step-decay schedule; the transformer and ConvNeXt models use AdamW with a cosine schedule and linear warmup, as these architectures optimize poorly under plain SGD. All models use weight decay and, for the AdamW models, automatic mixed precision.}
    \label{tab:training}
    \small
    \begin{adjustbox}{width=0.99\textwidth}
    \begin{tabular}{|p{3cm}|p{1.6cm}|p{1.5cm}|p{1.6cm}|p{1.6cm}|p{1.8cm}|p{1.6cm}|}
        \hline
        \textbf{Models} & \textbf{Optimizer} & \textbf{Batch Size} & \textbf{Momentum} & \textbf{Learning Rate} & \textbf{Weight Decay} & \textbf{Epochs} \\
        \hline \hline
         VGG19 & SGD & 32 & 0.9 & 0.001 & 0.0001 & 100, 219 \\ \hline
         ResNeXt-101 32x8d & SGD & 32 & 0.9 & 0.001 & 0.0001 & 100 \\ \hline
         MobileNet V3 Large & SGD & 128 & 0.9 & 0.064 & 0.0001 & 100 \\ \hline
         Vision Transformer Large 16 & AdamW & 512 & n/a & 0.0003 & 0.05 & 100, 131 \\ \hline
         ConvNeXt Large & AdamW & 128 & n/a & 0.001 & 0.05 & 90 \\ \hline
    \end{tabular}
    \end{adjustbox}
\end{table*}

We train 5 different machine learning models on the Hamming Weight scaled images. These machine learning models as well as their training configuration parameters can be seen in Table~\ref{tab:training}.
All models are trained on a hyper-computing cluster on a node that consists of 4 Nvidia A100 graphics cards with 80GB of video memory (VRAM) each.

We chose these five models  to demonstrate the robustness of the Hamming Weight classification attack across a variety of neural network architectures, from older CNNs (VGG19~\cite{simonyan2015deep}) to more modern designs (Vision Transformer~\cite{dosovitskiy2021image}, ConvNeXt~\cite{liu2022convnet}). We also evaluate ResNeXt-101~\cite{xie2017aggregated} and MobileNet V3~\cite{howard2019searching}. This shows the attack is not specific to one type of classifier. This choice covers a diverse range of architectures to test the generalizability of the attack.

The attacker's models are trained on the ImageNet-1k \emph{training} split (the standard training set of roughly 1.28M images), converted to the scaled Hamming Weight representation, to learn a general mapping from scaled Hamming Weight images to the 1000 ImageNet labels. The attack's goal is to classify a specific victim image that is not known beforehand. Concretely, each training image is first converted to its Hamming Weight version and then to its scaled Hamming Weight version, and is tagged with its original ImageNet label. Crucially, the victim (test) images used in our attack evaluation are drawn from the held-out ImageNet \emph{validation} split and are not part of the attacker's training set; there is no overlap between the attacker's training data and the images the attacker attempts to recover, so the reported results are not an artifact of the models having memorized the specific test images.

\paragraph{Assumption on the input distribution.} Training the attacker on ImageNet implicitly assumes that the victim's inputs are drawn from a similar distribution and share the same label space. This is a reasonable assumption when the victim runs a standard, publicly trained classifier, a very common case for benchmarking and for off-the-shelf accelerators, but it is a genuine limitation when the victim processes proprietary imagery or a private label set that the attacker cannot approximate. Two factors partially mitigate this. First, the scaled Hamming Weight transform is generic and label-agnostic, so the learned \emph{representation} transfers even when the exact class set does not; the attacker would, however, need labels representative of the victim's task. Second, this assumption is no stronger than the ``known data distribution'' assumption made throughout the prior side-channel literature on ML accelerators. We revisit this as an explicit limitation in Section~\ref{section:discussion}.

\subsection{Large-Scale Software Validation of Classifiability}
\label{sec:shw_software_validation}

The end-to-end hardware attack is necessarily evaluated on a small number of images, because each TDC capture is slow to acquire (Section~\ref{sec_run_times}). To nonetheless provide the larger, statistically meaningful evaluation that a rigorous assessment requires, we separate two distinct questions: (i)~\emph{are scaled Hamming Weight (sHW) images classifiable at all?}, and (ii)~\emph{can the sHW image be recovered from a real TDC trace?} Question~(i) does not depend on the hardware and can therefore be evaluated at full scale in software.

We evaluate question~(i) by converting the entire ImageNet-1k validation set ($50{,}000$ images) to its sHW representation \emph{in software}, with no TDC or FPGA in the loop, and classifying each image with the attacker's trained models. We report top-1 and top-5 accuracy with $95\%$ Wilson confidence intervals, and compare against the random-guess baselines of $0.1\%$ (top-1) and $0.5\%$ (top-5) for the $1000$-class problem. Results are shown in Table~\ref{tab:shw_software_accuracy}. This experiment establishes, on a large held-out set and with no train/test overlap, that the sHW representation carries substantial class information, i.e. that the fundamental premise of the Hamming-Weight attack is sound, while the smaller hardware evaluation in Section~\ref{sec_recoloring} then demonstrates that this representation can be recovered from real power traces for the images we were able to acquire.

\begin{table*}[t]
    \centering
    \caption{\small Large-scale, \emph{software-only} validation that scaled Hamming Weight (sHW) images remain classifiable. Each of the 50{,}000 ImageNet-1k validation images is converted to its sHW representation \emph{in software} and classified by the attacker's trained models. We report top-1 and top-5 accuracy with 95\% Wilson confidence intervals, against random-guess baselines of $0.1\%$ (top-1) and $0.5\%$ (top-5). This isolates the \emph{classifiability} of the sHW representation from the separate question of hardware trace recovery. All models far exceed the random baseline (final column: top-1 accuracy relative to $0.1\%$), confirming that the sHW transform preserves substantial class-discriminative information.}
    \label{tab:shw_software_accuracy}
    \small
    \begin{adjustbox}{width=0.99\textwidth}
    \begin{tabular}{|p{3.2cm}|p{1.4cm}|p{3.4cm}|p{3.4cm}|p{2.4cm}|}
        \hline
        \textbf{Model} & \textbf{Epoch} & \textbf{Top-1 Acc.\ (\%) [95\% CI]} & \textbf{Top-5 Acc.\ (\%) [95\% CI]} & \textbf{Top-1 / Random ($0.1\%$)} \\
        \hline \hline
        VGG19               & 219 & 63.57 [63.15, 63.99] & 85.30 [84.98, 85.61] & 636$\times$ \\ \hline
        VGG19               & 100 & 63.57 [63.15, 63.99] & 85.29 [84.97, 85.60] & 636$\times$ \\ \hline
        ResNeXt-101 32x8d   & 100 & 68.22 [67.81, 68.62] & 87.73 [87.44, 88.01] & 682$\times$ \\ \hline
        MobileNet V3 Large  & 100 & 58.60 [58.16, 59.03] & 81.77 [81.43, 82.11] & 586$\times$ \\ \hline
        Vision Transformer L/16 & 131 & 59.34 [58.91, 59.77] & 79.28 [78.92, 79.63] & 593$\times$ \\ \hline
        Vision Transformer L/16 & 100 & 58.67 [58.24, 59.10] & 78.88 [78.52, 79.24] & 587$\times$ \\ \hline
        ConvNeXt Large      & 90  & 74.22 [73.83, 74.60] & 91.49 [91.24, 91.73] & 742$\times$ \\ \hline
    \end{tabular}
    \end{adjustbox}
\end{table*}

Two clarifications about Table~\ref{tab:shw_software_accuracy}. First, we report two checkpoints (epochs~100 and~219) for VGG-19 and two (epochs~100 and~131) for the Vision Transformer only to show that accuracy has saturated: the near-identical values across checkpoints indicate the models have converged, and a single converged checkpoint per architecture would suffice; we retain both rows solely as evidence of convergence. Second, the scaled-Hamming-Weight scale factor and all training hyperparameters were fixed a priori (the scale factor of~$31$ follows directly from mapping the nine Hamming-Weight levels onto the $0$--$255$ range) and were \emph{not} selected using the validation set; the $50{,}000$-image validation split is therefore used purely as held-out test data, with no tuning performed on it.

\section{Attack Results using pixel value leakage}

This section presents the experimental results for our first attack methodology, which aims to achieve a high-fidelity visual reconstruction of a victim's input image by recovering its pixel values from the power side-channel. We first validate the fundamental image recovery capability of our TDC-based sensor on a series of simple, high-contrast test patterns. More details can be found in the appendix. We then demonstrate the effectiveness of our six ML-assisted attack scenarios (A-F) on complex, full-color images from the ImageNet dataset.

\subsection{Attack Results on ImageNet Images}

\begin{figure*}[!htbp]
    \centering
     \begin{subfigure}[b]{0.16\textwidth}
         \centering
         \includegraphics[width=2cm]{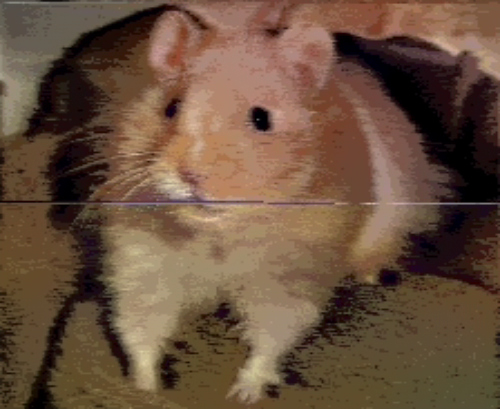}
         \caption{Image 1 Attack A}
        \label{recovered_grayscale_image_net_upscaled_first_fpga_recolored_1}
     \end{subfigure}
     \hfill
      \begin{subfigure}[b]{0.16\textwidth}
         \centering
         \includegraphics[width=2cm]{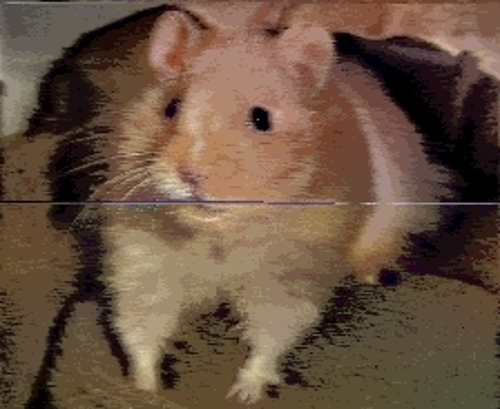}
         \caption{Image 1 Attack B}
         \label{recovered_grayscale_image_net_upscaled_first_ml_recolored_1}
     \end{subfigure}
     \hfill
      \begin{subfigure}[b]{0.16\textwidth}
         \centering
         \includegraphics[width=2cm]{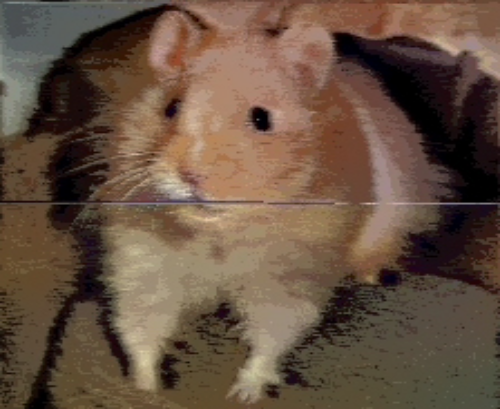}
         \caption{Image 1 Attack C}
        \label{recovered_grayscale_image_net_recolored_first_fpga_upscaled_1}
     \end{subfigure}
     \hfill
     \begin{subfigure}[b]{0.16\textwidth}
         \centering
         \includegraphics[width=2cm]{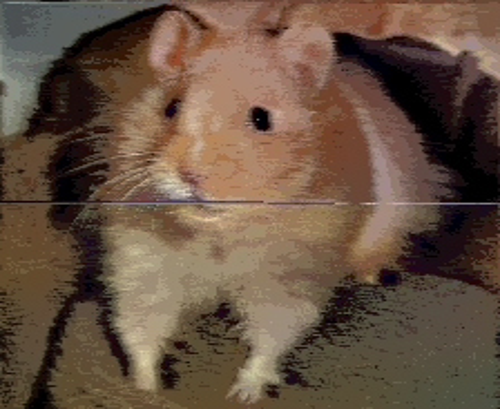}
         \caption{Image 1 Attack D}
         \label{recovered_grayscale_image_net_recolored_first_ml_upscaled_1}
     \end{subfigure}
     \hfill
     \begin{subfigure}[b]{0.16\textwidth}
         \centering
         \includegraphics[width=2cm]{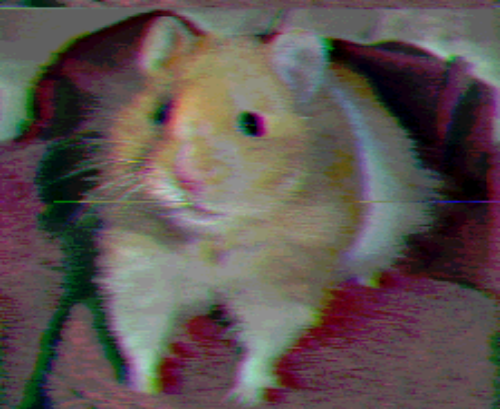}
         \caption{Image 1 Attack E}
         \label{recovered_color_image_net_upscaled_fpga_1}
     \end{subfigure}
     \hfill
     \begin{subfigure}[b]{0.16\textwidth}
         \centering
         \includegraphics[width=2cm]{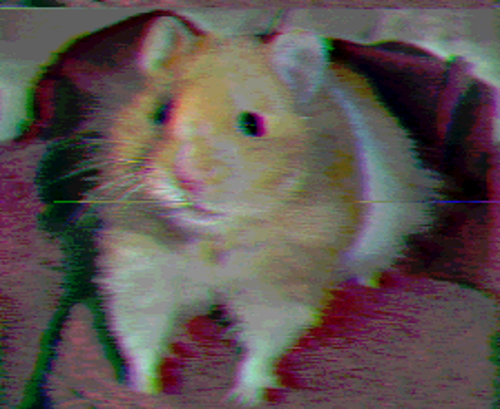}
         \caption{Image 1 Attack F}
         \label{recovered_color_image_net_upscaled_ml_1}
     \end{subfigure}
     \caption{\small ImageNet 1 images, size is 500x409 for pixel value leakage.}
     \label{reconstructed_images_image_net_1}
\end{figure*}

\begin{figure*}[!htbp]
    \centering
     \begin{subfigure}[b]{0.16\textwidth}
         \centering
         \includegraphics[width=2cm]{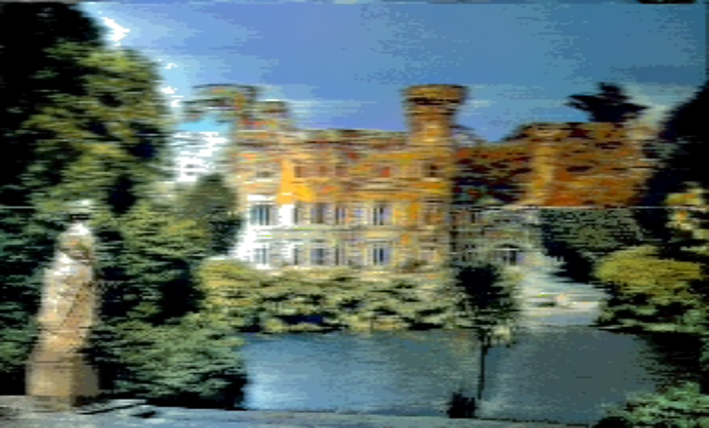}
         \caption{Image 2 Attack A}
        \label{recovered_grayscale_image_net_upscaled_first_fpga_recolored_2}
     \end{subfigure}
     \hfill
     \begin{subfigure}[b]{0.16\textwidth}
         \centering
         \includegraphics[width=2cm]{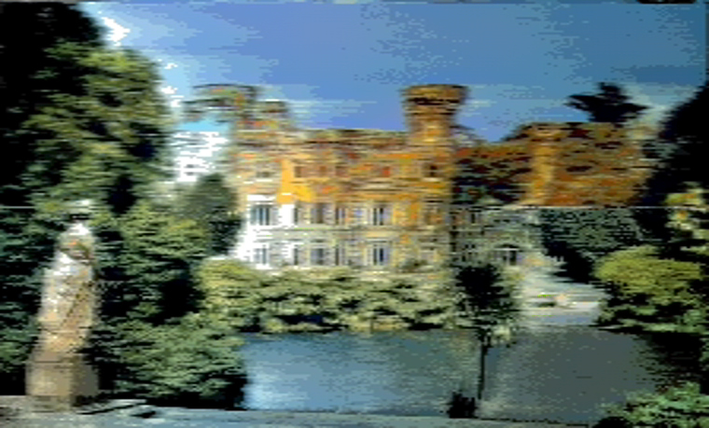}
         \caption{Image 2 Attack B}
         \label{recovered_grayscale_image_net_upscaled_first_ml_recolored_2}
     \end{subfigure}
     \hfill
     \begin{subfigure}[b]{0.16\textwidth}
         \centering
         \includegraphics[width=2cm]{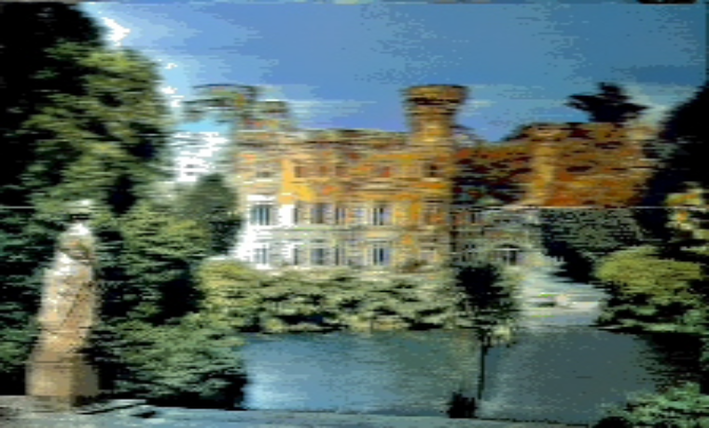}
         \caption{Image 2 Attack C}         \label{recovered_grayscale_image_net_recolored_first_fpga_upscaled_2}
     \end{subfigure}
     \hfill
     \begin{subfigure}[b]{0.16\textwidth}
         \centering
         \includegraphics[width=2cm]{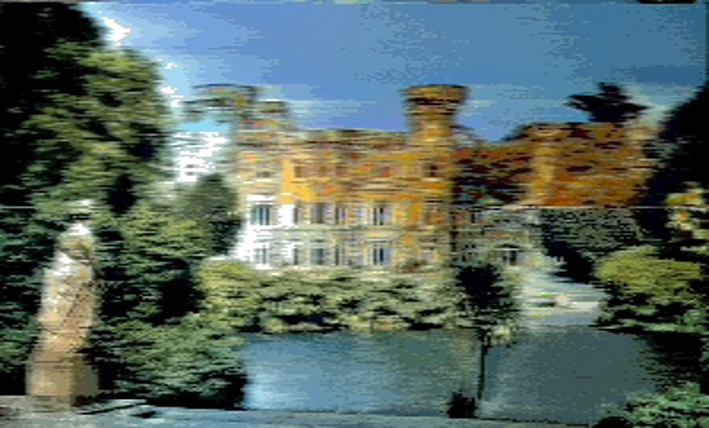}
         \caption{Image 2 Attack D}
         \label{recovered_grayscale_image_net_recolored_first_ml_upscaled_2}
     \end{subfigure}
     \hfill
     \begin{subfigure}[b]{0.16\textwidth}
         \centering
         \includegraphics[width=2cm]{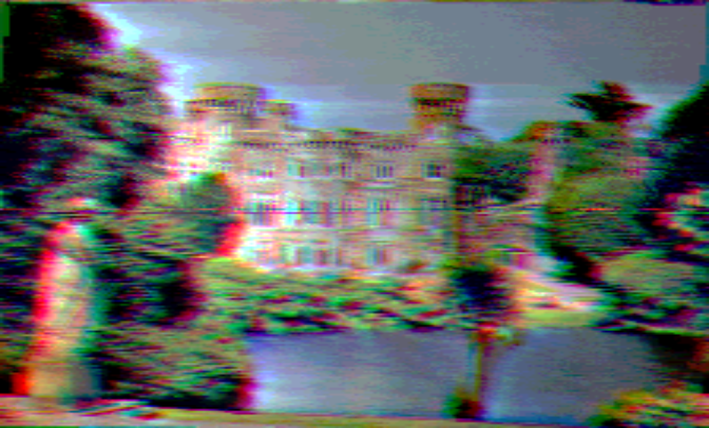}
         \caption{Image 2 Attack E}
         \label{recovered_color_image_net_upscaled_fpga_2}
     \end{subfigure}
     \hfill
     \begin{subfigure}[b]{0.16\textwidth}
         \centering
         \includegraphics[width=2cm]{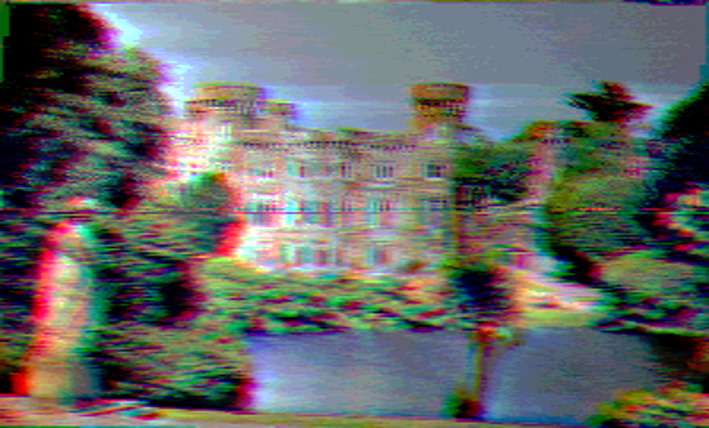}
         \caption{Image 2 Attack F}
         \label{recovered_color_image_net_upscaled_ml_2}
     \end{subfigure}
     \caption{\small ImageNet 2 images, size is 760x459 for pixel value leakage.}
     \label{reconstructed_images_image_net_2}
\end{figure*}

\begin{figure*}[!htbp]
    \centering
     \begin{subfigure}[b]{0.16\textwidth}
         \centering
         \includegraphics[width=2cm]{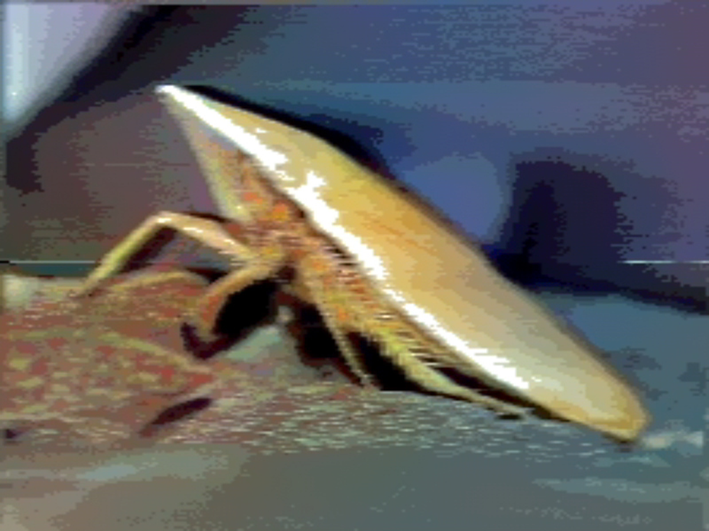}
         \caption{Image 3 Attack A}
        \label{recovered_grayscale_image_net_upscaled_first_fpga_recolored_3}
     \end{subfigure}
     \hfill
     \begin{subfigure}[b]{0.16\textwidth}
         \centering
         \includegraphics[width=2cm]{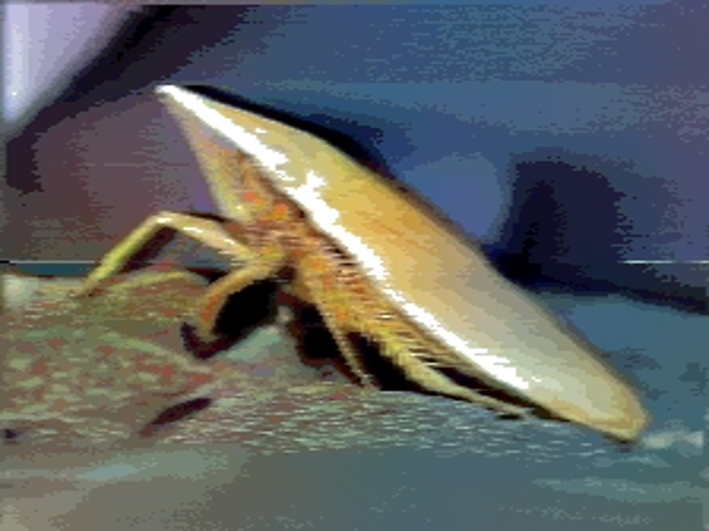}
         \caption{Image 3 Attack B}
         \label{recovered_grayscale_image_net_upscaled_first_ml_recolored_3}
     \end{subfigure}
     \hfill
     \begin{subfigure}[b]{0.16\textwidth}
         \centering
         \includegraphics[width=2cm]{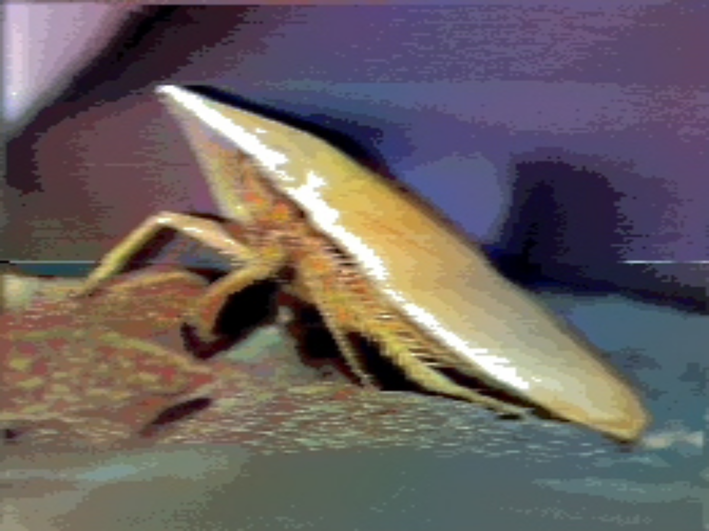}
         \caption{Image 3 Attack C}
        \label{recovered_grayscale_image_net_recolored_first_fpga_upscaled_3}
     \end{subfigure}
     \hfill
     \begin{subfigure}[b]{0.16\textwidth}
         \centering
         \includegraphics[width=2cm]{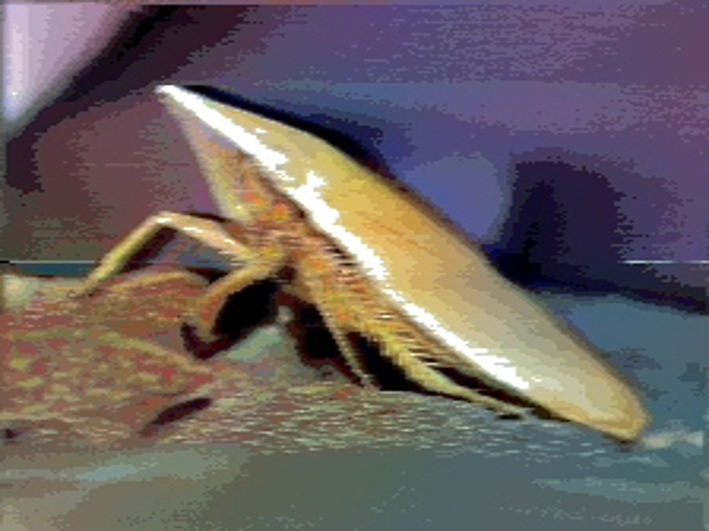}
         \caption{Image 3 Attack D}
         \label{recovered_grayscale_image_net_recolored_first_ml_upscaled_3}
     \end{subfigure}
     \hfill
     \begin{subfigure}[b]{0.16\textwidth}
         \centering
         \includegraphics[width=2cm]{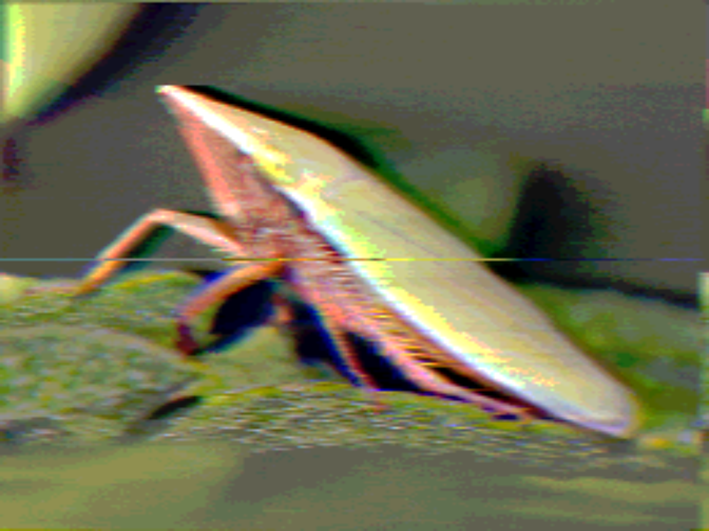}
         \caption{Image 3 Attack E}
         \label{recovered_color_image_net_upscaled_fpga_3}
     \end{subfigure}
     \hfill
     \begin{subfigure}[b]{0.16\textwidth}
         \centering
         \includegraphics[width=2cm]{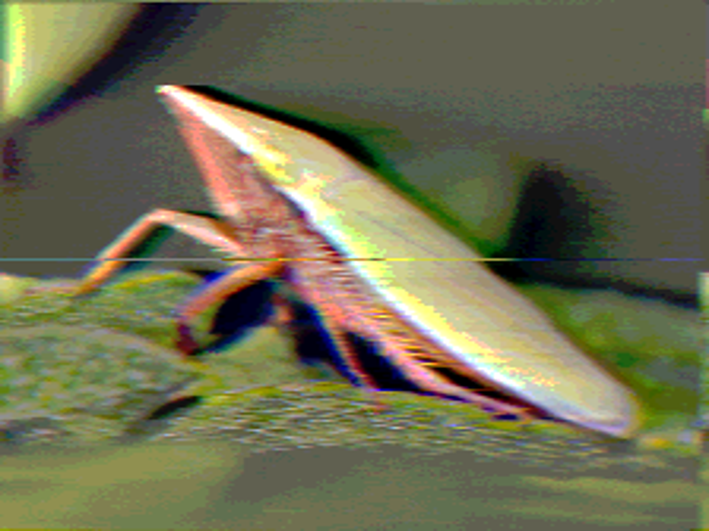}
         \caption{Image 3 Attack F}
         \label{recovered_color_image_net_upscaled_ml_3}
     \end{subfigure}
     \caption{\small ImageNet 3 images, size is 1200x899 for pixel value leakage.}
     \label{reconstructed_images_image_net_3}
\end{figure*}

\begin{figure*}[!htbp]
    \centering
     \begin{subfigure}[b]{0.16\textwidth}
         \centering
         \includegraphics[width=2cm]{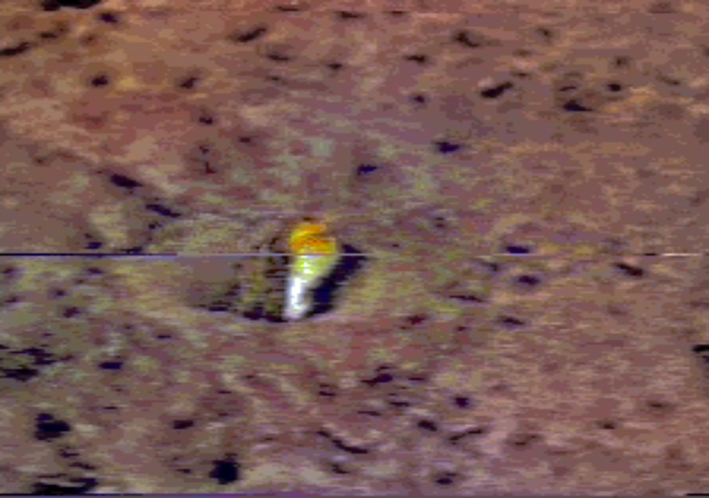}
         \caption{Image 4 Attack A}
        \label{recovered_grayscale_image_net_upscaled_first_fpga_recolored_4}
     \end{subfigure}
     \hfill
     \begin{subfigure}[b]{0.16\textwidth}
         \centering
         \includegraphics[width=2cm]{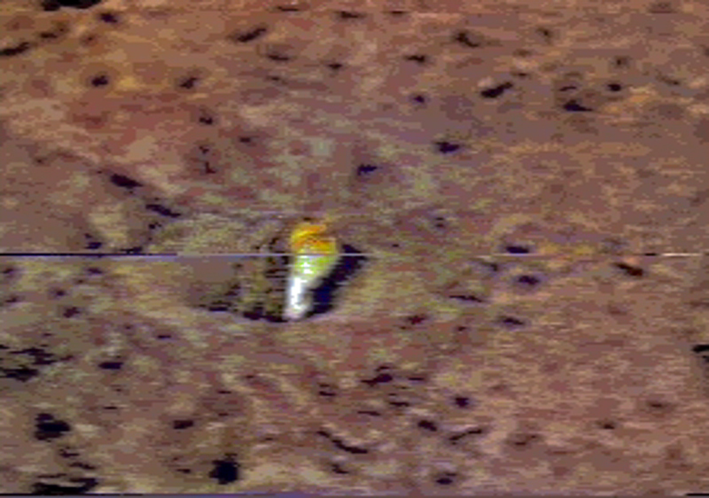}
         \caption{Image 4 Attack B}
         \label{recovered_grayscale_image_net_upscaled_first_ml_recolored_4}
     \end{subfigure}
     \hfill
     \begin{subfigure}[b]{0.16\textwidth}
         \centering
         \includegraphics[width=2cm]{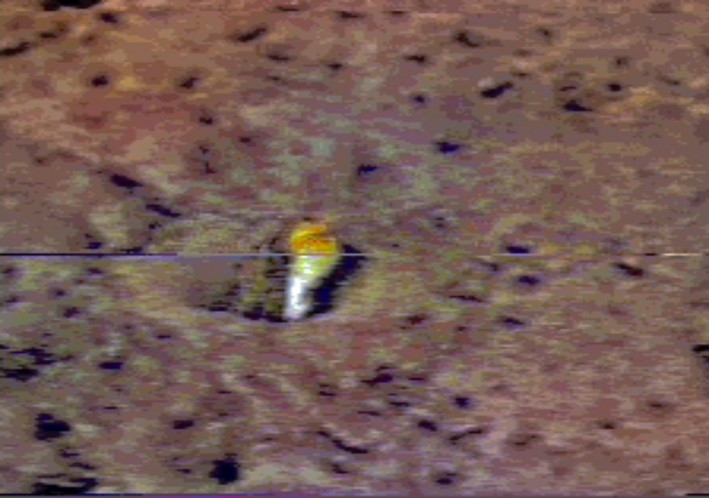}
         \caption{Image 4 Attack C}
\label{recovered_grayscale_image_net_recolored_first_fpga_upscaled_4}
     \end{subfigure}
     \hfill
     \begin{subfigure}[b]{0.16\textwidth}
         \centering
         \includegraphics[width=2cm]{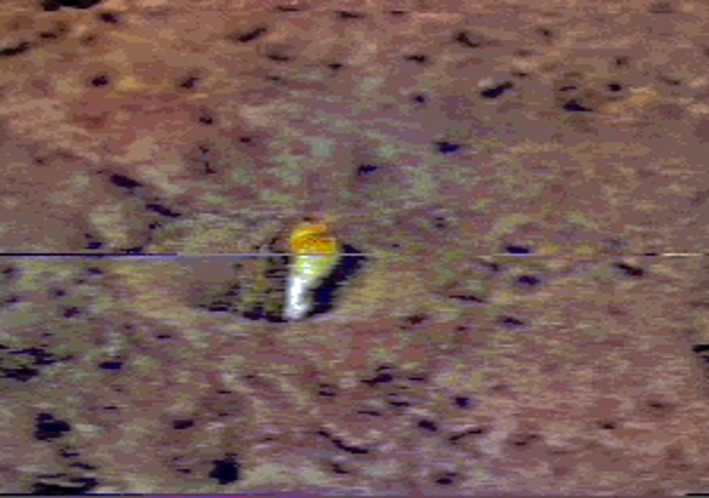}
         \caption{Image 4 Attack D}
\label{recovered_grayscale_image_net_recolored_first_ml_upscaled_4}
     \end{subfigure}
     \hfill
     \begin{subfigure}[b]{0.16\textwidth}
         \centering
         \includegraphics[width=2cm]{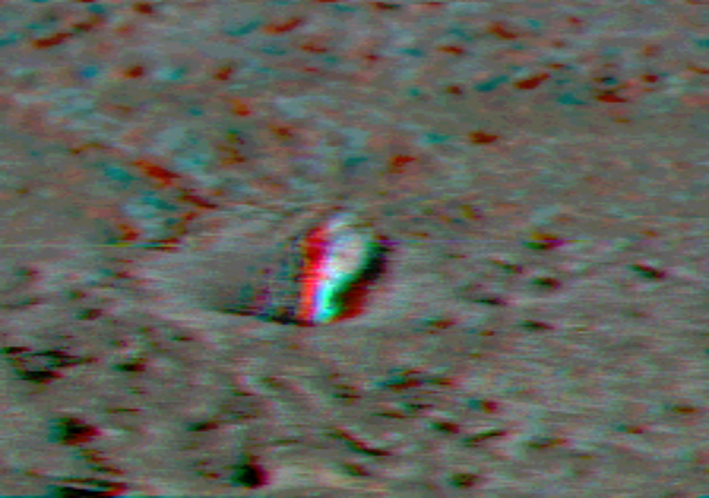}
         \caption{Image 4 Attack E}
         \label{recovered_color_image_net_upscaled_fpga_4}
     \end{subfigure}
     \hfill
     \begin{subfigure}[b]{0.16\textwidth}
         \centering
         \includegraphics[width=2cm]{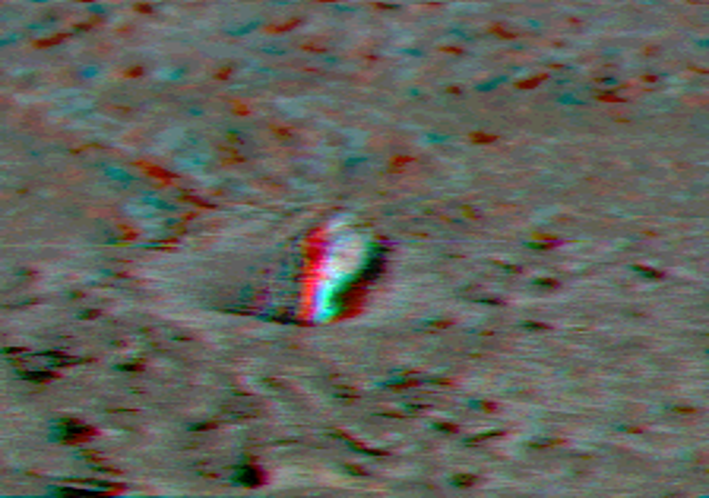}
         \caption{Image 4 Attack F}
         \label{recovered_color_image_net_upscaled_ml_4}
     \end{subfigure}
     \caption{\small ImageNet 4 images, size is 1600x1123 for pixel value leakage.}
     \label{reconstructed_images_image_net_4}
\end{figure*}

\begin{figure*}[!htbp]
    \centering
     \begin{subfigure}[b]{0.15\textwidth}
         \centering
         \includegraphics[width=2cm]{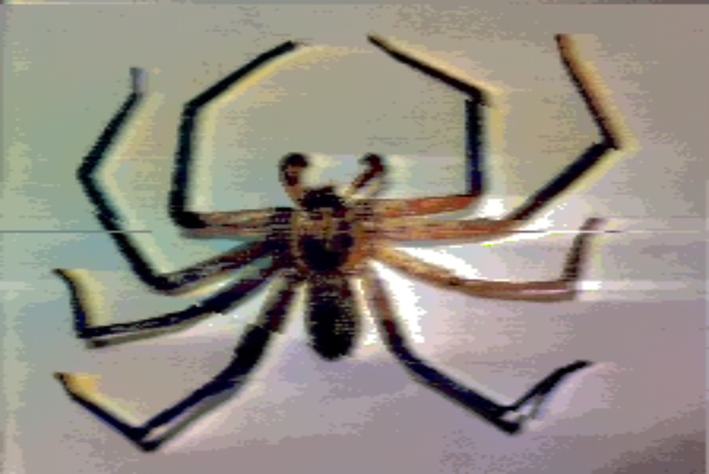}
         \caption{Image 5 \\Attack A}
        \label{recovered_grayscale_image_net_upscaled_first_fpga_recolored_5}
     \end{subfigure}
     \hfill
     \begin{subfigure}[b]{0.15\textwidth}
         \centering
         \includegraphics[width=2cm]{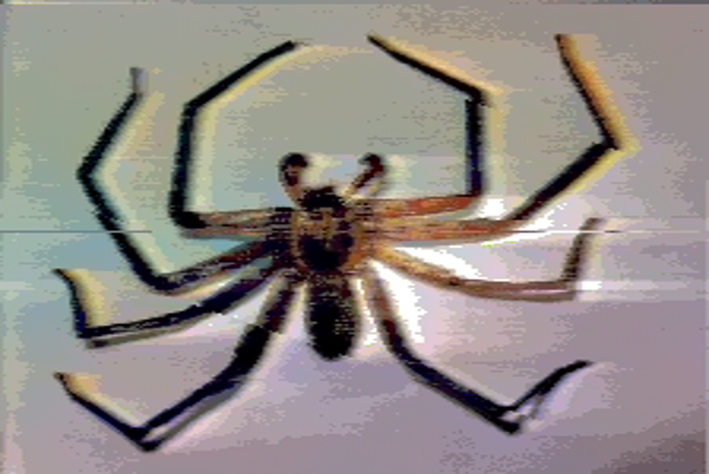}
         \caption{Image 5 \\ Attack B}
         \label{recovered_grayscale_image_net_upscaled_first_ml_recolored_5}
     \end{subfigure}
     \hfill
     \begin{subfigure}[b]{0.15\textwidth}
         \centering
         \includegraphics[width=2cm]{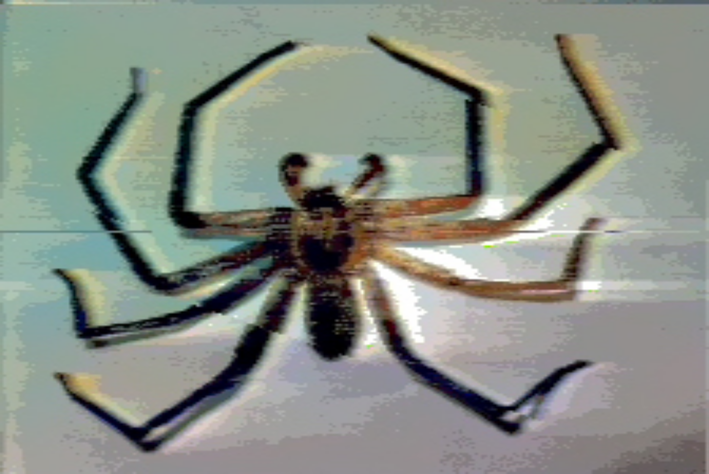}
         \caption{Image 5 \\Attack C}
        \label{recovered_grayscale_image_net_recolored_first_fpga_upscaled_5}
     \end{subfigure}
     \hfill
     \begin{subfigure}[b]{0.16\textwidth}
         \centering
         \includegraphics[width=2cm]{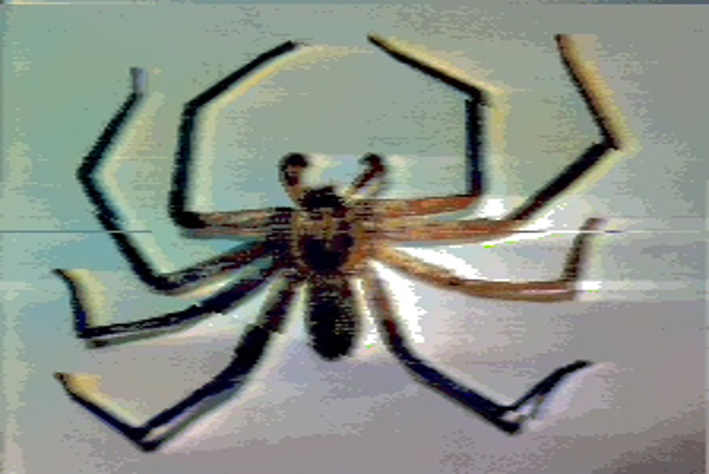}
         \caption{Image 5 \\ Attack D}
         \label{recovered_grayscale_image_net_recolored_first_ml_upscaled_5}
     \end{subfigure}
     \hfill
     \begin{subfigure}[b]{0.16\textwidth}
         \centering
         \includegraphics[width=2cm]{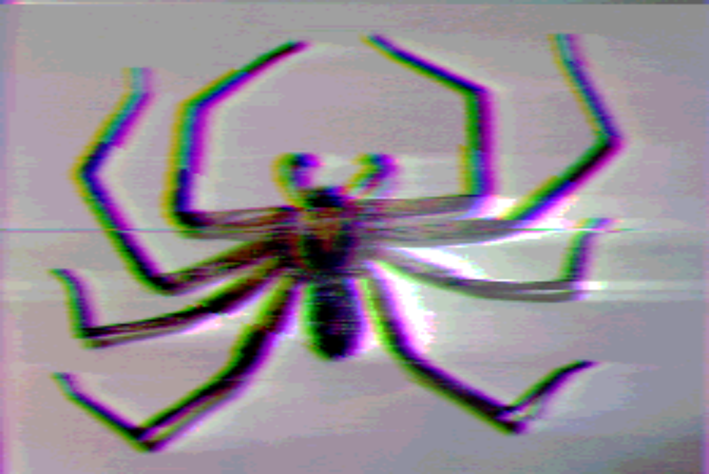}
         \caption{Image 5 \\Attack E}
         \label{recovered_color_image_net_upscaled_fpga_5}
     \end{subfigure}
     \hfill
     \begin{subfigure}[b]{0.16\textwidth}
         \centering
         \includegraphics[width=2cm]{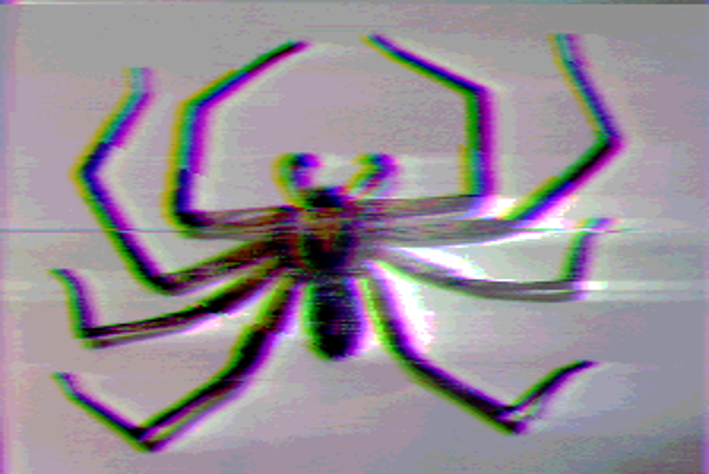}
         \caption{Image 5 \\Attack F}
         \label{recovered_color_image_net_upscaled_ml_5}
     \end{subfigure}
     \vspace{-0.5em}
     \caption{\small ImageNet 5 images, size is 2272x1520 for pixel value leakage.}
     \label{reconstructed_images_image_net_5}
\end{figure*}

After validating our TDC-based recovery on test patterns, we applied our six ML-assisted attack scenarios to complex images from the ImageNet dataset. The goal of these experiments is to evaluate the visual quality of the reconstructed images and to understand the impact of different enhancement pipelines. The Root Mean Square Error (RMSE) for each reconstruction is provided in Table~\ref{table_image_net_rms}, while the visual results are shown in Figures~\ref{recovered_grayscale_image_net_upscaled_first_fpga_recolored_1} through~\ref{recovered_color_image_net_upscaled_ml_5}.

\subsubsection{Analysis of Grayscale-Based Attacks (A, B, C, D)}
The first four attack scenarios all begin by recovering a $222\times222$ grayscale image, one value per output pixel of the first convolution layer, from an aggregate power trace. As shown in the corresponding figures, all four reconstructions suffer from similar artifacts, appearing pale or with unnatural coloring. This indicates that the recolorization ML model, while functional, is the primary source of error and visual infidelity in these pipelines. We observe only minor visual differences between the different orderings of upscaling and recoloring, suggesting that the initial lack of color information is a fundamental limitation that subsequent processing struggles to overcome.

\subsubsection{Analysis of Color-Based Attacks (E, F)}
Attacks E and F begin by recovering a full-color image, again at $222\times222$, a scenario that models a more powerful attacker who can isolate per-channel power traces. The results from these attacks are qualitatively superior to the grayscale-based methods. As seen in the figures, the images are significantly sharper, more defined, and more representative of the original input.

The primary difference between these two scenarios is the upscaling method. When comparing the results of Attack E (hardware bilinear interpolation) with Attack F (software bicubic interpolation), Attack F consistently produces a sharper final image with fewer aliasing artifacts. This is also supported by the slightly lower RMSE values for Attack F in Table~\ref{table_image_net_rms}.

\subsubsection{Summary of Findings}
Overall, these results demonstrate two key findings. First, the ability to recover per-channel color information (as in Attacks E and F) is the single most critical factor for achieving a high-fidelity visual reconstruction. Second, while enhancement techniques like ML-based recoloring and higher-order interpolation can improve results, their effectiveness is ultimately constrained by the quality of the initially recovered data.

\subsection{Reconstruction Quality: PSNR/SSIM and Recovered vs.\ Synthesized Information}
\label{sec:reconstruction_quality}

RMSE alone is difficult to interpret for visual reconstructions, so we additionally report the perceptual/quality metrics PSNR and SSIM. Table~\ref{tab:image_net_psnr_ssim} gives PSNR and SSIM for the final outputs of all six attacks, computed on exactly the same image pairs as the RMSE values in Table~\ref{table_image_net_rms}. Consistent with the qualitative discussion above, the color-recovery attacks (E, F) generally achieve higher PSNR and SSIM than the grayscale-based attacks (A--D) on the images that contain substantial color (Images 3--5), confirming that per-channel recovery is the dominant factor in reconstruction fidelity.

More importantly, these metrics let us make an explicit distinction: because the grayscale-based attacks colorize the image with a \emph{generative} ML model, their final color is not necessarily faithful to the victim's true colors. To separate genuinely recovered information from ML-synthesized detail, Table~\ref{tab:raw_recovery_metrics} reports the quality of the \emph{raw} side-channel recovery, the directly reconstructed grayscale image (aggregate leakage) and the directly reconstructed per-channel color image (before any upscaling or recoloring), measured against the original. Two observations follow. First, for the grayscale-based attacks (A--D) the final SSIM and PSNR are essentially unchanged from the raw grayscale-recovery values in Table~\ref{tab:raw_recovery_metrics}: the ML recoloring produces a plausible-looking image but adds little \emph{measured} fidelity, i.e., the color it introduces is largely synthesized rather than recovered. Second, for the color-based attacks (E, F) the final metrics closely track the raw per-channel recovery, because those pipelines only upscale (they do not hallucinate color); their fidelity therefore reflects genuinely recovered per-channel structure. We consequently caution the reader to interpret the visually convincing color images from Attacks A--D as \emph{plausible renderings} conditioned on a recovered grayscale structure, and only the per-channel results (E, F) as reconstructions whose color is actually recovered from the side channel.

To separate the error introduced by the side channel from the resampling loss that any $222\times222$ reconstruction would incur, we compute a downsample--upsample baseline (Table~\ref{tab:resize_baseline}). Because the first convolution layer uses a $3\times3$ kernel with stride $1$ and no padding, a $224\times224$ input yields a $222\times222$ output feature map, and the TDC observes one multiply-accumulate burst per output pixel; each capture provides $2^{20}$ samples for one half-frame ($111\times222$ output pixels), i.e., roughly $42$ samples per output pixel. The attacker therefore recovers a $222\times222$ field, which is upscaled to the original photograph's dimensions before the metrics of Table~\ref{tab:image_net_psnr_ssim} are computed. The baseline isolates the cost of that resampling alone: downsampling each original to $222\times222$ and upsampling it back yields a mean PSNR of $28.25$~dB and mean SSIM of $0.781$, which is the ceiling any reconstruction at this resolution could attain. Averaged over Attacks A--F, our reconstructions reach $14.40$~dB PSNR and $0.420$ SSIM, leaving a mean residual gap of $13.85$~dB and $0.361$ SSIM below that ceiling. This decomposition makes the interpretation precise: a non-trivial share of the absolute error in Table~\ref{tab:image_net_psnr_ssim} is an unavoidable consequence of reconstructing a multi-megapixel photograph from a $222\times222$ field, while the remaining gap, the majority of the SSIM deficit, is attributable to side-channel noise and, for Attacks A--D, to the generative recoloring. Reported values should therefore be read against the baseline in Table~\ref{tab:resize_baseline}, not against an idealized lossless reference.

A second methodological point concerns the comparison between the raw recovery of Table~\ref{tab:raw_recovery_metrics} and the recolored results of Table~\ref{tab:image_net_psnr_ssim}: the former scores a grayscale reconstruction against a grayscale reference, while the latter scores a colored reconstruction against an RGB reference, so the two sets of values are not directly comparable and cannot on their own establish how much the recoloring stage contributes. Table~\ref{tab:recolor_samedomain} therefore repeats the comparison in a single domain: the raw grayscale recovery is replicated across the three color channels and scored against the original RGB photograph, exactly as the recolored attacks are. Measured this way, recoloring changes fidelity by $-0.04$~dB PSNR and $0.000$ SSIM on average, and it reduces both metrics on three of the five images. This same-domain evidence supports the interpretation given above: for Attacks A--D the color is visually plausible but largely synthesized by the generative model, and only the per-channel attacks (E, F) recover color information from the side channel itself.

\begin{table*}[t]
\centering
\small
\caption{\small PSNR (dB, higher is better) and SSIM (higher is better) between each ImageNet input and the final reconstruction produced by Attacks A--F, computed on the same image pairs as the RMSE of Table~\ref{table_image_net_rms} (reconstruction resized to the original resolution; SSIM computed over RGB). For the color-recovery attacks (E, F) these metrics reflect genuinely recovered per-channel information; for the grayscale-based attacks (A--D) the color is ML-synthesized, so their PSNR/SSIM should be read together with the raw-recovery metrics of Table~\ref{tab:raw_recovery_metrics}.}
\label{tab:image_net_psnr_ssim}
\begin{adjustbox}{width=1\textwidth}
\small
\begin{tabular}{|l|l|c|c|c|c|c|c|}
\hline
\textbf{Image} & \textbf{Metric} & \textbf{A} & \textbf{B} & \textbf{C} & \textbf{D} & \textbf{E} & \textbf{F} \\ \hline\hline
\multirow{2}{*}{Image 1} & PSNR & 18.12 & 18.03 & 18.18 & 18.03 & 16.83 & 16.73 \\ \cline{2-8}
 & SSIM & 0.320 & 0.307 & 0.322 & 0.306 & 0.285 & 0.272 \\ \hline
\multirow{2}{*}{Image 2} & PSNR & 12.91 & 12.82 & 12.89 & 12.77 & 12.78 & 12.67 \\ \cline{2-8}
 & SSIM & 0.209 & 0.199 & 0.208 & 0.199 & 0.223 & 0.215 \\ \hline
\multirow{2}{*}{Image 3} & PSNR & 13.63 & 13.36 & 13.54 & 13.48 & 14.93 & 14.85 \\ \cline{2-8}
 & SSIM & 0.519 & 0.499 & 0.518 & 0.499 & 0.534 & 0.517 \\ \hline
\multirow{2}{*}{Image 4} & PSNR & 17.86 & 17.54 & 18.17 & 17.98 & 20.20 & 19.97 \\ \cline{2-8}
 & SSIM & 0.473 & 0.452 & 0.479 & 0.457 & 0.496 & 0.475 \\ \hline
\multirow{2}{*}{Image 5} & PSNR & 8.76 & 8.89 & 8.64 & 8.66 & 9.35 & 9.34 \\ \cline{2-8}
 & SSIM & 0.606 & 0.589 & 0.605 & 0.583 & 0.629 & 0.611 \\ \hline
\end{tabular}
\end{adjustbox}
\end{table*}

\begin{table}[t]
    \centering
    \caption{\small Downsample--upsample baseline separating unavoidable resampling loss from side-channel reconstruction error. Because the first convolution layer uses a $3\times3$ kernel with stride $1$ and no padding, it produces a $222\times222$ output feature map; the attacker recovers a $222\times222$ field that is upscaled to the original photograph's dimensions before the metrics of Table~\ref{tab:image_net_psnr_ssim} are computed. This table reports the ceiling imposed by that resampling alone: each original is downsampled to $222\times222$ and upsampled back to its native size, with no side channel involved. The attack columns give the mean over Attacks A--F. The residual gap is the portion of the error attributable to side-channel noise and generative recoloring rather than to resizing.}
    \label{tab:resize_baseline}
    \small
    \begin{adjustbox}{width=0.99\columnwidth}
    \begin{tabular}{|c|c|c|c|c|c|c|c|}
        \hline
        \multirow{2}{*}{\textbf{Image}} & \multirow{2}{*}{\textbf{Size}} & \multicolumn{2}{c|}{\textbf{Resize ceiling}} & \multicolumn{2}{c|}{\textbf{Attacks (mean A--F)}} & \multicolumn{2}{c|}{\textbf{Residual gap}} \\ \cline{3-8}
         & & \textbf{PSNR} & \textbf{SSIM} & \textbf{PSNR} & \textbf{SSIM} & \textbf{PSNR} & \textbf{SSIM} \\
        \hline \hline
        1 & $500\times409$   & 28.47 & 0.721 & 17.65 & 0.302 & 10.82 & 0.419 \\ \hline
        2 & $760\times459$   & 19.77 & 0.588 & 12.81 & 0.209 &  6.96 & 0.379 \\ \hline
        3 & $1200\times899$  & 30.58 & 0.894 & 13.96 & 0.514 & 16.62 & 0.380 \\ \hline
        4 & $1600\times1123$ & 30.99 & 0.774 & 18.62 & 0.472 & 12.37 & 0.302 \\ \hline
        5 & $1136\times760$  & 31.43 & 0.927 &  8.94 & 0.604 & 22.49 & 0.323 \\ \hline
        \textbf{Mean} & --- & \textbf{28.25} & \textbf{0.781} & \textbf{14.40} & \textbf{0.420} & \textbf{13.85} & \textbf{0.361} \\ \hline
    \end{tabular}
    \end{adjustbox}
\end{table}

\begin{table}[t]
    \centering
    \caption{\small Same-domain comparison isolating the measured contribution of the ML recoloring stage. To avoid scoring a grayscale reconstruction against a grayscale reference while scoring a recolored reconstruction against an RGB reference, both are evaluated here in the \emph{same} RGB domain: the raw grayscale recovery is replicated across the three channels and compared against the original RGB photograph, exactly as the recolored attacks are. Values for the recolored column are the mean over Attacks A--D. Averaged over the five images, recoloring changes fidelity by $-0.04$~dB PSNR and $0.000$ SSIM, and it \emph{reduces} both metrics on three of the five images: the color it produces is visually plausible but essentially entirely synthesized, adding no measurable fidelity over the grayscale information actually recovered from the side channel.}
    \label{tab:recolor_samedomain}
    \small
    \begin{adjustbox}{width=0.99\columnwidth}
    \begin{tabular}{|c|c|c|c|c|c|c|}
        \hline
        \multirow{2}{*}{\textbf{Image}} & \multicolumn{2}{c|}{\textbf{Raw grayscale recovery}} & \multicolumn{2}{c|}{\textbf{Recolored (mean A--D)}} & \multicolumn{2}{c|}{\textbf{Change}} \\ \cline{2-7}
         & \textbf{PSNR} & \textbf{SSIM} & \textbf{PSNR} & \textbf{SSIM} & \textbf{PSNR} & \textbf{SSIM} \\
        \hline \hline
        1 & 17.21 & 0.303 & 18.09 & 0.314 & $+$0.88 & $+$0.011 \\ \hline
        2 & 12.49 & 0.200 & 12.85 & 0.204 & $+$0.36 & $+$0.004 \\ \hline
        3 & 13.66 & 0.512 & 13.50 & 0.509 & $-$0.16 & $-$0.003 \\ \hline
        4 & 18.91 & 0.470 & 17.89 & 0.465 & $-$1.02 & $-$0.005 \\ \hline
        5 &  8.98 & 0.601 &  8.74 & 0.596 & $-$0.24 & $-$0.005 \\ \hline
        \textbf{Mean} & \textbf{14.25} & \textbf{0.417} & \textbf{14.21} & \textbf{0.417} & \textbf{$-$0.04} & \textbf{0.000} \\ \hline
    \end{tabular}
    \end{adjustbox}
\end{table}

\begin{table*}[t]
\centering
\small
\caption{\small \emph{Raw side-channel recovery} quality, isolating what the TDC actually recovers from the ML post-processing. The grayscale columns compare the directly reconstructed grayscale image (aggregate leakage, before any upscaling/recoloring) to the grayscale original; the color columns compare the directly reconstructed per-channel image (before upscaling) to the color original. RMSE is normalized to $[0,1]$; PSNR in dB. Comparing these values with Table~\ref{tab:image_net_psnr_ssim} shows that for Attacks A--D the ML recoloring adds little to SSIM/PSNR over the raw grayscale recovery, i.e. the added color is plausible but is largely synthesized rather than recovered.}
\label{tab:raw_recovery_metrics}
\begin{adjustbox}{width=1\textwidth}
\small
\begin{tabular}{|l|ccc|ccc|}
\hline
& \multicolumn{3}{c|}{\textbf{Grayscale recovery (aggregate)}} & \multicolumn{3}{c|}{\textbf{Per-channel color recovery}} \\ \cline{2-7}
\textbf{Image} & PSNR & SSIM & RMSE & PSNR & SSIM & RMSE \\ \hline\hline
Image 1 & 18.47 & 0.319 & 0.119 & 16.73 & 0.272 & 0.146 \\ \hline
Image 2 & 13.16 & 0.205 & 0.220 & 12.67 & 0.215 & 0.233 \\ \hline
Image 3 & 16.49 & 0.609 & 0.150 & 14.85 & 0.517 & 0.181 \\ \hline
Image 4 & 19.15 & 0.480 & 0.110 & 19.97 & 0.475 & 0.100 \\ \hline
Image 5 & 9.03 & 0.604 & 0.354 & 9.34 & 0.612 & 0.341 \\ \hline
\end{tabular}
\end{adjustbox}
\end{table*}

\subsection{Comparing Recoloring Methods}

\subsubsection{ DeepAI Image Colorization  API}
We leverage DeepAI Image Colorization API~\cite{deep_ai} to recolor grayscale images. While not perfect, the  DeepAI Image Colorization API allows for obtaining reasonable, colored images from grayscale images.
We observe that for smaller images, the recoloring tends to work better, and also it is less noticeable. As expected, for larger images, the recoloring artifacts are more obvious. Note that we used publicly available DeepAI Image Colorization API that was not trained specifically on our images. Much better results can be obtained if the recoloring algorithm is trained on the same set of images as the ones that we want to~recolor.

\subsubsection{Py Simple GUI Colorizer}
We also leverage Py Simple GUI Colorizer~\cite{py_simple_gui} that transforms black and white images into colored images using deep learning. It uses OpenCV~\cite{open_cv} and Numpy~\cite{numpy} to colorize the input image. The colorization algorithm developed by Zhang, et al.~\cite{zhang2016colorful}, is used for the colorization.

\begin{figure*}[!htbp]
    \centering
     \begin{subfigure}[b]{0.15\textwidth}
         \centering
         \includegraphics[width=1.5cm]{figures/tdc_images/image_net_1/recovered_grayscale_image_net_upscaled_first_fpga_recolored_1.png}
         \caption{Img. 1, Attack A, Deep AI}         \label{recovered_grayscale_image_net_upscaled_first_fpga_recolored_1_compare}
     \end{subfigure}
     \hfill
     \begin{subfigure}[b]{0.15\textwidth}
         \centering
         \includegraphics[width=1.5cm]{figures/tdc_images/image_net_1/recovered_grayscale_image_net_upscaled_first_ml_recolored_1.png}
          \caption{Img. 1, Attack B, Deep AI}
\label{recovered_grayscale_image_net_recolored_first_ml_upscaled_1_compare}
     \end{subfigure}
     \hfill
     \begin{subfigure}[b]{0.15\textwidth}
         \centering
         \includegraphics[width=1.5cm]{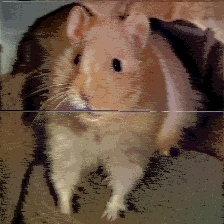}
          \caption{Img. 1, Attack C or D, Deep AI}
         \label{recovered_grayscale_image_net_recolored_first_1}
     \end{subfigure}
     \hfill
     \begin{subfigure}[b]{0.15\textwidth}
         \centering
         \includegraphics[width=1.5cm]{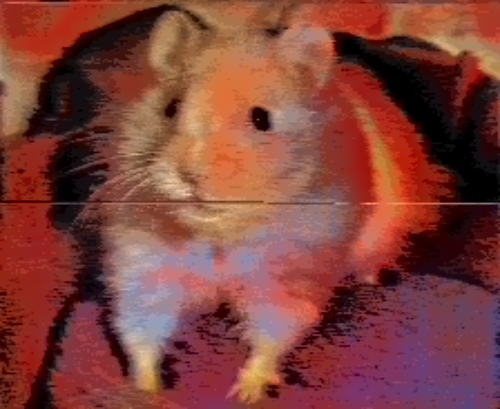}
         \caption{Img. 1, Attack A, Py Simple GUI}
         \label{color_recovered_grayscale_image_net_upscaled_first_fpga_1}
     \end{subfigure}
     \hfill
     \begin{subfigure}[b]{0.15\textwidth}
         \centering
         \includegraphics[width=1.5cm]{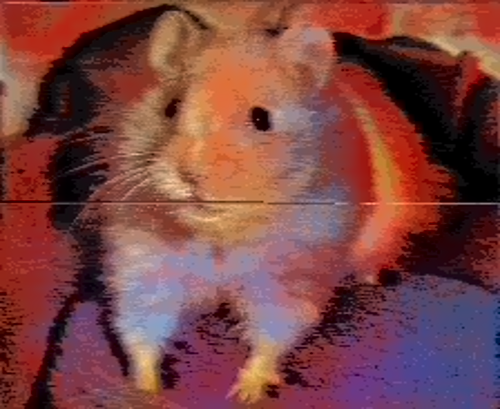}
         \caption{Img. 1, Attack B, Py Simple GUI}
         \label{color_recovered_grayscale_image_net_upscaled_first_ml_1}
     \end{subfigure}
     \hfill
     \begin{subfigure}[b]{0.16\textwidth}
         \centering
         \includegraphics[width=1.5cm]{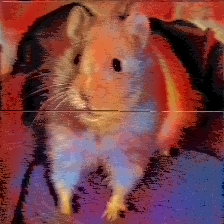}
         \caption{Img. 1, Attack C or D, Py Simple GUI}
\label{py_color_image_net_grayscale_reconstructed_1}
     \end{subfigure}
     \vspace{-0.5em}
     \caption{\small Recolored grayscale ImageNet 1 images using DeepAI and Py Simple GUI image colorizer.}
     \label{deepai_vs_py_simple_gui_1}
\end{figure*}

\begin{figure*}[!htbp]
    \centering
     \begin{subfigure}[b]{0.15\textwidth}
         \centering
         \includegraphics[width=1.5cm]{figures/tdc_images/image_net_2/recovered_grayscale_image_net_upscaled_first_fpga_recolored_2.png}
         \caption{Img. 2, Attack A, Deep AI}
\label{recovered_grayscale_image_net_upscaled_first_fpga_recolored_2_compare}
     \end{subfigure}
     \hfill
     \begin{subfigure}[b]{0.15\textwidth}
         \centering
         \includegraphics[width=1.5cm]{figures/tdc_images/image_net_2/recovered_grayscale_image_net_upscaled_first_ml_recolored_2.png}
         \caption{Img. 2, Attack B, Deep AI}
\label{recovered_grayscale_image_net_recolored_first_ml_upscaled_2_compare}
     \end{subfigure}
     \hfill
     \begin{subfigure}[b]{0.15\textwidth}
         \centering
         \includegraphics[width=1.5cm]{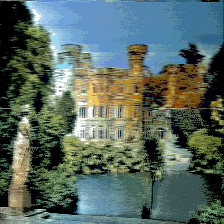}
         \caption{Img. 2, Attack C or D, Deep AI}
         \label{recovered_grayscale_image_net_recolored_first_2}
     \end{subfigure}
     \hfill
     \begin{subfigure}[b]{0.15\textwidth}
         \centering
         \includegraphics[width=1.5cm]{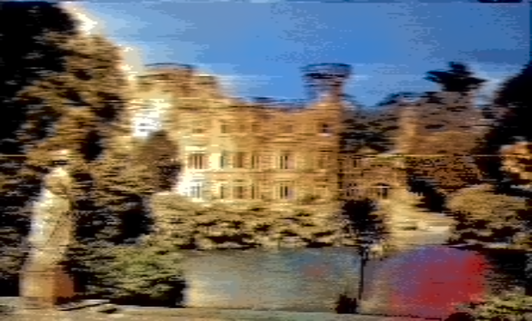}
         \caption{Img. 2, Attack A, Py Simple GUI}
         \label{color_recovered_grayscale_image_net_upscaled_first_fpga_2}
     \end{subfigure}
     \hfill
     \begin{subfigure}[b]{0.15\textwidth}
         \centering
         \includegraphics[width=1.5cm]{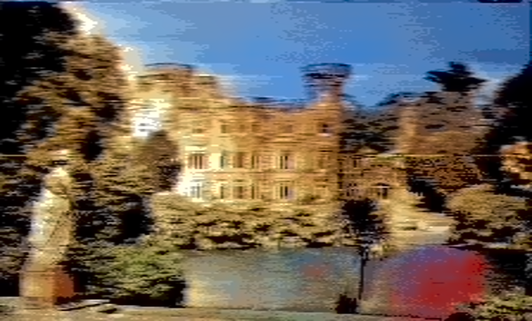}
         \caption{Img. 2, Attack B, Py Simple GUI}
         \label{color_recovered_grayscale_image_net_upscaled_first_ml_2}
     \end{subfigure}
     \hfill
     \begin{subfigure}[b]{0.16\textwidth}
         \centering
         \includegraphics[width=1.5cm]{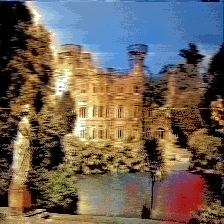}
         \caption{Img. 2, Attack C or D, Py Simple GUI}
         \label{py_color_image_net_grayscale_reconstructed_2}
     \end{subfigure}
     \caption{\small Recolored Grayscale ImageNet 2 images using DeepAI and Py Simple GUI image colorizer.}
     \label{deepai_vs_py_simple_gui_2}
\end{figure*}

\begin{figure*}[!htbp]
    \centering
     \begin{subfigure}[b]{0.15\textwidth}
         \centering
         \includegraphics[width=1.5cm]{figures/tdc_images/image_net_3/recovered_grayscale_image_net_upscaled_first_fpga_recolored_3.png}
         \caption{Img. 3, Attack A, Deep AI}
\label{recovered_grayscale_image_net_upscaled_first_fpga_recolored_3_compare}
     \end{subfigure}
     \hfill
     \begin{subfigure}[b]{0.15\textwidth}
         \centering
         \includegraphics[width=1.5cm]{figures/tdc_images/image_net_3/recovered_grayscale_image_net_upscaled_first_ml_recolored_3.png}
         \caption{Img. 3, Attack B, Deep AI}
\label{recovered_grayscale_image_net_recolored_first_ml_upscaled_3_compare}
     \end{subfigure}
     \hfill
     \begin{subfigure}[b]{0.15\textwidth}
         \centering
         \includegraphics[width=1.5cm]{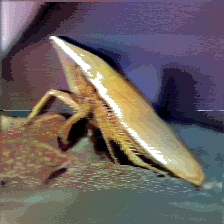}
         \caption{Img. 3, Attack C or D, Deep AI}
         \label{recovered_grayscale_image_net_recolored_first_3}
     \end{subfigure}
     \hfill
     \begin{subfigure}[b]{0.15\textwidth}
         \centering
         \includegraphics[width=1.5cm]{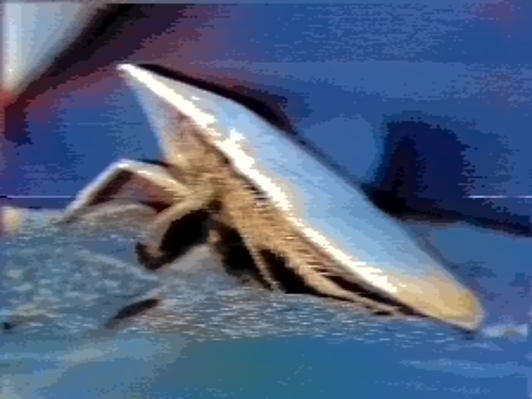}
         \caption{Img. 3, Attack A, Py Simple GUI}
         \label{color_recovered_grayscale_image_net_upscaled_first_fpga_3}
     \end{subfigure}
     \hfill
     \begin{subfigure}[b]{0.15\textwidth}
         \centering
         \includegraphics[width=1.5cm]{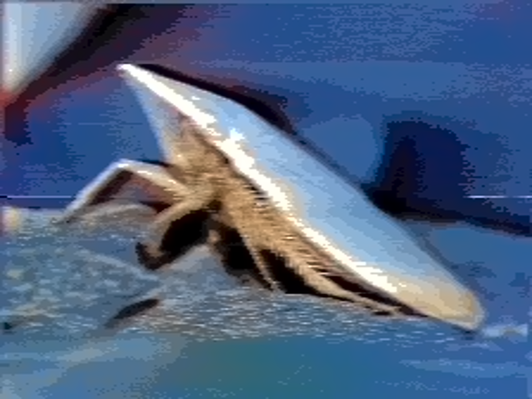}
         \caption{Img. 3, Attack B, Py Simple GUI}
         \label{color_recovered_grayscale_image_net_upscaled_first_ml_3}
     \end{subfigure}
     \hfill
     \begin{subfigure}[b]{0.16\textwidth}
         \centering
         \includegraphics[width=1.5cm]{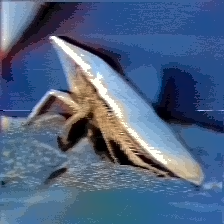}
         \caption{Img. 3, Attack C or D, Py Simple GUI}
         \label{py_color_image_net_grayscale_reconstructed_3}
     \end{subfigure}
     \hfill
     \caption{\small Recolored Grayscale ImageNet 3 images using DeepAI and Py Simple GUI image colorizer.}
     \label{deepai_vs_py_simple_gui_3}
\end{figure*}

\begin{figure*}[!htbp]
    \centering
     \begin{subfigure}[b]{0.15\textwidth}
         \centering
         \includegraphics[width=1.5cm]{figures/tdc_images/image_net_4/recovered_grayscale_image_net_upscaled_first_fpga_recolored_4.png}
         \caption{Img. 4, Attack A, Deep AI}
\label{recovered_grayscale_image_net_upscaled_first_fpga_recolored_4_compare}
     \end{subfigure}
     \hfill
     \begin{subfigure}[b]{0.15\textwidth}
         \centering
         \includegraphics[width=1.5cm]{figures/tdc_images/image_net_4/recovered_grayscale_image_net_upscaled_first_ml_recolored_4.png}
         \caption{Img. 4, Attack B, Deep AI}
\label{recovered_grayscale_image_net_recolored_first_ml_upscaled_4_compare}
     \end{subfigure}
     \hfill
     \begin{subfigure}[b]{0.15\textwidth}
         \centering
         \includegraphics[width=1.5cm]{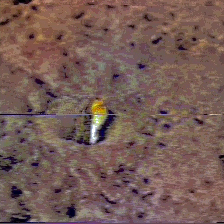}
          \caption{Img. 4, Attack C or D, Deep AI}
         \label{recovered_grayscale_image_net_recolored_first_4}
     \end{subfigure}
     \hfill
     \begin{subfigure}[b]{0.15\textwidth}
         \centering
         \includegraphics[width=1.5cm]{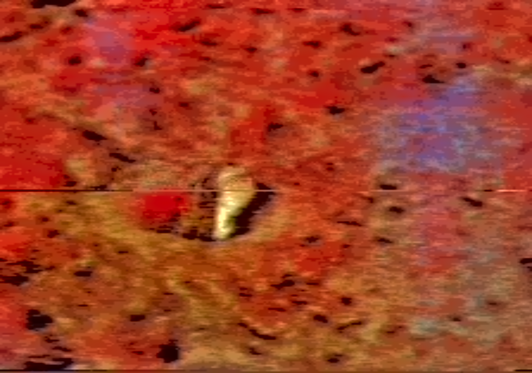}
         \caption{Img. 4, Attack A, Py Simple GUI}
         \label{color_recovered_grayscale_image_net_upscaled_first_fpga_4}
     \end{subfigure}
     \hfill
     \begin{subfigure}[b]{0.15\textwidth}
         \centering
         \includegraphics[width=1.5cm]{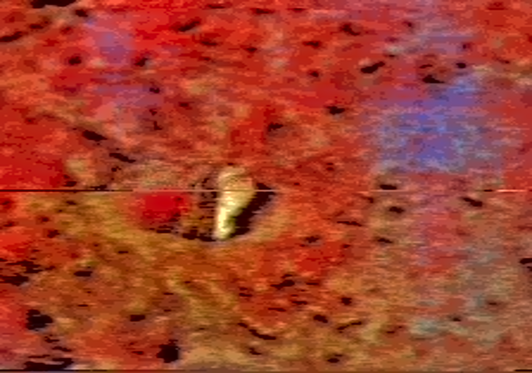}
         \caption{Img. 4, Attack B, Py Simple GUI}
         \label{color_recovered_grayscale_image_net_upscaled_first_ml_4}
     \end{subfigure}
     \hfill
     \begin{subfigure}[b]{0.16\textwidth}
         \centering
         \includegraphics[width=1.5cm]{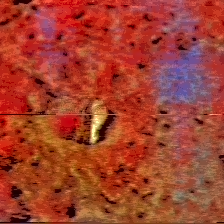}
         \caption{Img. 4, Attack C or D, Py Simple GUI}
         \label{py_color_image_net_grayscale_reconstructed_4}
     \end{subfigure}
     \caption{\small Recolored Grayscale ImageNet 4 images using DeepAI and Py Simple GUI image colorizer.}
     \label{deepai_vs_py_simple_gui_4}
\end{figure*}

\begin{figure*}[!htbp]
    \centering
     \begin{subfigure}[b]{0.15\textwidth}
         \centering
         \includegraphics[width=1.5cm]{figures/tdc_images/image_net_5/recovered_grayscale_image_net_upscaled_first_fpga_recolored_5.png}
         \caption{Img. 5, Attack A, Deep AI}
\label{recovered_grayscale_image_net_upscaled_first_fpga_recolored_5_compare}
     \end{subfigure}
     \hfill
     \begin{subfigure}[b]{0.15\textwidth}
         \centering
         \includegraphics[width=1.5cm]{figures/tdc_images/image_net_5/recovered_grayscale_image_net_upscaled_first_ml_recolored_5.png}
         \caption{Img. 5, Attack B, Deep AI}
\label{recovered_grayscale_image_net_recolored_first_ml_upscaled_5_compare}
     \end{subfigure}
     \hfill
     \begin{subfigure}[b]{0.15\textwidth}
         \centering
         \includegraphics[width=1.5cm]{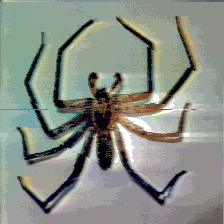}
         \caption{Img. 5, Attack C or D, Deep AI}
         \label{recovered_grayscale_image_net_recolored_first_5}
     \end{subfigure}
     \hfill
     \begin{subfigure}[b]{0.15\textwidth}
         \centering
         \includegraphics[width=1.5cm]{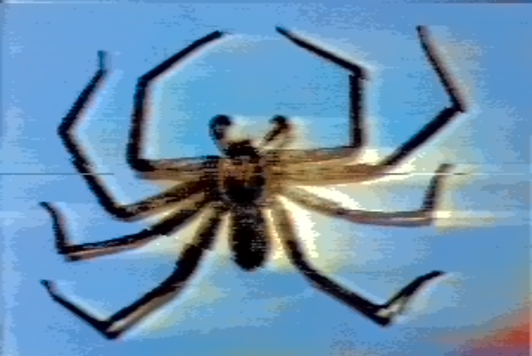}
         \caption{Img. 5, Attack A, Py Simple GUI}
         \label{color_recovered_grayscale_image_net_upscaled_first_fpga_5}
     \end{subfigure}
     \hfill
     \begin{subfigure}[b]{0.15\textwidth}
         \centering
         \includegraphics[width=1.5cm]{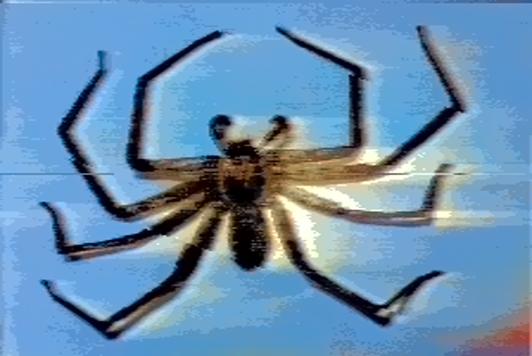}
         \caption{Img. 5, Attack B, Py Simple GUI}
         \label{color_recovered_grayscale_image_net_upscaled_first_ml_5}
     \end{subfigure}
     \hfill
     \begin{subfigure}[b]{0.16\textwidth}
         \centering
         \includegraphics[width=1.5cm]{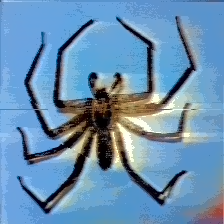}
         \caption{Img. 5, Attack C or D, Py Simple GUI}
         \label{py_color_image_net_grayscale_reconstructed_5}
     \end{subfigure}
     \caption{\small Recolored Grayscale ImageNet 5 images using DeepAI and Py Simple GUI image colorizer.}
     \label{deepai_vs_py_simple_gui_5}
\end{figure*}

Comparison of the recolored images colored using the DeepAI colorizer and Py Simple GUI colorizer is shown in Figures~\ref{deepai_vs_py_simple_gui_1} to~\ref{deepai_vs_py_simple_gui_5}. The RMSE is also shown in Table~\ref{table_deepa-ai_vs_py_simple_rms}. From the results, we can observe the Deep AI colorizer produces results more similar to the original images than those of the Py Simple GUI colorizer.

\begin{table*}[t]
\centering
\caption{\small RMSE between ImageNet inputs and reconstructed images (pixel-value leakage). RMSE is computed as $\sqrt{\mathrm{mean}((\hat I - I)^2)}$ over all pixels and channels, with pixel values normalized to $[0,1]$; lower is better. Complementary PSNR/SSIM for the same image pairs are given in Table~\ref{tab:image_net_psnr_ssim}, and the quality of the raw side-channel recovery (before ML post-processing) in Table~\ref{tab:raw_recovery_metrics}.}
\begin{adjustbox}{width=1\textwidth}
\label{table_image_net_rms}
\small
\begin{tabular}{|p{1.5cm}|p{1.5cm}|p{1.5cm}|p{1.5cm}|p{1.5cm}|p{1.5cm}|p{1.5cm}|p{1.5cm}|p{1.5cm}|}
\hline
\textbf{ImageNet Image} &  \textbf{Resolution} & \textbf{RMSE of Original Image } & \textbf{RMSE of Attack A} & \textbf{RMSE of Attack B} & \textbf{RMSE of Attack C} & \textbf{RMSE of Attack D} & \textbf{RMSE of Attack E} & \textbf{RMSE of Attack F}
\\ \hline \hline
Image 1 & 500x409 & 0
&  0.12 & 0.13 & 0.12 & 0.13 & 0.14 & 0.15 \\ \hline
Image 2 & 760x459 & 0  & 0.23   &  0.23 &0.23 & 0.23 & 0.23 & 0.23\\ \hline
Image 3  & 1200x899 & 0   & 0.21    &   0.21 & 0.21 & 0.21 & 0.18 & 0.18\\ \hline
Image 4 & 1600x1123 & 0     &  0.13  &   0.13   &0.12 & 0.13 & 0.10 & 0.10\\ \hline
Image 5  & 2272x1520 & 0    &  0.37 &   0.36  &0.37  & 0.37 & 0.34 & 0.34\\ \hline
\end{tabular}
\end{adjustbox}
\end{table*}

\begin{table*}[t]
\centering
\small
\caption{\small RMSE between ImageNet reference images and colorized outputs (DeepAI; PySimpleGUI Image Colorizer) to assess pixel-value leakage. RMSE is computed as $\sqrt{\mathrm{mean}((\hat I - I)^2)}$ over all pixels and channels, with pixel values normalized to $[0,1]$; lower is better.}

\begin{adjustbox}{width=1\textwidth}
\label{table_deepa-ai_vs_py_simple_rms}
\small
\begin{tabular}{|p{1.5cm}|p{1.5cm}|p{1.5cm}|p{1.5cm}|p{1.5cm}|p{1.5cm}|p{1.5cm}|p{1.5cm}|p{1.5cm}|}
\hline
\textbf{ImageNet Image} &  \textbf{Resolution} & \textbf{RMSE of Original Image } & \textbf{RMSE of Attack A recolored using DeepAI colorizer} & \textbf{RMSE of Attack B recolored using DeepAI colorizer} & \textbf{RMSE of Attack C or D recolored  using DeepAI colorizer} & \textbf{RMSE of Attack A recolored using Py Simple GUI colorizer} & \textbf{RMSE of Attack B recolored using Py Simple GUI colorizer} & \textbf{RMSE of Attack C or D  recolored using Py Simple GUI colorizer }
\\ \hline \hline
Image 1  & 500x409 & 0
&  0.12 & 0.13 & 0.13 & 0.16& 0.16 & 0.17 \\ \hline
Image 2 & 760x459 & 0  & 0.23  &  0.23 &0.23 & 0.23 & 0.23 & 0.23\\ \hline
Image 3  & 1200x899 & 0   & 0.21   &   0.21  & 0.21 & 0.27 & 0.28 & 0.28\\ \hline
Image 4  & 1600x1123 & 0    &  0.13  &   0.13  &0.13 & 0.24& 0.24 & 0.24\\ \hline
Image 5  & 2272x1520 & 0     &  0.37 &   0.36  &0.37  & 0.38 & 0.38 & 0.37\\ \hline

\end{tabular}
\end{adjustbox}
\end{table*}

\section{End-to-End Attack}
\label{sec_recoloring}

\begin{figure*}[t]
    \centering
    \begin{subfigure}[b]{0.31\textwidth}
        \centering
        \includegraphics[width=0.5\linewidth]{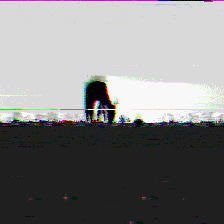}
        \caption{\footnotesize Reconstructed Hamming Weight Image, Top Image 1}
        \label{fig:reconstructed_top_image_1}
    \end{subfigure}
    \hfill
    \begin{subfigure}[b]{0.31\textwidth}
        \centering
        \includegraphics[width=0.5\linewidth]{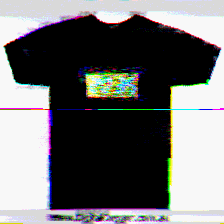}
        \caption{\footnotesize Reconstructed Hamming Weight Image, Top Image 2}
        \label{fig:reconstructed_top_image_2}
    \end{subfigure}
    \hfill
    \begin{subfigure}[b]{0.31\textwidth}
        \centering
        \includegraphics[width=0.5\linewidth]{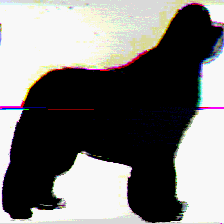}
        \caption{\footnotesize Reconstructed Hamming Weight Image, Top Image 3}
        \label{fig:reconstructed_top_image_3}
    \end{subfigure}

    \vspace{1em} %

    \begin{subfigure}[b]{0.31\textwidth}
        \centering
        \includegraphics[width=0.5\linewidth]{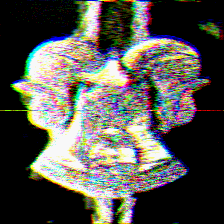}
        \caption{\footnotesize Reconstructed Hamming Weight Image, Top Image Colored 1}
        \label{fig:reconstructed_top_image_color_1}
    \end{subfigure}
    \hfill
    \begin{subfigure}[b]{0.31\textwidth}
        \centering
        \includegraphics[width=0.5\linewidth]{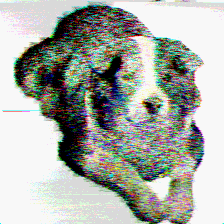}
        \caption{\footnotesize Reconstructed Hamming Weight Image, Top Image Colored 2}
        \label{fig:reconstructed_top_image_color_2}
    \end{subfigure}
    \hfill
    \begin{subfigure}[b]{0.31\textwidth}
        \centering
        \includegraphics[width=0.5\linewidth]{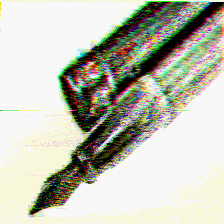}
        \caption{\footnotesize Reconstructed Hamming Weight Image, Top Image Colored 3}
        \label{fig:reconstructed_top_image_color_3}
    \end{subfigure}

    \vspace{1em} %

    \begin{subfigure}[b]{0.31\textwidth}
        \centering
        \includegraphics[width=0.5\linewidth]{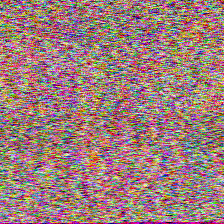}
        \caption{\footnotesize Reconstructed Hamming Weight Image, Bottom Image Colored 1}
        \label{fig:reconstructed_bottom_image_color_1}
    \end{subfigure}
    \hfill
    \begin{subfigure}[b]{0.31\textwidth}
        \centering
        \includegraphics[width=0.5\linewidth]{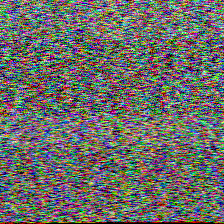}
        \caption{\footnotesize Reconstructed Hamming Weight Image, Bottom Image Colored 2}
        \label{fig:reconstructed_bottom_image_color_2}
    \end{subfigure}
    \hfill
    \begin{subfigure}[b]{0.31\textwidth}
        \centering
        \includegraphics[width=0.5\linewidth]{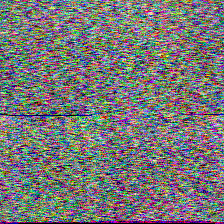}
        \caption{\footnotesize Reconstructed Hamming Weight Image, Bottom Image Colored 3}
        \label{fig:reconstructed_bottom_image_color_3}
    \end{subfigure}

    \caption{\small Reconstructed Hamming Weight images.}
    \label{fig:reconstructed_image_net_images_first}
\end{figure*}

\begin{figure*}[t]
    \centering
        \includegraphics[width=0.2\linewidth]{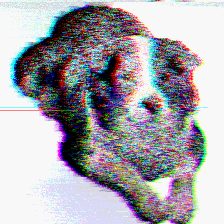}
    \caption{\small Reconstructed Hamming Weight image for VGG-16. The same image for VGG-19 is shown in Figure~\ref{fig:reconstructed_top_image_color_2}.}
    \label{fig:reconstructed_image_net_images_first_vgg_16}
\end{figure*}

Our end-to-end attack involves recovering the Hamming Weight scaled images and performing inferences using our trained models. The recovered Hamming Weight images for VGG-19 can be seen in Figure~\ref{fig:reconstructed_image_net_images_first}. Figure~\ref{fig:reconstructed_image_net_images_first_vgg_16} shows the recovered Hamming Weight image for VGG-16. We observe that the top 3 contrast images and top 3 colored contrast images reasonably match the reference Hamming Weight images computed from the original images, as shown before in Figure~\ref{fig:original_image_net_images_hw_scaled}. However, the bottom 3 images clearly do not match the reference Hamming Weight images.

\begin{table*}[t]
    \centering
    \caption{\small Attacker's ML model performance on different images without our defense being used. A checkmark means attacker is able to classify the victim image correctly and learn its class. We report all nine test images, including the three low-contrast (``Bottom Image'') cases; consistent with the failed reconstructions discussed in Section~\ref{sec_recoloring}, none of the low-contrast images is classified correctly by any model, so these rows make the method's failure cases explicit rather than reporting only the successful high-contrast cases. The two VGG19 columns and the two ViT-L/16 columns are not duplicates: they are two different training checkpoints (epochs) of the same architecture, as indicated by the epoch labels in the column headers and the ``Epoch No'' row, and as summarized in Table~\ref{tab:training}.}
    \begin{adjustbox}{width=0.99\textwidth}
    \label{tab:model_performance_no_defense}
    \begin{tabular}{|p{3cm}|p{1.3cm}|p{1.3cm}|p{1.75cm}|p{1.75cm}|p{2.2cm}|p{2.2cm}|p{1.75cm}|}
        \hline
        \textbf{ImageNet Image} & \textbf{VGG19 (ep.~219)} & \textbf{VGG19 (ep.~100)} & \textbf{ResNeXt-101 32x8d} & \textbf{MobileNet V3 Large} & \textbf{ViT-L/16 (ep.~131)} & \textbf{ViT-L/16 (ep.~100)} & \textbf{ConvNeXt Large} \\
        \hline
        Epoch No & 219 & 100 & 100 & 100 & 131 & 100 & 90 \\ \hline
        \hline
         Reconstructed Image, Top Image 1 & $\times$ & $\times$ & $\times$ & $\times$  &  \checkmark & $\times$ & $\times$ \\ \hline
         Reconstructed Image, Top Image 2 & $\times$ & $\times$ & $\times$ & $\times$  &  \checkmark & $\times$ &  \checkmark\\ \hline
         Reconstructed Image, Top Image 3 & \checkmark & \checkmark & \checkmark & $\times$  & $\times$  & $\times$ & $\times$ \\ \hline
         Reconstructed Image, Top Image Colored 1 & \checkmark & \checkmark & \checkmark & $\times$  & $\times$  & \checkmark & \checkmark \\ \hline
         Reconstructed Image, Top Image Colored 2 & $\times$ & $\times$ & \checkmark & $\times$  &  $\times$ & $\times$ & $\times$ \\ \hline
         Reconstructed Image, Top Image Colored 3 &\checkmark  & \checkmark & \checkmark & \checkmark & $\times$  & $\times$ & $\times$ \\ \hline
        Reconstructed Image, Bottom Image 1 & $\times$ & $\times$ & $\times$ & $\times$ & $\times$ & $\times$ & $\times$ \\ \hline
        Reconstructed Image, Bottom Image 2 & $\times$ & $\times$ & $\times$ & $\times$ & $\times$ & $\times$ & $\times$ \\ \hline
         Reconstructed Image, Bottom Image 3 & $\times$ & $\times$ & $\times$ & $\times$ & $\times$ & $\times$ & $\times$ \\ \hline
    \end{tabular}
    \end{adjustbox}
\end{table*}

Our observations are quantified in Table~\ref{tab:model_performance_no_defense}. The table shows the inference results for the different images used in our attack. The images were fed into different machine learning models, and whether the model was able to correctly classify the image. This table demonstrates the effectiveness of the attack on various test images across five neural network models. We observe that the top color contrast images are better predicted than the highest contrast images. Additionally, older neural network models seem to perform better than state-of-the-art transformers and convolutional neural networks. Thus, an  attacker could accurately predict the recovered~images.

\section{Defense by Pixel Perturbation}

We first present how our defense works by pixel substitution. We then evaluate it on the selected test images and at full scale on the ImageNet validation set, measuring both how well the attacker can still classify the defended Hamming Weight images and how much accuracy the victim's own models lose.

\subsection{Pixel Perturbation by 30, 60, and 90 absolute pixel difference.}

\begin{figure*}[t]
    \centering
    \begin{subfigure}[b]{0.12\textwidth}
        \centering
        \includegraphics[width=\linewidth]{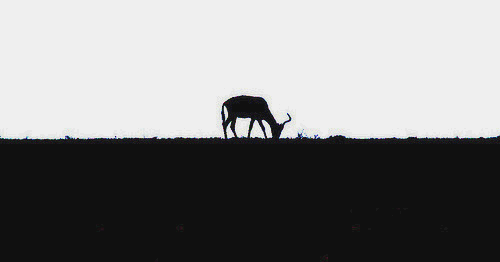}
        \caption{\footnotesize Perturbed Image, Image~1}
        \label{fig:altered_image_1}
    \end{subfigure}
    \hfill
    \begin{subfigure}[b]{0.12\textwidth}
        \centering
        \includegraphics[width=\linewidth]{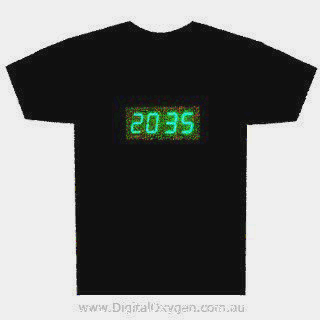}
        \caption{\footnotesize Perturbed Image, Image~2}
        \label{fig:altered_image_2}
    \end{subfigure}
    \hfill
    \begin{subfigure}[b]{0.12\textwidth}
        \centering
        \includegraphics[width=\linewidth]{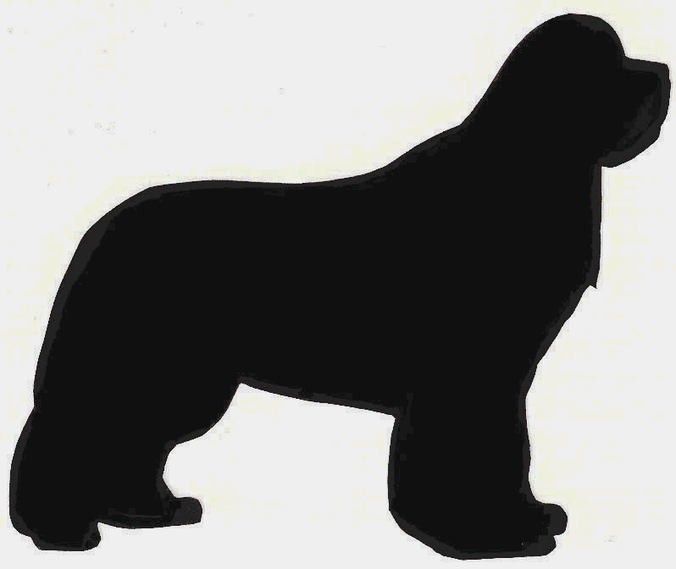}
        \caption{\footnotesize Perturbed Image, Image~3}
        \label{fig:altered_image_3}
    \end{subfigure}
    \hfill
    \begin{subfigure}[b]{0.12\textwidth}
        \centering
        \includegraphics[width=\linewidth]{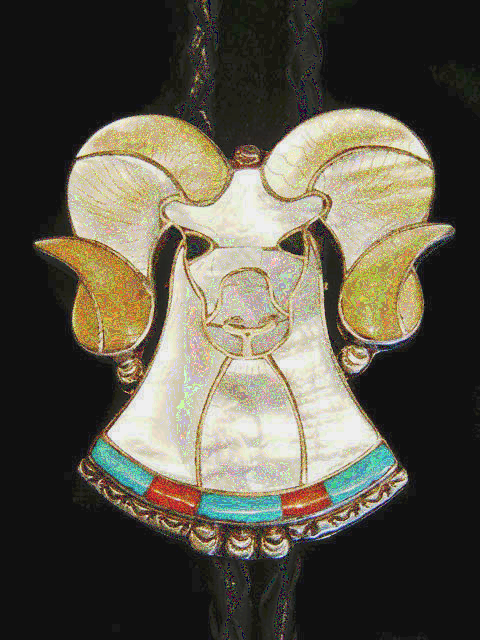}
        \caption{\footnotesize Perturbed Image, Image~4}
        \label{fig:altered_image_4}
    \end{subfigure}
    \hfill
    \begin{subfigure}[b]{0.12\textwidth}
        \centering
        \includegraphics[width=\linewidth]{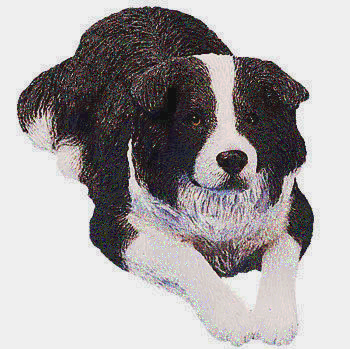}
        \caption{\footnotesize Perturbed Image, Image~5}
        \label{fig:altered_image_5}
    \end{subfigure}
    \hfill
    \begin{subfigure}[b]{0.12\textwidth}
        \centering
        \includegraphics[width=\linewidth]{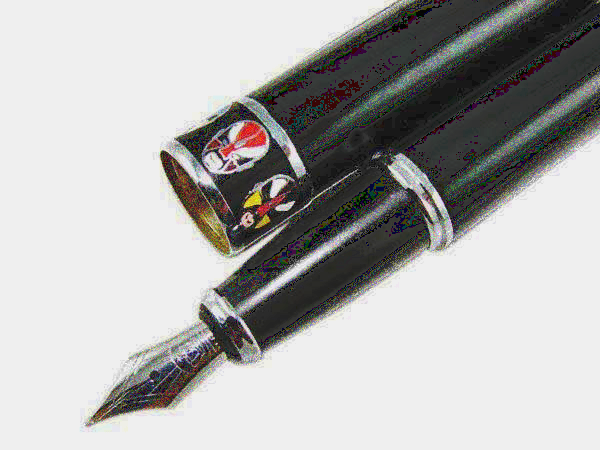}
        \caption{\footnotesize Perturbed Image, Image~6}
        \label{fig:altered_image_6}
    \end{subfigure}

    \caption{\small Images perturbed with our defense, images can be compared to original images with no defense shown in Figure~\ref{fig:original_image_net_images_first}; they are visually very similar to originals, so humans do not perceive major differences, while the perturbations help defend the attacks.}
    \label{fig:altered_image_net_images}
\end{figure*}

As we have discovered, low contrasting Hamming Weight images are difficult to classify. As a result, to mitigate our attack, we devised a potential defense strategy that perturbs the victim's input pixel values so that the Hamming Weight image observed by the attacker changes. Because Hamming Weight is not monotonic in intensity, nearby integer values can have very different Hamming Weights; for example, 127 and 128 differ by one intensity level but by six in Hamming Weight. The substitution is a fixed lookup table over the 256 possible 8-bit values, parameterized by a budget $B \in \{30, 60, 90\}$: for each value $i$, among all values $j$ with $|j-i| \le B$ and $\mathrm{HW}(j) \neq \mathrm{HW}(i)$, we select the one that maximizes $|j-i| / |\mathrm{HW}(j)-\mathrm{HW}(i)|$. In practice, this rule changes the Hamming Weight of every pixel by exactly one level while moving its intensity by close to the full budget (on average $29.0$, $58.4$, and $87.5$ levels over the 256 values for $B = 30$, $60$, and $90$). Because black pixels (Hamming Weight $0$) are always mapped to Hamming Weight $1$ and white pixels (Hamming Weight $8$) to $7$, the substitution compresses the Hamming Weight contrast of the image while preserving its spatial structure. This approach aims to hinder the attacker's ability to accurately recover the original images and perform inferences using the trained DNN models.
The most likely scenario is that the victim tenant applies this perturbation as a pre-processing step inside their own secure logic on the FPGA before feeding the image to their ML accelerator. This protects the data during the vulnerable computation phase.

Figure~\ref{fig:altered_image_net_images} shows the perturbed test images with our defense applied when perturbing by 30 absolute pixel difference. The perturbed images remain visually similar to the originals: their content and spatial structure are preserved, and the most visible change is a reduction in contrast as black and white regions shift toward gray.

\begin{figure*}[h]
    \centering
    \begin{subfigure}[b]{0.1\textwidth}
        \centering
        \includegraphics[width=\linewidth]{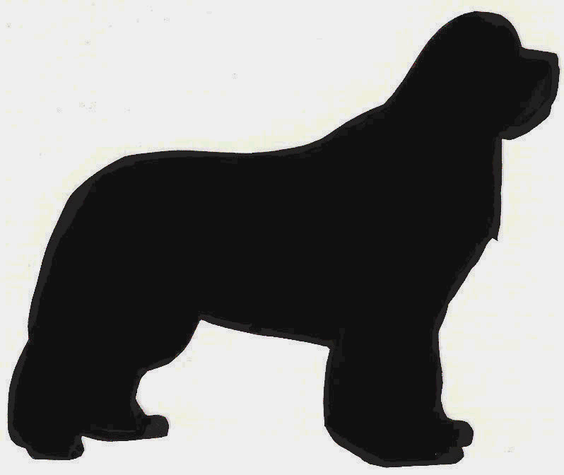}
        \caption{\footnotesize Perturbed Image, Image~3 with 30}
        \label{fig:altered_image_8_30}
    \end{subfigure}
    \hfill
    \begin{subfigure}[b]{0.1\textwidth}
        \centering
        \includegraphics[width=\linewidth]{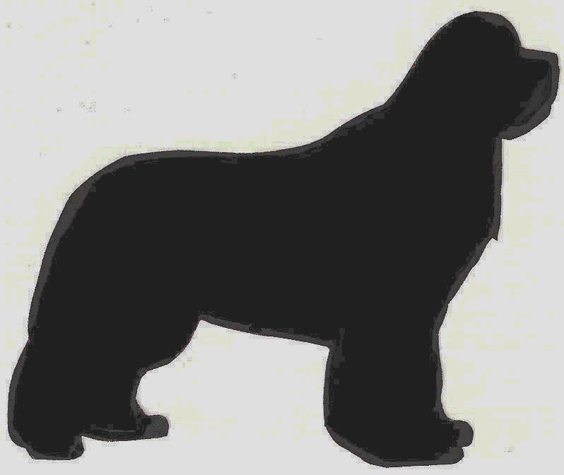}
        \caption{\footnotesize Perturbed Image, Image~3 with 60}
        \label{fig:altered_image_8_60}
    \end{subfigure}
    \hfill
    \begin{subfigure}[b]{0.1\textwidth}
        \centering
        \includegraphics[width=\linewidth]{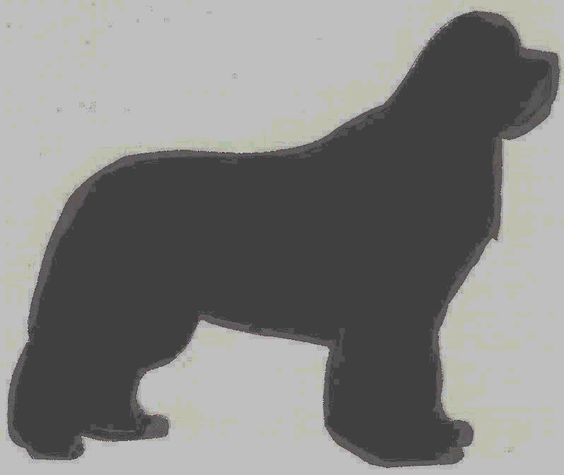}
        \caption{\footnotesize Perturbed Image, Image~3 with 90}
        \label{fig:altered_image_8_90}
    \end{subfigure}
    \hfill
    \begin{subfigure}[b]{0.1\textwidth}
        \centering
        \includegraphics[width=\linewidth]{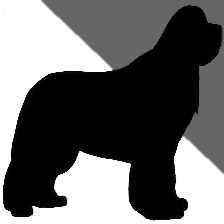}
        \caption{\footnotesize Perturbed Image, Image~3 with shadow}
        \label{fig:altered_image_8_shadow}
    \end{subfigure}
    \hfill
    \begin{subfigure}[b]{0.1\textwidth}
        \centering
        \includegraphics[width=\linewidth]{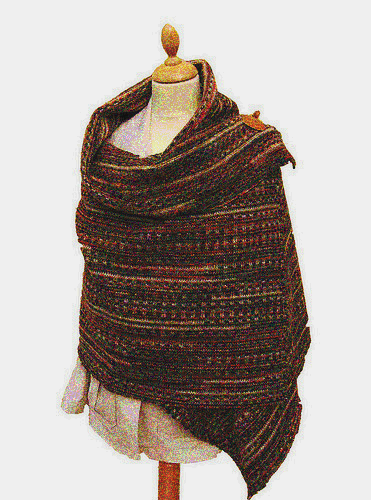}
        \caption{\footnotesize Perturbed Image, Image~2000 with 30}
        \label{fig:altered_image_2000_30}
    \end{subfigure}
    \hfill
    \begin{subfigure}[b]{0.1\textwidth}
        \centering
        \includegraphics[width=\linewidth]{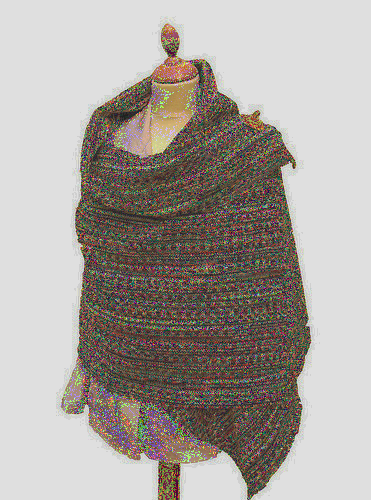}
        \caption{\footnotesize Perturbed Image, Image~2000 with 60}
        \label{fig:altered_image_2000_60}
    \end{subfigure}
    \hfill
    \begin{subfigure}[b]{0.1\textwidth}
        \centering
        \includegraphics[width=\linewidth]{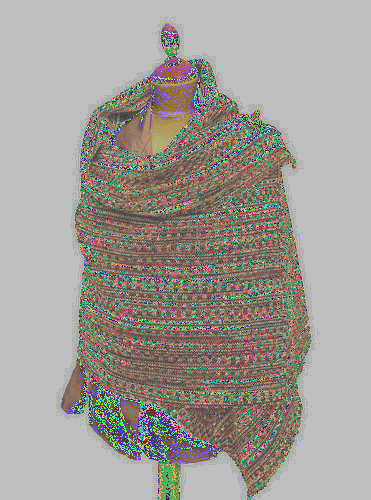}
        \caption{\footnotesize Perturbed Image, Image~2000 with 90}
        \label{fig:altered_image_2000_90}
    \end{subfigure}
    \hfill
    \begin{subfigure}[b]{0.1\textwidth}
        \centering
        \includegraphics[width=\linewidth]{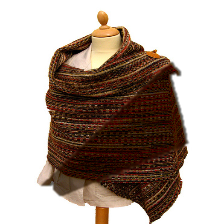}
        \caption{\footnotesize Perturbed Image, Image~2000 with shadow}
        \label{fig:altered_image_2000_shadow}
    \end{subfigure}

    \caption{\small Sample Images perturbed with our 4 defenses.}
\label{fig:altered_image_net_images_30_60_90}
\end{figure*}

Figure~\ref{fig:altered_image_net_images_30_60_90} shows images perturbed by 60 absolute pixel difference. This figure includes an additional image used to test the method’s robustness, enhancing the comprehensiveness of the evaluation. The added image introduces variability, enabling an assessment of the model's performance under different conditions and offering further insights into its adaptability and reliability in practical applications.
Lastly, Figure~\ref{fig:altered_image_net_images_30_60_90} also shows various images perturbed by 90 absolute pixel difference.

\subsection{Adding Adversarial Shadow to Image}

We start with the shadow-based adversarial attack, originally proposed by Zhong et al.~\cite{zhong2022shadows}, which utilizes shadows as a natural and stealthy form of perturbation. This attack exploits the fact that shadows, a common phenomenon in real-world images, can subtly alter the pixel values of an image without raising suspicion. We apply this attack to the ImageNet dataset, targeting its large-scale, diverse set of images to demonstrate the effectiveness of this technique in a more generalized and challenging scenario. By leveraging shadows as perturbations, we aim to create imperceptible changes that deceive machine learning models while maintaining visual realism.

The shadow attack, leveraging natural phenomena of light shadows, can subtly manipulate machine learning models in ways that are difficult to detect. The shadow attack leverages a Particle
Swarm Optimization (PSO)~\cite{kennedy1995particle} approach to find the optimal location to add a shadow to an image. A shadow is a polygon of dark or gray color that is meant to look like a shadow cast by a tree, building, or another object. The attack analyzes different shadow shapes and locations in order to determine the optimal type of shadow. Once the best shadow is determined, the attacker can reproduce it by applying the same shadow to a digital image. Note that using PSO to find the optimal shadow is computationally expensive and is typically done offline. For a practical defense, one would have to pre-compute a set of ``adversarial shadows'' and apply one at random during inference, which would be a much lower-cost operation.

The limitation of the existing attack is that for each image, there is a need to define a mask, or region where the shadow is applied. For ImageNet images, the mask would simply be a binary mask of the same size as the image itself, as generating specific masks for each image would be impractical due to varying content and structures. Figure~\ref{fig:altered_image_net_images_30_60_90} shows the adversarial shadow added to images.

\subsection{Reconstructed Images without and with Defense}

\begin{figure*}[t]
    \centering
    \begin{subfigure}[b]{0.12\textwidth}
        \centering
        \includegraphics[width=\linewidth]{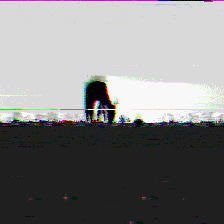}
        \caption{\footnotesize Reconstructed Original Hamming Weight Image, Image~1}
        \label{fig:reconstructed_image_1}
    \end{subfigure}
    \hfill
    \begin{subfigure}[b]{0.12\textwidth}
        \centering
        \includegraphics[width=\linewidth]{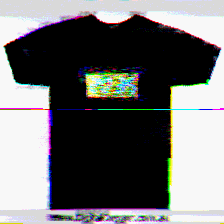}
        \caption{\footnotesize Reconstructed Original Hamming Weight Image, Image~2}
        \label{fig:reconstructed_image_2}
    \end{subfigure}
    \hfill
    \begin{subfigure}[b]{0.12\textwidth}
        \centering
        \includegraphics[width=\linewidth]{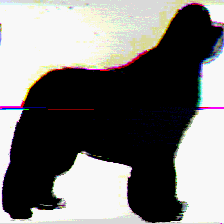}
        \caption{\footnotesize Reconstructed Original Hamming Weight Image, Image~3}
        \label{fig:reconstructed_image_3}
    \end{subfigure}
    \hfill
    \begin{subfigure}[b]{0.12\textwidth}
        \centering
        \includegraphics[width=\linewidth]{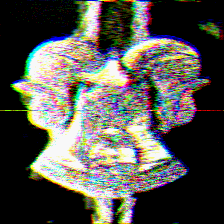}
        \caption{\footnotesize Reconstructed Original Hamming Weight Image, Image~4}
        \label{fig:reconstructed_image_4}
    \end{subfigure}
    \hfill
    \begin{subfigure}[b]{0.12\textwidth}
        \centering
        \includegraphics[width=\linewidth]{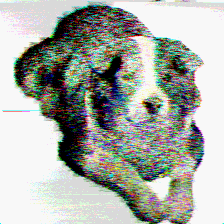}
        \caption{\footnotesize Reconstructed Original Hamming Weight Image, Image~5}
        \label{fig:reconstructed_image_5}
    \end{subfigure}
    \hfill
    \begin{subfigure}[b]{0.12\textwidth}
        \centering
        \includegraphics[width=\linewidth]{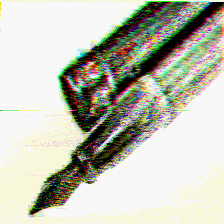}
        \caption{\footnotesize Reconstructed Original Hamming Weight Image, Image~6}
        \label{fig:reconstructed_image_6}
    \end{subfigure}

    \caption{\small Reconstructed Hamming Weight images, these are same as in Figure~\ref{fig:reconstructed_image_net_images_first}.}
    \label{fig:reconstructed_image_net_images}
\end{figure*}

\begin{figure*}[t]
    \centering
    \begin{subfigure}[b]{0.12\textwidth}
        \centering
        \includegraphics[width=\linewidth]{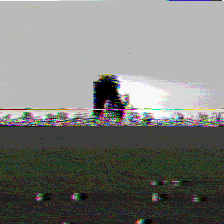}
        \caption{\footnotesize Reconstructed Perturbed Hamming Weight Image, Image~1}
        \label{fig:reconstructed_altered_image_1}
    \end{subfigure}
    \hfill
    \begin{subfigure}[b]{0.12\textwidth}
        \centering
        \includegraphics[width=\linewidth]{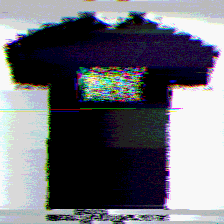}
        \caption{\footnotesize Reconstructed Perturbed Hamming Weight Image, Image~2}
        \label{fig:reconstructed_altered_image_2}
    \end{subfigure}
    \hfill
    \begin{subfigure}[b]{0.12\textwidth}
        \centering
        \includegraphics[width=\linewidth]{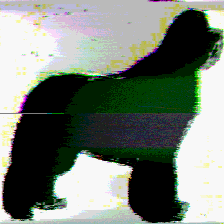}
        \caption{\footnotesize Reconstructed Perturbed Hamming Weight Image, Image~3}
        \label{fig:reconstructed_altered_image_3}
    \end{subfigure}
    \hfill
    \begin{subfigure}[b]{0.12\textwidth}
        \centering
        \includegraphics[width=\linewidth]{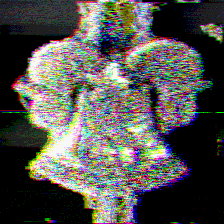}
        \caption{\footnotesize Reconstructed Perturbed Hamming Weight Image, Image~4}
        \label{fig:reconstructed_altered_image_4}
    \end{subfigure}
    \hfill
    \begin{subfigure}[b]{0.12\textwidth}
        \centering
        \includegraphics[width=\linewidth]{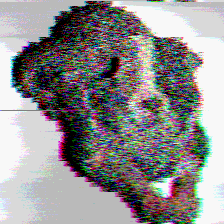}
        \caption{\footnotesize Reconstructed Perturbed Hamming Weight Image, Image~5}
        \label{fig:reconstructed_altered_image_5}
    \end{subfigure}
    \hfill
    \begin{subfigure}[b]{0.12\textwidth}
        \centering
        \includegraphics[width=\linewidth]{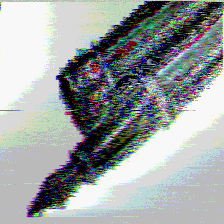}
        \caption{\footnotesize Reconstructed Perturbed Hamming Weight Image, Image~6}
        \label{fig:reconstructed_altered_image_6}
    \end{subfigure}

    \caption{\small Reconstructed Perturbed Hamming Weight images.}
    \label{fig:reconstructed_altered_image_net_images}
\end{figure*}

Here we compare how the Hamming Weight images look when they are recovered from the TDC measurements.
Figure~\ref{fig:reconstructed_altered_image_net_images} shows the Hamming Weight images obtained from TDC power measurements when the defense was applied to the input images. These recovered
images exhibit differences compared to those without the defense, as seen in Figure~\ref{fig:reconstructed_image_net_images}. This indicates that our defense strategy introduces enough noise to obscure the original
image content from the attacker.

\subsection{Quantifying Effectiveness of the Defense}

\begin{table*}[t]
    \centering
     \caption{\small Attacker's ML model performance on different images with our defense being used. Checkmark means attacker is able to classify the victim image correctly and learn its~class. ``Epoch No'' refers to the checkpoint of the attacker's trained model that achieved the best validation accuracy during its training. }
    \begin{adjustbox}{width=0.99\textwidth}
    \label{tab:model_performance_defense}
    \begin{tabular}{|p{3cm}|p{1.3cm}|p{1.3cm}|p{1.75cm}|p{1.75cm}|p{2.2cm}|p{2.2cm}|p{1.75cm}|}
        \hline
        \textbf{ImageNet Image} & \textbf{VGG19 (ep.~219)} & \textbf{VGG19 (ep.~100)} & \textbf{ResNeXt-101 32x8d} & \textbf{MobileNet V3 Large} & \textbf{ViT-L/16 (ep.~131)} & \textbf{ViT-L/16 (ep.~100)} & \textbf{ConvNeXt Large} \\
        \hline
        Epoch No & 219 & 100 & 100 & 100 & 131 & 100 & 90 \\ \hline
        \hline
         Reconstructed Perturbed Image, Top Image 1 & $\times$ & $\times$ & $\times$ & $\times$  &  $\times$ & $\times$ & $\times$ \\ \hline
         Reconstructed Perturbed Image, Top Image 2 & $\times$ & \checkmark & $\times$ & $\times$  &  \checkmark & $\times$&  $\times$ \\ \hline
         Reconstructed Perturbed Image, Top Image 3 & \checkmark & \checkmark & \checkmark & $\times$  & $\times$  & $\times$ & $\times$ \\ \hline
         Reconstructed Perturbed Image, Top Image Colored 1 & $\times$ & $\times$ & $\times$ & $\times$  & $\times$  & $\times$ & \checkmark \\ \hline
         Reconstructed Perturbed Image, Top Image Colored 2 & $\times$ & $\times$ & $\times$ & $\times$  &  $\times$ & $\times$ & $\times$ \\ \hline
         Reconstructed Perturbed Image, Top Image Colored 3 & $\times$ & $\times$ & $\times$ & $\times$  &  $\times$ & $\times$ & $\times$ \\ \hline
    \end{tabular}
   \end{adjustbox}
\end{table*}

\begin{table*}[t]
    \centering
     \caption{\small Attacker's ML model performance on Top Image 3 with our 60 and 90 pixel perturbation being used. Checkmark means attacker is able to classify the victim image correctly and learn its~class. ``Epoch No'' refers to the checkpoint of the attacker's trained model that achieved the best validation accuracy during its training.}
     \label{tab:model_performance_defense_top_image_3}
    \begin{adjustbox}{width=0.99\textwidth}
    \begin{tabular}{|p{3cm}|p{1.3cm}|p{1.3cm}|p{1.75cm}|p{1.75cm}|p{2.2cm}|p{2.2cm}|p{1.75cm}|}
        \hline
        \textbf{ImageNet Image} & \textbf{VGG19 (ep.~219)} & \textbf{VGG19 (ep.~100)} & \textbf{ResNeXt-101 32x8d} & \textbf{MobileNet V3 Large} & \textbf{ViT-L/16 (ep.~131)} & \textbf{ViT-L/16 (ep.~100)} & \textbf{ConvNeXt Large} \\
        \hline
        Epoch No & 219 & 100 & 100 & 100 & 131 & 100 & 90 \\ \hline
        \hline
         Reconstructed Perturbed Image with 60, Top Image 3 &  $\times$  &  $\times$  & \checkmark & $\times$  & $\times$  & $\times$ & $\times$ \\ \hline
          Reconstructed Perturbed Image with 90, Top Image 3 & \checkmark & \checkmark & \checkmark & $\times$  & $\times$  & $\times$ & $\times$ \\ \hline

    \end{tabular}
   \end{adjustbox}
\end{table*}

\begin{table*}[t]
    \centering
     \caption{\small Attacker's ML model performance on 2000 Image  with our 60 and 90 pixel perturbation being used. Checkmark means attacker is able to classify the victim image correctly and learn its~class. ``Epoch No'' refers to the checkpoint of the attacker's trained model that achieved the best validation accuracy during its training. }
    \label{tab:model_performance_defense_2000_image}
    \begin{adjustbox}{width=0.99\textwidth}
    \begin{tabular}{|p{3cm}|p{1.3cm}|p{1.3cm}|p{1.75cm}|p{1.75cm}|p{2.2cm}|p{2.2cm}|p{1.75cm}|}
        \hline
        \textbf{ImageNet Image} & \textbf{VGG19 (ep.~219)} & \textbf{VGG19 (ep.~100)} & \textbf{ResNeXt-101 32x8d} & \textbf{MobileNet V3 Large} & \textbf{ViT-L/16 (ep.~131)} & \textbf{ViT-L/16 (ep.~100)} & \textbf{ConvNeXt Large} \\
        \hline
        Epoch No & 219 & 100 & 100 & 100 & 131 & 100 & 90 \\ \hline
        \hline
         Reconstructed Perturbed Image with 60, 2000 Image &  $\times$  &  $\times$  &  $\times$  & $\times$  & $\times$  & $\times$ & $\times$ \\ \hline
          Reconstructed Perturbed Image with 90, 2000 Image &  $\times$  &  $\times$ &  $\times$  & $\times$  & $\times$  & $\times$ & $\times$ \\ \hline

    \end{tabular}
   \end{adjustbox}
\end{table*}

\begin{table*}[t]
    \centering
     \caption{\small Attacker's ML model performance on Top Image 3 and 2000 Image with  shadow adversarial defense being used. Checkmark means attacker is able to classify the victim image correctly and learn its~class. ``Epoch No'' refers to the checkpoint of the attacker's trained model that achieved the best validation accuracy during its training. }
    \label{tab:model_performance_defense_shadow}
    \begin{adjustbox}{width=0.99\textwidth}
    \begin{tabular}{|p{3cm}|p{1.3cm}|p{1.3cm}|p{1.75cm}|p{1.75cm}|p{2.2cm}|p{2.2cm}|p{1.75cm}|}
        \hline
        \textbf{ImageNet Image} & \textbf{VGG19 (ep.~219)} & \textbf{VGG19 (ep.~100)} & \textbf{ResNeXt-101 32x8d} & \textbf{MobileNet V3 Large} & \textbf{ViT-L/16 (ep.~131)} & \textbf{ViT-L/16 (ep.~100)} & \textbf{ConvNeXt Large} \\
        \hline
        Epoch No & 219 & 100 & 100 & 100 & 131 & 100 & 90 \\ \hline
        \hline
        Reconstructed Perturbed Image with shadow, Top Img. 3 &  $\times$  & \checkmark & \checkmark & $\times$  & $\times$  & $\times$ & $\times$ \\ \hline
           Reconstructed Perturbed Image with shadow, 2000 Image &  \checkmark  & \checkmark & \checkmark & $\times$  & $\times$  & $\times$ & $\times$ \\ \hline

    \end{tabular}
   \end{adjustbox}
\end{table*}

Table~\ref{tab:model_performance_defense} illustrates the impact of implementing this defense strategy on the selected test images. On these images, the attacker's original models fail to predict the correct labels for most inputs once the defense is applied, particularly the colored inputs, and the number of correctly predicted images drops across the different neural network models. These per-image results show specific instances where the perturbation prevents a non-adaptive attacker from classifying the input; the full-scale evaluation below shows how far this effect generalizes.

On the selected test images, this perturbation substantially reduces a non-adaptive attacker's prediction accuracy. However, as we quantify below on the full validation set, the reduction is much smaller at scale, it comes at a significant cost to the victim model's own accuracy, and the perturbation does not withstand an adaptive attacker that retrains on the perturbed data.

We evaluate the defense on two held-out test cases drawn from the ImageNet validation set: ``Top Image 3'', one of the high-contrast validation images used in the earlier reconstruction experiments, and ``Image 2000'', an additional validation image used to evaluate the defense.
Tables~\ref{tab:model_performance_defense_top_image_3},~\ref{tab:model_performance_defense_2000_image}, and~\ref{tab:model_performance_defense_shadow} show the effect of the stronger 60 and 90 perturbations and of the shadow defense, respectively, on the test images. We observe that the perturbation prevents the non-adaptive attacker from classifying Image 2000 at both budgets, whereas Top Image 3 is still classified by some models, including three of the seven at budget 90. However, the images are still predicted even with the shadow defense.
Surprisingly in some cases, the results show that the shadow defense sometimes failed or helped the attacker. This could be perhaps the shadow, while fooling a standard CNN, creates high-contrast edges that make the Hamming Weight signature
more distinct and thus easier for the attacker's specialized model to classify.

\begin{table*}[t]
    \centering
    \caption{\small Victim utility cost of the pixel-substitution defense at three perturbation budgets, computed in software over the 50{,}000-image ImageNet validation set. For each victim model we report top-1 and top-5 accuracy (\%) on clean inputs and on inputs transformed by the defense at budgets $30$, $60$, and $90$; the top-1 drop relative to clean is given in parentheses and is computed from unrounded accuracies. The defense imposes a large, budget-dependent cost on the victim model's \emph{own} accuracy, and this cost is strongly architecture-dependent: VGG-19 falls below $3\%$ top-1 at budgets $60$ and $90$, while the modern ConvNeXt-L and ViT-L/16 are far more robust, particularly at budget $30$.}
    \label{tab:defense_complete}
    \small
    \begin{adjustbox}{max width=\textwidth}
    \begin{tabular}{|l|c|c|c|c|c|}
        \hline
        \textbf{Victim Model} & \textbf{Metric} & \textbf{Clean} & \textbf{Budget 30} & \textbf{Budget 60} & \textbf{Budget 90} \\
        \hline \hline
        \multirow{2}{*}{VGG19}
          & Top-1 & 72.4 & 28.5 ($-$43.9) & 2.9 ($-$69.5)  & 2.4 ($-$70.0) \\ \cline{2-6}
          & Top-5 & 90.9 & 51.0 & 7.4 & 6.6 \\ \hline
        \multirow{2}{*}{ResNeXt-101 32x8d}
          & Top-1 & 79.3 & 55.3 ($-$24.0) & 10.9 ($-$68.5) & 5.8 ($-$73.6) \\ \cline{2-6}
          & Top-5 & 94.5 & 77.8 & 23.1 & 13.3 \\ \hline
        \multirow{2}{*}{MobileNet V3 Large}
          & Top-1 & 74.1 & 43.4 ($-$30.7) & 14.0 ($-$60.1) & 14.4 ($-$59.7) \\ \cline{2-6}
          & Top-5 & 91.3 & 67.3 & 29.2 & 29.8 \\ \hline
        \multirow{2}{*}{ViT-L/16}
          & Top-1 & 79.7 & 67.6 ($-$12.1) & 25.2 ($-$54.5) & 23.5 ($-$56.2) \\ \cline{2-6}
          & Top-5 & 94.5 & 86.8 & 43.7 & 41.4 \\ \hline
        \multirow{2}{*}{ConvNeXt Large}
          & Top-1 & 84.1 & 73.2 ($-$10.9) & 45.7 ($-$38.5) & 49.8 ($-$34.3) \\ \cline{2-6}
          & Top-5 & 96.9 & 90.7 & 66.9 & 71.0 \\ \hline
    \end{tabular}
    \end{adjustbox}
\end{table*}

\begin{table*}[t]
    \centering
    \caption{\small Adaptive-attacker evaluation of the pixel-substitution defense across all five attacker architectures and three perturbation budgets. For each attacker we report top-1 accuracy on defended scaled-Hamming-Weight images from the 50{,}000-image ImageNet validation set when the attacker is \emph{non-adaptive} (trained only on undefended data) versus \emph{adaptive} (fine-tuned on defended data, following Kerckhoffs's principle); ``Recovery'' is the adaptive gain, computed from unrounded accuracies. At budget $90$, where the defense lowers the non-adaptive attacker's accuracy the most, every adaptive attacker recovers accuracy, most of all ViT-L/16 (from $45.5\%$ to $56.5\%$). At budgets $30$ and $60$ the defense barely affects the non-adaptive attacker, so there is little to recover: adaptation slightly lowers the accuracy of VGG19 and ResNeXt and slightly raises that of the other three models. Because the perturbation is deterministic and publicly reconstructable, it provides no durable protection against an adversary that retrains on the transformed data.}
    \label{tab:adaptive}
    \small
    \begin{adjustbox}{width=0.86\textwidth}
    \begin{tabular}{|p{3.4cm}|p{1.5cm}|p{2.6cm}|p{2.6cm}|p{2.0cm}|}
        \hline
        \multirow{2}{*}{\textbf{Attacker Model}} & \multirow{2}{*}{\textbf{Budget}} & \multicolumn{2}{c|}{\textbf{Attacker Top-1 on Defended (\%)}} & \multirow{2}{*}{\textbf{Recovery}} \\ \cline{3-4}
         & & \textbf{Non-adaptive} & \textbf{Adaptive} & \\
        \hline \hline
        \multirow{3}{*}{VGG19}
          & 30 & 60.6 & 57.4 & $-$3.2 \\ \cline{2-5}
          & 60 & 60.4 & 57.2 & $-$3.1 \\ \cline{2-5}
          & 90 & 52.9 & 54.7 & $+$1.8 \\ \hline
        \multirow{3}{*}{ResNeXt-101 32x8d}
          & 30 & 65.7 & 65.1 & $-$0.7 \\ \cline{2-5}
          & 60 & 66.1 & 64.8 & $-$1.3 \\ \cline{2-5}
          & 90 & 58.8 & 62.0 & $+$3.3 \\ \hline
        \multirow{3}{*}{MobileNet V3 Large}
          & 30 & 55.9 & 57.4 & $+$1.5 \\ \cline{2-5}
          & 60 & 56.0 & 57.2 & $+$1.2 \\ \cline{2-5}
          & 90 & 48.4 & 54.1 & $+$5.6 \\ \hline
        \multirow{3}{*}{ViT-L/16}
          & 30 & 54.0 & 58.4 & $+$4.4 \\ \cline{2-5}
          & 60 & 55.1 & 57.7 & $+$2.6 \\ \cline{2-5}
          & 90 & 45.5 & 56.5 & $+$11.1 \\ \hline
        \multirow{3}{*}{ConvNeXt Large}
          & 30 & 72.2 & 73.4 & $+$1.3 \\ \cline{2-5}
          & 60 & 72.3 & 73.4 & $+$1.1 \\ \cline{2-5}
          & 90 & 66.6 & 71.5 & $+$4.9 \\ \hline
    \end{tabular}
    \end{adjustbox}
\end{table*}

Tables~\ref{tab:defense_complete}~and~\ref{tab:adaptive} together characterize the pixel-substitution defense across all three perturbation budgets and all five attacker architectures. The perturbation imposes a substantial, budget-dependent cost on the victim's own accuracy: at budget~$30$ the most robust model, ConvNeXt-L, falls only modestly from $84.1\%$ to $73.2\%$, whereas at budget~$90$ VGG-19 collapses from $72.4\%$ to $2.4\%$. The defense is therefore far from fidelity-preserving except at the smallest budget on the most robust architectures. Its cost also falls mostly on the victim rather than the attacker. At budget~$30$, the non-adaptive attackers still reach $54.0$--$72.2\%$ top-1 accuracy on defended inputs (Table~\ref{tab:adaptive}), only slightly below their undefended accuracies in Table~\ref{tab:shw_software_accuracy}; for example, the VGG-19 attacker still reaches $60.6\%$ (versus $63.6\%$ undefended), while the VGG-19 victim falls from $72.4\%$ to $28.5\%$. This follows from the substitution rule: it changes each pixel's Hamming Weight, which is all the attacker observes, by a single level, but moves the pixel's intensity, which is what the victim classifies, by close to the full budget. More importantly, the defense does not withstand an adaptive attacker. Following Kerckhoffs's principle, we retrain each attacker on scaled-Hamming-Weight images to which the defense has already been applied: each attacker is initialized from its scaled-Hamming-Weight checkpoint, fine-tuned for three epochs on the defended ImageNet training set (AdamW, learning rate $10^{-4}$, cosine schedule), and evaluated on the defended $50{,}000$-image validation set. As Table~\ref{tab:adaptive} shows, adaptation helps most where the defense is otherwise strongest: at the largest budget ($90$), every attacker improves through retraining. The gain is largest for the transformer, whose top-1 accuracy rises from $45.5\%$ (non-adaptive) to $56.5\%$ (adaptive), and smallest for VGG-19, which improves from $52.9\%$ to $54.7\%$. At budgets $30$ and $60$, the defense barely affects the non-adaptive attacker, so there is little accuracy to regain; with this short fine-tuning schedule, the VGG-19 and ResNeXt attackers end slightly below their non-adaptive accuracy at these budgets, while the other three improve. We therefore conclude that the deterministic perturbation is best understood as a non-adaptive obfuscation technique rather than a durable defense: because the same fixed mapping is applied to every image and can be publicly reconstructed, an adversary who anticipates it can simply retrain against it. A defense that resists adaptation would need to be randomized or input-dependent, which we identify as important future work. Our finding is consistent with the broader literature on defending FPGA accelerators against ML-based power side-channel attacks. AIAShield~\cite{aiashield}, for example, uses adversarial-attack techniques to craft fine-grained noise that obfuscates the adversary's side-channel observation with low overhead; unlike our deterministic pixel substitution, such an approach is input-dependent and therefore harder for an adaptive attacker to learn. A systematic comparison of our Hamming-Weight-targeted perturbation against adversarial-noise countermeasures such as AIAShield is a promising direction for designing a defense that is both utility-preserving and robust to adaptation.

\section{Related Work}
\label{sec:related_work}

Gongye et al. developed a  successful reverse-engineering attack on a commercially available advanced encrypted FPGA DNN accelerator intellectual property (IP)~\cite{sp_side}. This attack relied on use of electromagnetic (EM)
side-channel analysis to reveal the data flow and scheduling of the DNN accelerators. Their work is particularly relevant to our threat model, as it demonstrates that the microarchitectural knowledge our attack requires (data scheduling, parallelism) can be obtained even from encrypted accelerator designs. Our attack focuses on the subsequent data-extraction phase, assuming that such reverse-engineering has already been performed or that the victim's design is based on a public architecture.

Luo et al. were able to precisely inject faults in a cloud-based FPGA DNN accelerator to decrease the accuracy of the inference~\cite{deep_strike}. Our work does not focus on fault injection, but the information we recover, i.e. the image being processed, could be used to help determine when to trigger the fault, e.g., to decrease accuracy of inference for only selected inputs.

Moini et al. used power traces from a TDC targeting a binarized convolutional neural network implemented on an FPGA to recover the input image data~\cite{moini_remote_2021}. The extended journal version of their work~\cite{moini_journal_2021} provides a detailed analysis of the TDC implementation, the power leakage mechanism through the shared PDN, and achieves high-accuracy reconstruction of input MNIST~\cite{mnist} images. Their work provides the foundational demonstration that TDC-based power side channels can leak input data from neural network accelerators. However, their attack is limited to 28$\times$28 grayscale MNIST images processed by a binarized neural network (BNN), where the binary weights create particularly strong and distinct power signatures. Our work extends this to the significantly more challenging setting of 224$\times$224 full-color ImageNet images processed by a standard (non-binarized) CNN, which produces far subtler and more complex power traces.

Wei et al. executed an  attack on a CNN accelerator implemented on an FPGA, and successfully retrieved the input image from the gathered power traces without having knowledge of the detailed parameters within the neural network~\cite{wei2018know}. However, their attack required physical access to an oscilloscope, making it impractical for remote cloud settings. Huegle et al.~\cite{huegle2023power2picture} demonstrated the use of generative CNNs to recover input images from power side channels captured by on-chip TDC sensors, but their work was also limited to small grayscale MNIST and Fashion-MNIST images.

Tian et al. showcased that TDCs can be used to construct covert channels to leak information from the FPGA~\cite{tian_remote_2021}.
Drewes et al. implemented a tuneable TDC that was effectively able to recognize co-tenant computation in a multi-tenant FPGA~\cite{tuneable_tdc}. Schellenberg et al.~\cite{schellenberg_remote_2018} demonstrated remote power analysis attacks from within FPGAs, showing that side-channel information can be extracted without physical access.
Beyond power-based side channels, Etim et al.~\cite{etim2024extending} demonstrated that long-wire crosstalk within an FPGA can be exploited by a Trojan phantom circuit to spy on a victim and then exfiltrate the stolen information across FPGAs via ring-oscillator stressors. Their work broadens the threat landscape for cloud FPGAs by showing that information leakage is not limited to a single FPGA or a single side-channel type, and that even cross-FPGA exfiltration is feasible.

Several recent works further develop power side-channel attacks on FPGA-based DNN accelerators along directions complementary to ours. Recent work has been done that recovers inputs using a three-dimensional power surface representation of the leakage~\cite{host25_3dpower}, and another also demonstrates a practical remote power attack against machine learning accelerators in cloud FPGAs~\cite{date23_remotepower}, underscoring the growing practicality of this threat. Most recently, the authors in~\cite{host26_talkingtraces} extended power side-channel input recovery from images to \emph{audio} reconstruction from a neural-network FPGA accelerator, showing that the leakage of input data is not specific to the visual domain. Our contribution is orthogonal to these: we target full-color, ImageNet-scale \emph{visual} inputs and introduce ML-assisted color recovery and scaled-Hamming-Weight classification.

Our work builds upon this body of research, and in particular on Power2Picture~\cite{huegle2023power2picture}, which first applied generative CNNs to input recovery from on-chip FPGA power side channels but was limited to small grayscale MNIST and Fashion-MNIST images. We extend this direction along three axes that, in combination, are novel: we demonstrate an attack on complex, full-color \emph{ImageNet} images rather than MNIST; we introduce ML-assisted \emph{color} recovery via upscaling and recoloring together with a scaled-Hamming-Weight classification attack; and we propose and evaluate a defense that specifically targets Hamming-Weight leakage.

\section{Discussion and Limitations}
\label{section:discussion}

Our work demonstrates the feasibility of ML-assisted power side-channel attacks on ML accelerators in shared FPGAs. However, it is important to acknowledge the assumptions and limitations of our current study, which define the boundaries of our results and highlight critical directions for future research. We discuss each limitation candidly below.

\textbf{TDC Synchronization and Victim Targeting.}
The attack assumes the adversary can synchronize their TDC measurements with the victim's inference operation to isolate the power trace of the first convolution layer. This is the same synchronization assumption made by all closely related prior works on TDC-based FPGA side-channel attacks~\cite{moini_remote_2021, moini_journal_2021, huegle2023power2picture, tian_remote_2021}. In practice, an attacker might achieve synchronization by monitoring shared resources for activity indicative of the victim's operation, e.g., PCIe bus traffic for large data transfers~\cite{zhao_fpga_2018}, or by detecting distinctive power signatures that mark the start of inference. Implementing a robust trigger in a dynamic, real-world cloud environment, and reliably co-locating with a target victim, remain significant practical challenges that no existing work has fully solved; we consider these common limitations of the research area rather than unique weaknesses of our approach.

\textbf{Convolution Frequency.}
Our experiments were conducted with the victim circuit operating at 25~MHz to improve the signal-to-noise ratio (SNR) of TDC measurements. The TDC sensor operates at 50~MHz, so slowing the victim's clock yields approximately two TDC samples per victim clock cycle, which is essential for reliable per-pixel measurements. This approach of reducing the victim's clock frequency relative to the sensor's sampling rate is a standard methodology in academic side-channel research, employed by Moini et al.~\cite{moini_remote_2021, moini_journal_2021} (who used 1--5~MHz victim clocks), and Huegle et al.~\cite{huegle2023power2picture} (who used similarly reduced frequencies).

Real-world ML accelerators in cloud FPGAs typically operate at 100--300~MHz, which is significantly faster than in our experimental setting. At these frequencies, the number of TDC samples per computational operation would be dramatically reduced, degrading the attacker's ability to resolve per-pixel power consumption. A successful attack at realistic speeds would require either: (a)~a higher-frequency TDC sensor, which is feasible as TDC designs operating at hundreds of MHz have been demonstrated~\cite{tuneable_tdc}; (b)~advanced signal processing or ML-based denoising techniques to extract information from lower-SNR traces; or (c)~averaging measurements over multiple inferences of the same image, which may be possible if the victim processes images repeatedly (e.g., in a video stream). Whether these techniques can fully bridge the gap to realistic operating frequencies remains an open question. We note that varying the victim clock is a supported parameter of our platform rather than a fundamental limitation; a systematic frequency sweep is a natural and feasible next experiment, which we leave to future work as it is orthogonal to the leakage and reconstruction questions that are the focus of this study.

\textbf{Effect of Other Tenants on the Victim.}
This work establishes a baseline in a controlled setting with only the attacker and victim present on the FPGA. This is consistent with the experimental methodology of all closely related works~\cite{moini_remote_2021, moini_journal_2021, huegle2023power2picture}, none of which evaluate attacks in the presence of additional co-tenants.

In a real multi-tenant FPGA, other tenants' computations would introduce substantial noise into the shared power distribution network. This additional noise would degrade the quality of the attacker's TDC measurements, potentially making it difficult or impossible to isolate the victim's power signature without sophisticated filtering. The severity of this effect depends on factors such as the FPGA's power delivery architecture, the computational intensity of the other tenants' workloads, and their physical placement on the FPGA relative to the attacker and victim. In current cloud FPGA offerings, including both Amazon EC2 F1 and the second-generation EC2 F2 family~\cite{aws_fpga_2021, aws_f2_2024}, FPGAs are allocated to a single tenant, so multi-tenant noise is not yet a practical concern and, indeed, the shared-fabric co-location our attack requires is not yet available on these platforms. This makes the single-victim, single-attacker setting a reasonable baseline for today, but it also means our results should be read as characterizing leakage in an anticipated multi-tenant future rather than a present-day deployment. As multi-tenant FPGA platforms mature~\cite{multi_tenant_2025_survey, 10.1145/3474058}, quantifying the attack's robustness in the presence of multiple noisy co-tenants becomes essential. We consider this a primary area for future work.

\textbf{Known Architecture and Microarchitecture.}
Our attack relies on the attacker knowing both the victim's CNN architecture (e.g., VGG-19) and, to a significant extent, its hardware microarchitecture (the degree of parallelism, scheduling, and memory access patterns that determine cycle-by-cycle switching activity). Two implementations of the same CNN with different parallelism would produce substantially different power signatures. For accelerators built on public, open-source frameworks (e.g., Vitis AI or published RTL), this is effectively known; for proprietary designs, a separate reverse-engineering phase would be required. Prior work~\cite{sp_side} demonstrates that even encrypted DNN accelerator IPs can be reverse-engineered via EM side-channel analysis, but such reverse-engineering is itself a complex attack. If the attacker's microarchitectural knowledge is imprecise, the correlation between TDC measurements and pixel values degrades; characterizing this degradation is left to future work.

\textbf{First Convolution Layer.} This work focuses exclusively on the first convolution layer because it directly processes the raw input image pixels, making the correlation between power consumption and pixel data the strongest. Subsequent layers process intermediate feature maps that are data-dependent transformations of the input, which would make the correlation between power traces and original input pixels significantly more complex. Extending the attack to deeper layers, or combining information from multiple layers, could be a direction for future work. We also note the related, and easier-to-test question: even without implementing a full CNN, having \emph{other operations active on the FPGA}, the remaining network layers, or unrelated logic, would add competing switching activity to the shared PDN and thus stress the attack in the same way that additional co-tenants do. In our current setup the first convolution layer runs essentially in isolation, so our results do not capture this interference; evaluating the attack with the rest of the network (or other active circuits) present is a concrete intermediate step between our isolated proof-of-concept and a fully realistic deployment, and it overlaps with the noisy co-tenant question discussed above.

\textbf{Power Amplification.}
In our experimental setup, we amplify the power side-channel signal from the victim's convolution engine to make the leakage more prominent for the TDC. We emphasize that this amplification increases the magnitude of an existing leakage signal; it does not create information that is not present. Without amplification, the fundamental leakage still exists but with a lower signal-to-noise ratio, meaning more measurements or a more sensitive sensor would be needed.

FPGAs in cloud settings (like those on Amazon EC2 F1 and F2 instances) are more complex and have more sophisticated power delivery networks with voltage regulators, decoupling capacitors, and potentially separate power domains. These could either \emph{dampen} the leakage signal (making the attack harder) or, if the attacker and victim share the same power domain, could \emph{isolate} the victim's signal from other noise sources (potentially making the attack easier). The net effect in any particular commercial FPGA platform would need to be empirically characterized.

\textbf{Limited Number of Test Images.}
Due to the time-intensive nature of our data acquisition process (approximately 33 minutes per image due to I/O bottlenecks, as detailed in Section~\ref{sec_run_times}), the \emph{end-to-end hardware} evaluation is limited to 9 test images for the Hamming Weight experiments and 5 for the pixel value experiments. We want to be explicit that these images were selected to span the contrast spectrum---high contrast, high color contrast, and low contrast---and were \emph{not} chosen to flatter the method: the low-contrast group is deliberately included precisely because it is where the attack performs \emph{worst} (the bottom-contrast Hamming Weight reconstructions in Section~\ref{sec_recoloring} clearly fail to match their references), so the reported set contains the method's failure cases as well as its successes. Nonetheless, we acknowledge that this small hardware sample size limits the statistical confidence and generalizability of the end-to-end results. To compensate, the \emph{classifiability} of the scaled Hamming Weight representation, the core premise of the Hamming-Weight attack, is validated at full scale in software on all $50{,}000$ ImageNet validation images with top-1/top-5 accuracy and confidence intervals (Section~\ref{sec:shw_software_validation}), which does not require the slow hardware acquisition. A larger-scale \emph{end-to-end} evaluation, made possible by a faster data acquisition pipeline (e.g., a PCIe-connected FPGA or the on-chip BRAM buffering described in Section~\ref{sec:image_prep_splicing}), would strengthen the conclusions further.

\textbf{Impact of Defense on Victim's Classification Accuracy.}
A key consideration for the proposed pixel perturbation defense is whether modified inputs remain correctly classified by the victim's network. We measure this in software on the full $50{,}000$-image ImageNet validation set (Table~\ref{tab:defense_complete}). The cost is large and architecture-dependent: even at budget $30$, VGG-19's top-1 accuracy falls from $72.4\%$ to $28.5\%$, and the most robust models, ConvNeXt-L and ViT-L/16, still fall from $84.1\%$ to $73.2\%$ and from $79.7\%$ to $67.6\%$. This cost follows directly from the substitution rule, which moves each pixel's intensity by close to the full budget in exchange for a one-level change in its Hamming Weight. A rule that instead maximizes the Hamming Weight change for a small intensity change (for example, mapping $128$ to $127$) would alter the attacker's view more while perturbing the victim's input less: at budget $30$ it moves intensity by $15.6$ levels on average over the 256 values while changing Hamming Weight by $3.5$ levels, compared with $29.0$ and $1.0$ for the rule evaluated here. Evaluating such a variant, together with perceptual-quality metrics over the full dataset, is future work.

\textbf{Adaptive Attacker Against the Defense.}
A \emph{non-adaptive} evaluation, in which the attacker's models are trained on unperturbed scaled Hamming Weight images and then tested on perturbed inputs, measures whether the perturbation shifts inputs away from the attacker's learned distribution, but not whether it withstands an attacker who knows that the defense is in use. Because our perturbation is deterministic and its rule is public, the appropriate threat model under Kerckhoffs's principle is an adaptive attacker who applies the same transformation to the training data and re-trains on \emph{perturbed} scaled Hamming Weight images. We evaluate this adaptive attacker for all five architectures and all three budgets (Table~\ref{tab:adaptive}). At budget $90$, where the defense is strongest, adaptation recovers accuracy for every architecture (for example, ViT-L/16 rises from $45.5\%$ to $56.5\%$), and at budgets $30$ and $60$ the defense barely affects even the non-adaptive attacker. We therefore do not claim security against an adaptive adversary. Two implications follow. First, a deterministic perturbation whose rule is public should be regarded as obfuscation rather than a robust defense. Second, a randomized perturbation---e.g., drawing a fresh, secret perturbation pattern per inference so that the attacker cannot train against a fixed mapping---is more likely to resist adaptation, and the shadow-based variant we explore is a step in this direction.

\textbf{Attacker's Knowledge of the Input Distribution.}
As noted in Section~\ref{sec:shw_software_validation}, the attacker trains on ImageNet and thus implicitly assumes the victim's inputs share a similar distribution and label space. This holds for victims running standard, publicly trained classifiers but is a genuine limitation for proprietary imagery or private label sets. The scaled Hamming Weight \emph{representation} is generic and transfers, but obtaining labels representative of a truly private victim task is a separate obstacle we do not solve here.

\section{Conclusion}
\label{sec_conclusion}

This paper examined the security implications of multi-tenant FPGAs and demonstrated that a co-located attacker can extract information from a victim neural network using ML-assisted side-channel analysis. We introduced a new class of attacks where Time-to-Digital Converters (TDCs) capture power variations that are then processed with machine learning to infer the victim’s input images. Two complementary attack methodologies were presented. The first reconstructs recognizable visual representations of victim inputs through pixel recovery, interpolation-based upscaling, and ML-based recoloring. The second classifies input images by reconstructing their Hamming-weight signatures and employing a novel scaled representation to train ML models. Both approaches were validated on the ImageNet dataset using a PYNQ-Z2 FPGA platform.
We further studied a lightweight pixel-perturbation obfuscation as a preliminary countermeasure against the Hamming-weight attack. Our evaluation shows that, while it degrades the attacker's non-adaptive performance, it does so at a significant cost to the victim model's own accuracy and is largely circumvented by an adaptive attacker that retrains on the perturbed data. We therefore present it as non-adaptive obfuscation rather than an effective defense, and identify randomized or input-dependent perturbations as a direction for robust countermeasures.

We acknowledge that our current experimental setup involves several simplifications, including a reduced convolution frequency, power signal amplification, and a controlled two-tenant environment, as discussed in detail in Section~\ref{section:discussion}. These are common across related work in this area and serve to establish a proof-of-concept baseline. Future work should investigate the attack's effectiveness at realistic operating frequencies, in the presence of noisy co-tenants, and with imprecise knowledge of the victim's microarchitecture. Additionally, exploring stronger countermeasures such as hardware-level power noise injection, randomized data scheduling, or runtime input masking would further advance the security of multi-tenant FPGA platforms.

\section*{Acknowledgements}

This work was supported in part by National Science Foundation grants \nsf{2245344} and \nsf{2332406}.

\bibliographystyle{ACM-Reference-Format}
\bibliography{bibtex/references}

@inproceedings{giechaskiel2018leaky,
  title={Leaky wires: Information leakage and covert communication between FPGA long wires},
  author={Giechaskiel, Ilias and Rasmussen, Kasper B and Eguro, Ken},
  booktitle={Proceedings of the 2018 on Asia Conference on Computer and Communications Security},
  pages={15--27},
  year={2018}
}

@inproceedings{8457631,
  author={Ramesh, Chethan and Patil, Shivukumar B. and Dhanuskodi, Siva Nishok and Provelengios, George and Pillement, Sebastien and Holcomb, Daniel and Tessier, Russell},
  booktitle={2018 IEEE 26th Annual International Symposium on Field-Programmable Custom Computing Machines (FCCM)}, 
  title={FPGA Side Channel Attacks without Physical Access}, 
  year={2018},
  volume={},
  number={},
  pages={45-52},
  doi={10.1109/FCCM.2018.00016}}

@InProceedings{schellenberg_remote_2018,
  author    = {Schellenberg, Falk and Gnad, Dennis R. E. and Moradi, Amir
               and Tahoori, Mehdi B.},
  title     = {Remote Inter-Chip Power Analysis Side-channel Attacks at Board-Level},
  booktitle = {International Conference on Computer-Aided Design (ICCAD)},
  year      = {2018},
}

@InProceedings{zhao_fpga_2018,
  author    = {Zhao, Mark and Suh, G. Edward},
  title     = {{FPGA}-Based Remote Power Side-Channel Attacks},
  booktitle = {IEEE Symposium on Security and Privacy (S\&P)},
  year      = {2018},
}

@InProceedings{islam_exploiting_2017,
  author    = {Islam, Mohammad A. and Ren, Shaolei and Wierman, Adam},
  title     = {Exploiting a Thermal Side Channel for Power Attacks in Multi-Tenant Data Centers},
  booktitle = {ACM Conference on Computer and Communications Security (CCS)},
  year      = {2017},
}

@Misc{aws_fpga_2021,
    author  = {{Amazon Web Services}},
    title   = {{FPGA} Developer {AMI}},
    year    = {2022},
    howpublished = {\url{https://aws.amazon.com/marketplace/pp/prodview-gimv3gqbpe57k}},
    note    = {{A}ccessed: 2022-01-15}
}

@InProceedings{tian_remote_2021,
  author    = {Tian, Shanquan and Xiong, Wenjie and Giechaskiel, Ilias
              and Szefer, Jakub},
  title     = {Remote Power Attacks on the Versatile Tensor Accelerator in Multi-Tenant {FPGA}s},
  booktitle = {IEEE Symposium on Field-Programmable Custom Computing Machines (FCCM)},
  year      = {2021},
}

@InProceedings{moini_remote_2021,
  author    = {Moini, Shayan and Tian, Shanquan and Holcomb, Daniel and Szefer, Jakub and Tessier, Russell},
  title     = {Remote Power Side-Channel Attacks on {BNN} Accelerators in {FPGA}s},
  booktitle = {Design, Automation \& Test in Europe Conference \& Exhibition (DATE)},
  year      = {2021},
}

@InProceedings{boutros_neighbors_2020,
  author    = {Boutros, Andrew and Hall, Matthew and Papernot, Nicolas and Betz, Vaughn},
  title     = {Neighbors From Hell: {V}oltage Attacks Against Deep Learning Accelerators on Multi-Tenant {FPGA}s},
  booktitle = {International Conference on Field-Programmable Technology (FPT)},
  year      = {2020},
}

@Article{provelengios_power_2021,
  author    = {Provelengios, George and Holcomb, Daniel and Tessier, Russell},
  title     = {Power Distribution Attacks in Multi-Tenant FPGAs},
  journal   = {IEEE Transactions on Very Large Scale Integration (VLSI) Systems (TVLSI)},
  year      = {2020}
}

@misc{amazonf1, 
title={Amazon EC2 F1 Instances},         
note = {\url{https://aws.amazon.com/ec2/instance-types/f1/}},
publisher={Amazon},
year = {2026}
}

@misc{vmaccel,
  title={VMAccel},         
  note = {\url{https://www.vmaccel.com/}},
  publisher={VMAccel},
  year = {2026}
}

@misc{pynq-z2,
    title={TUL PYNQ-Z2 board},
    note ={\url{https://www.tulembedded.com/FPGA/ProductsPYNQ-Z2.html}},
    publisher={TUL Technology Unlimited},
    year = {2026}
}

@article{10.1145/3474058,
author = {Mbongue, Joel Mandebi and Kwadjo, Danielle Tchuinkou and Shuping, Alex and Bobda, Christophe},
title = {Deploying Multi-Tenant FPGAs within Linux-Based Cloud Infrastructure},
year = {2021},
issue_date = {June 2022},
address = {New York, NY, USA},
volume = {15},
number = {2},
issn = {1936-7406},
url = {https://doi.org/10.1145/3474058},
doi = {10.1145/3474058},
journal = {ACM Trans. Reconfigurable Technol. Syst.},
month = {dec},
articleno = {19},
numpages = {31}
}

@inproceedings{istvan2018providing,
  title={Providing multi-tenant services with FPGAs: Case study on a key-value store},
  author={Istv{\'a}n, Zsolt and Alonso, Gustavo and Singla, Ankit},
  booktitle={2018 28th International Conference on Field Programmable Logic and Applications (FPL)},
  pages={119--1195},
  year={2018},
  organization={IEEE}
}

@inproceedings{multi_defense_attack,
  author={Elnaggar, Rana and Karri, Ramesh and Chakrabarty, Krishnendu},
  booktitle={2019 Design, Automation \& Test in Europe Conference \& Exhibition (DATE)}, 
  title={Multi-Tenant FPGA-based Reconfigurable Systems: Attacks and Defenses}, 
  year={2019},
  volume={},
  number={},
  pages={7-12},
  doi={10.23919/DATE.2019.8714904}}

@misc{smartssd, title={Samsung SmartSSD},         
    note = {\url{https://www.xilinx.com/applications/data-center/computational-storage/smartssd.html}},
     publisher={Xilinx},
     year = {2026}
}

@misc{simonyan2015deep,
      title={Very Deep Convolutional Networks for Large-Scale Image Recognition}, 
      author={Karen Simonyan and Andrew Zisserman},
      year={2015},
      eprint={1409.1556},
      archivePrefix={arXiv},
      primaryClass={cs.CV}
}

@INPROCEEDINGS{imagenet,

  author={Deng, Jia and Dong, Wei and Socher, Richard and Li, Li-Jia and Kai Li and Li Fei-Fei},

  booktitle={2009 IEEE Conference on Computer Vision and Pattern Recognition}, 

  title={ImageNet: A large-scale hierarchical image database}, 

  year={2009},

  volume={},

  number={},

  pages={248-255},

  doi={10.1109/CVPR.2009.5206848}}

@ARTICLE{dnn_survey,

  author={Sze, Vivienne and Chen, Yu-Hsin and Yang, Tien-Ju and Emer, Joel S.},

  journal={Proceedings of the IEEE}, 

  title={Efficient Processing of Deep Neural Networks: A Tutorial and Survey}, 

  year={2017},

  volume={105},

  number={12},

  pages={2295-2329},

  doi={10.1109/JPROC.2017.2761740}}

@misc{deep_ai,
  title={DeepAI Image Colorization API },         
  note = {\url{https://deepai.org/machine-learning-model/colorizer}},
  publisher={DeepAI },
  year = {2026}
}

@misc{py_simple_gui,
  title={PySimpleGUI Photo Colorizer},         
  note = {\url{https://github.com/PySimpleGUI/PySimpleGUI-Photo-Colorizer}},
  publisher={PySimpleGUI},
  year = {2026}
}

@misc{open_cv,
  title={OpenCV},         
  note = {\url{https://opencv.org}},
  publisher={OpenCV},
  year = {2026}
}

@misc{numpy,
  title={NumPy},         
  note = {\url{https://numpy.org/}},
  publisher={NumPy},
  year = {2026}
}

@ARTICLE{mnist,

  author={Deng, Li},

  journal={IEEE Signal Processing Magazine}, 

  title={The MNIST Database of Handwritten Digit Images for Machine Learning Research [Best of the Web]}, 

  year={2012},

  volume={29},

  number={6},

  pages={141-142},

  doi={10.1109/MSP.2012.2211477}}

@inproceedings{sp_side,
  title={Side-channel-assisted reverse-engineering of encrypted dnn hardware accelerator ip and attack surface exploration},
  author={Gongye, Cheng and Luo, Yukui and Xu, Xiaolin and Fei, Yunsi},
  booktitle={2024 IEEE Symposium on Security and Privacy (SP)},
  pages={4678--4695},
  year={2024},
  organization={IEEE}
}

@INPROCEEDINGS{deep_strike,

  author={Luo, Yukui and Gongye, Cheng and Fei, Yunsi and Xu, Xiaolin},

  booktitle={2021 58th ACM/IEEE Design Automation Conference (DAC)}, 

  title={DeepStrike: Remotely-Guided Fault Injection Attacks on DNN Accelerator in Cloud-FPGA}, 

  year={2021},

  volume={},

  number={},

  pages={295-300},

  doi={10.1109/DAC18074.2021.9586262}}

@inproceedings{tuneable_tdc,
author = {Drewes, Colin and Weng, Olivia and Ryan, Keegan and Hunter, Bill and McCarty, Christopher and Kastner, Ryan and Richmond, Dustin},
title = {Turn on, Tune in, Listen up: Maximizing Side-Channel Recovery in Time-to-Digital Converters},
year = {2023},
isbn = {9781450394178},
address = {New York, NY, USA},
url = {https://doi.org/10.1145/3543622.3573193},
doi = {10.1145/3543622.3573193},
booktitle = {Proceedings of the 2023 ACM/SIGDA International Symposium on Field Programmable Gate Arrays},
pages = {111–122},
numpages = {12},
location = {Monterey, CA, USA},
series = {FPGA '23}
}

@misc{zhang2016colorful,
      title={Colorful Image Colorization}, 
      author={Richard Zhang and Phillip Isola and Alexei A. Efros},
      year={2016},
      eprint={1603.08511},
      archivePrefix={arXiv},
      primaryClass={cs.CV}
}

@inproceedings{zhong2022shadows,
  title={Shadows can be dangerous: Stealthy and effective physical-world adversarial attack by natural phenomenon},
  author={Zhong, Yiqi and Liu, Xianming and Zhai, Deming and Jiang, Junjun and Ji, Xiangyang},
  booktitle={Proceedings of the IEEE/CVF Conference on Computer Vision and Pattern Recognition},
  pages={15345--15354},
  year={2022}
}

@inproceedings{kennedy1995particle,
  title={Particle swarm optimization},
  author={Kennedy, James and Eberhart, Russell},
  booktitle={Proceedings of International Conference on Neural Networks},
  volume={4},
  pages={1942--1948},
  year={1995},
}

@inproceedings{huegle2023power2picture,
  title={Power2Picture: Using generative CNNs for input recovery of neural network accelerators through power side-channels on FPGAs},
  author={Huegle, Lukas and Gotthard, Martin and Meyers, Vincent and Krautter, Jonas and Gnad, Dennis RE and Tahoori, Mehdi B},
  booktitle={2023 IEEE 31st Annual International Symposium on Field-Programmable Custom Computing Machines (FCCM)},
  pages={155--161},
  year={2023},
  organization={IEEE}
}

@article{maji2021leaky,
  title={Leaky nets: Recovering embedded neural network models and inputs through simple power and timing side-channels—Attacks and defenses},
  author={Maji, Saurav and Banerjee, Utsav and Chandrakasan, Anantha P},
  journal={IEEE Internet of Things Journal},
  volume={8},
  number={15},
  pages={12079--12092},
  year={2021},
  publisher={IEEE}
}

@inproceedings{wei2018know,
  title={I know what you see: Power side-channel attack on convolutional neural network accelerators},
  author={Wei, Lingxiao and Luo, Bo and Li, Yu and Liu, Yannan and Xu, Qiang},
  booktitle={Proceedings of the 34th Annual Computer Security Applications Conference},
  pages={393--406},
  year={2018}
}

@article{multi_tenant_2025_survey,
author = {Kawser Ahmed, Muhammed and Panoff Kealoha, Maximillian and Mandebi Mbongue, Joel and Saha, Sujan Kumar and Nghonda Tchinda, Erman and Mbua, Peter Esenju and Bobda, Christophe},
title = {Multi-Tenant Cloud FPGA: A Survey on Security, Trust, and Privacy},
year = {2025},
issue_date = {June 2025},
publisher = {Association for Computing Machinery},
address = {New York, NY, USA},
volume = {18},
number = {2},
issn = {1936-7406},
url = {https://doi.org/10.1145/3713078},
doi = {10.1145/3713078},
journal = {ACM Trans. Reconfigurable Technol. Syst.},
month = apr,
articleno = {23},
numpages = {44}
}

@incollection{kirkland2010bilinear,
  title={Bilinear interpolation},
  author={Kirkland, Earl J},
  booktitle={Advanced computing in electron microscopy},
  pages={261--263},
  year={2010},
  publisher={Springer}
}

@article{probst2025remote,
  title={Remote side-channel analysis of the loop PUF using a TDC-based voltage sensor},
  author={Probst, Matthias and Tebelmann, Lars and Wettermann, Moritz and Pehl, Michael},
  journal={Journal of Cryptographic Engineering},
  volume={15},
  number={1},
  pages={1},
  year={2025},
  publisher={Springer}
}

@article{moini_journal_2021,
  title={Power Side-Channel Attacks on BNN Accelerators in Remote FPGAs},
  author={Moini, Shayan and Tian, Shanquan and Holcomb, Daniel and Szefer, Jakub and Tessier, Russell},
  journal={IEEE Journal on Emerging and Selected Topics in Circuits and Systems},
  volume={11},
  number={2},
  pages={357--370},
  year={2021},
  doi={10.1109/JETCAS.2021.3074608},
  publisher={IEEE}
}

@inproceedings{etim2024extending,
  title={Extending FPGA Information Leaks with Trojan Phantom Circuits},
  author={Etim, Anthony and Tian, Shanquan and Szefer, Jakub},
  booktitle={2024 International Symposium on Secure and Private Execution Environment Design (SEED)},
  pages={1--10},
  year={2024},
  organization={IEEE}
}

@inproceedings{xie2017aggregated,
  title={Aggregated Residual Transformations for Deep Neural Networks},
  author={Xie, Saining and Girshick, Ross and Doll{\'a}r, Piotr and Tu, Zhuowen and He, Kaiming},
  booktitle={2017 IEEE Conference on Computer Vision and Pattern Recognition (CVPR)},
  pages={5987--5995},
  year={2017},
  organization={IEEE}
}

@inproceedings{howard2019searching,
  title={Searching for MobileNetV3},
  author={Howard, Andrew and Sandler, Mark and Chu, Grace and Chen, Liang-Chieh and Chen, Bo and Tan, Mingxing and Wang, Weijun and Zhu, Yukun and Pang, Ruoming and Vasudevan, Vijay and others},
  booktitle={2019 IEEE/CVF International Conference on Computer Vision (ICCV)},
  pages={1314--1324},
  year={2019},
  organization={IEEE}
}

@inproceedings{dosovitskiy2021image,
  title={An Image is Worth 16x16 Words: Transformers for Image Recognition at Scale},
  author={Dosovitskiy, Alexey and Beyer, Lucas and Kolesnikov, Alexander and Weissenborn, Dirk and Zhai, Xiaohua and Unterthiner, Thomas and Dehghani, Mostafa and Minderer, Matthias and Heigold, Georg and Gelly, Sylvain and others},
  booktitle={International Conference on Learning Representations (ICLR)},
  year={2021}
}

@inproceedings{liu2022convnet,
  title={A ConvNet for the 2020s},
  author={Liu, Zhuang and Mao, Hanzi and Wu, Chao-Yuan and Feichtenhofer, Christoph and Darrell, Trevor and Xie, Saining},
  booktitle={2022 IEEE/CVF Conference on Computer Vision and Pattern Recognition (CVPR)},
  pages={11976--11986},
  year={2022},
  organization={IEEE}
}

@article{gao2011bilinear,
  title={Bilinear and bicubic interpolation methods for division of focal plane polarimeters},
  author={Gao, Shengkui and Gruev, Viktor},
  journal={Optics express},
  volume={19},
  number={27},
  pages={26161--26173},
  year={2011},
  publisher={Optical Society of America}
}

@misc{pillow_lib,
  title={Pillow (PIL Fork) Documentation},
  author={Clark, Alex},
  year={2015},
  publisher={readthedocs},
 url={https://buildmedia.readthedocs.org/media/pdf/pillow/latest/pillow.pdf}
}

@misc{aws_f2_2024,
    author = {{Amazon Web Services}},
    title = {Amazon {EC2} {F2} Instances},
    year = {2024},
    note = {Second-generation FPGA-powered instances, generally available December 2024. \url{https://aws.amazon.com/ec2/instance-types/f2/}},
    publisher = {Amazon}
}

@inproceedings{host25_3dpower,
  title={An input recovery side-channel attack on dnn accelerator with three-dimensional power surface},
  author={Wu, Le and Wu, Liji and Ba, Zhiwei and Zhang, Xiangmin},
  booktitle={2025 IEEE International Symposium on Hardware Oriented Security and Trust (HOST)},
  pages={1--11},
  year={2025},
  organization={IEEE}
}

@inproceedings{date23_remotepower,
  title={A practical remote power attack on machine learning accelerators in cloud FPGAs},
  author={Tian, Shanquan and Moini, Shayan and Holcomb, Daniel and Tessier, Russell and Szefer, Jakub},
  booktitle={2023 Design, Automation \& Test in Europe Conference \& Exhibition (DATE)},
  pages={1--6},
  year={2023},
  organization={IEEE}
}

@inproceedings{host26_talkingtraces,
  title={Talking Traces: Audio Reconstruction Power Side-Channel Attack on Neural Network FPGA Accelerators},
  author={Reibold, Johannes and Meyers, Vincent and Tahoori, Mehdi},
  booktitle={2026 IEEE International Symposium on Hardware Oriented Security and Trust (HOST)},
  pages={297--308},
  year={2026},
  organization={IEEE}
}

@article{aiashield,
  title={Defense against ML-based power side-channel attacks on DNN accelerators with adversarial attacks},
  author={Yan, Xiaobei and Chang, Chip Hong and Zhang, Tianwei},
  journal={arXiv preprint arXiv:2312.04035},
  year={2023}
}
\appendix
\section*{Appendix: Attack Results on Test Image Patterns}
\label{sec:appendix}

\begin{figure*}[h]
     \centering
     \begin{subfigure}[b]{0.08\textwidth}
         \centering
         \includegraphics[width=1.6cm]{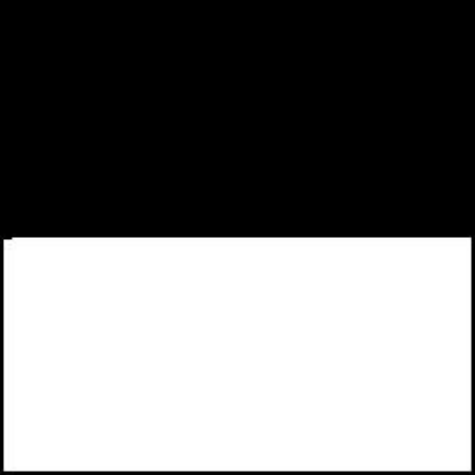}
         \caption{Half \\ Black \\White \\Horizontal\\ }
\label{image_half_black_white_horizontal_org}
     \end{subfigure}
     \hfill
     \begin{subfigure}[b]{0.08\textwidth}
         \centering
         \includegraphics[width=1.6cm]{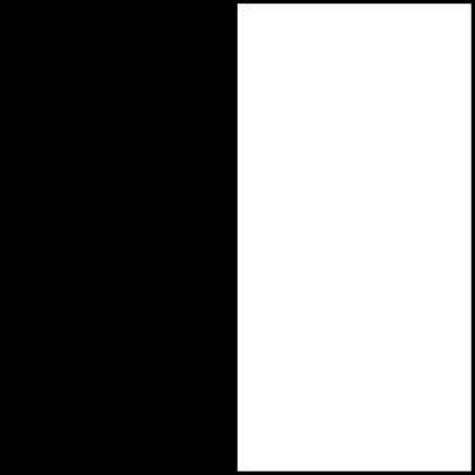}
         \caption{Half \\ Black \\White \\Vertical\\}
        \label{image_half_black_white_vertical_org}
     \end{subfigure}
     \hfill
     \begin{subfigure}[b]{0.08\textwidth}
         \centering
         \includegraphics[width=1.6cm]{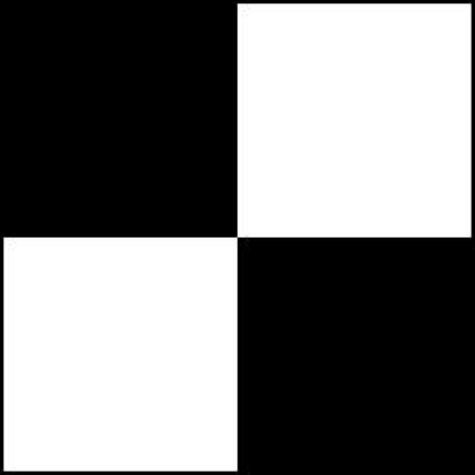}
         \caption{2x2 \\ Checkered \\\\\\}
         \label{image_2x2_checkered_one_org}
     \end{subfigure}
     \hfill
      \begin{subfigure}[b]{0.08\textwidth}
         \centering
         \includegraphics[width=1.6cm]{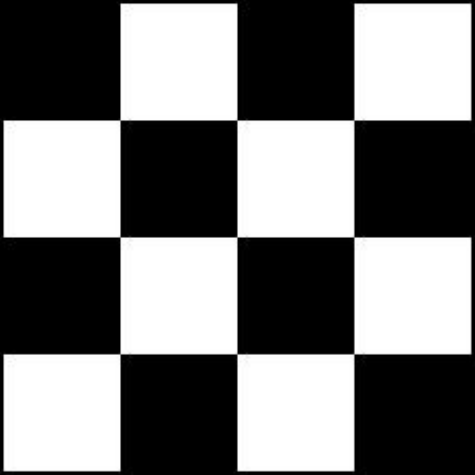}
         \caption{4x4\\ Checkered \\\\ \\}
         \label{image_4x4_checkered_one_org}
     \end{subfigure}
     \hfill
      \begin{subfigure}[b]{0.08\textwidth}
         \centering
         \includegraphics[width=1.6cm]{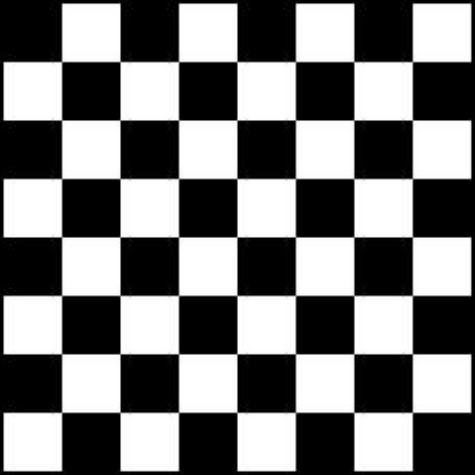}
         \caption{8x8\\ Checkered \\ \\\\}
         \label{image_8x8_checkered_one_org}
     \end{subfigure}
     \hfill
      \begin{subfigure}[b]{0.08\textwidth}
         \centering
         \includegraphics[width=1.6cm]{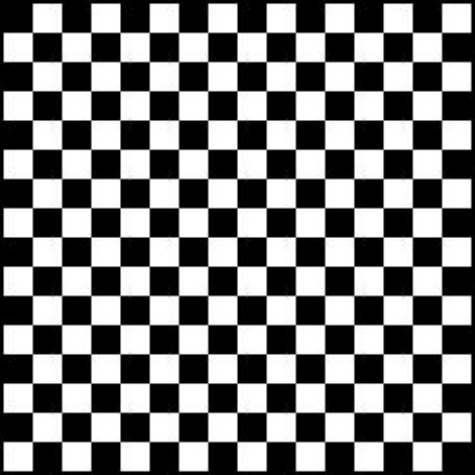}
         \caption{16x16 \\ Checkered \\\\\\}
         \label{image_16x16_checkered_one_org}
     \end{subfigure}
     \hfill
      \begin{subfigure}[b]{0.08\textwidth}
         \centering
         \includegraphics[width=1.6cm]{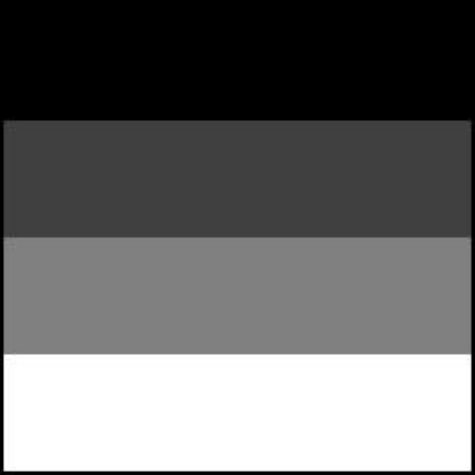}
         \caption{Horizontal \\ Gradient \\ \\}
         \label{image_gradient_horizontal_org}
     \end{subfigure}
     \hfill
      \begin{subfigure}[b]{0.08\textwidth}
         \centering
         \includegraphics[width=1.6cm]{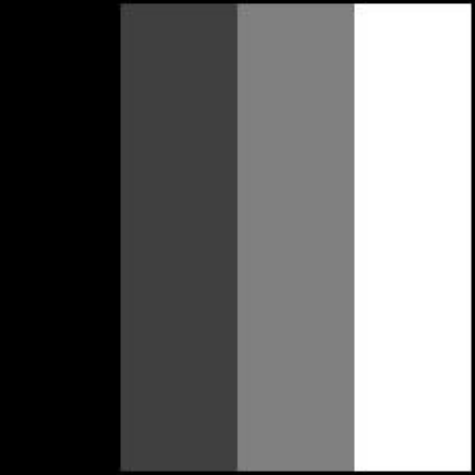}
         \caption{Vertical \\ Gradient \\ \\}
\label{image_gradient_vertical_org}
     \end{subfigure}
    \hfill
        \caption{\small Original test image patterns for pixel value leakage.}
        \label{original_input_images_checkered}
\end{figure*}

We first developed a set of simple test patterns to perform a baseline validation of our image recovery capability. For these tests, we generated 224x224 pixel images with high-contrast checkerboards of varying sizes, as well as horizontal and vertical stripes and grayscale gradients, as shown in Figure~\ref{original_input_images_checkered}. The goal was to analyze how well our TDC-based side-channel could recover patterns with different spatial frequencies.

\begin{figure*}[h]
     \centering
     \begin{subfigure}[b]{0.08\textwidth}
         \centering
         \includegraphics[width=1.6cm]{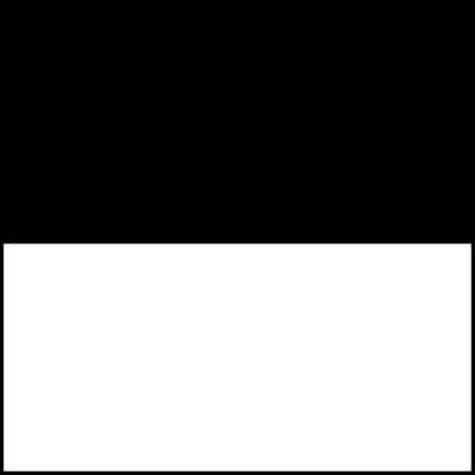}
         \caption{Half \\ Black \\White \\Horizontal\\ }
\label{half_black_white_horizontal_reconstructed}
     \end{subfigure}
     \hfill
     \begin{subfigure}[b]{0.08\textwidth}
         \centering
         \includegraphics[width=1.6cm]{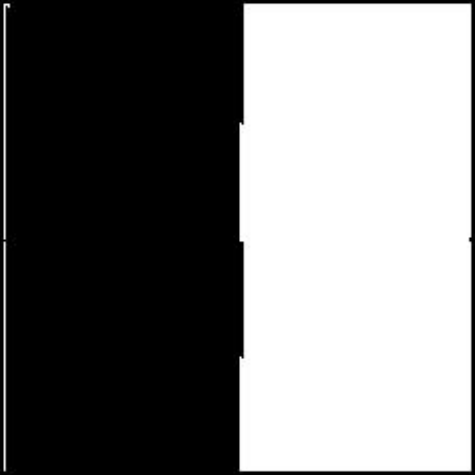}
         \caption{Half \\ Black \\ White \\Vertical\\}
         \label{half_black_white_vertical_reconstructed}
     \end{subfigure}
     \hfill
     \begin{subfigure}[b]{0.08\textwidth}
         \centering
         \includegraphics[width=1.6cm]{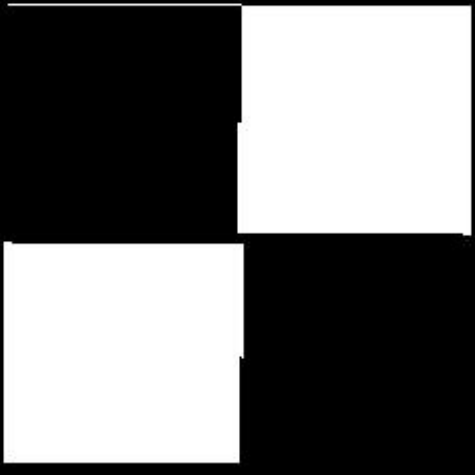}
         \caption{2x2 \\ Checkered \\ \\\\}
\label{2x2_checkered_one_reconstructed}
     \end{subfigure}
     \hfill
     \begin{subfigure}[b]{0.08\textwidth}
         \centering
         \includegraphics[width=1.6cm]{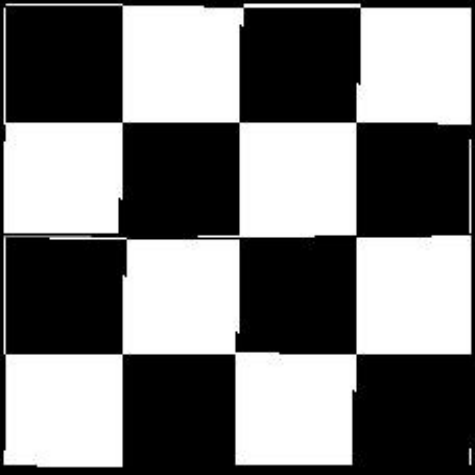}
         \caption{4x4\\ Checkered \\ \\\\}
         \label{4x4_checkered_one_reconstructed}
     \end{subfigure}
     \hfill
     \begin{subfigure}[b]{0.08\textwidth}
         \centering
         \includegraphics[width=1.6cm]{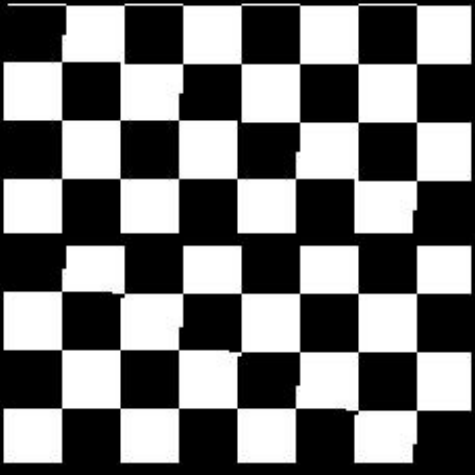}
         \caption{8x8 \\ Checkered \\ \\ \\}
         \label{8x8_checkered_one_reconstructed}
     \end{subfigure}
     \hfill
     \begin{subfigure}[b]{0.08\textwidth}
         \centering
         \includegraphics[width=1.6cm]{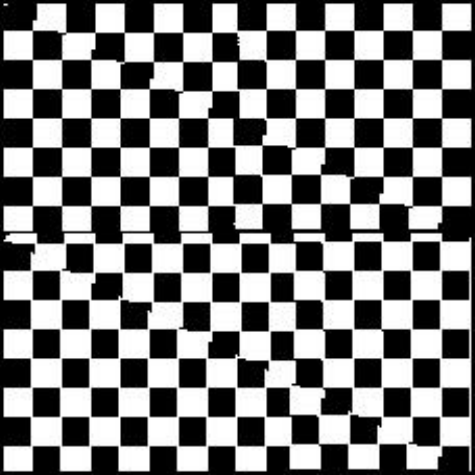}
         \caption{16x16 \\ Checkered \\ \\\\}
         \label{16x16_checkered_one_reconstructed}
     \end{subfigure}
     \hfill
     \begin{subfigure}[b]{0.08\textwidth}
         \centering
         \includegraphics[width=1.6cm]{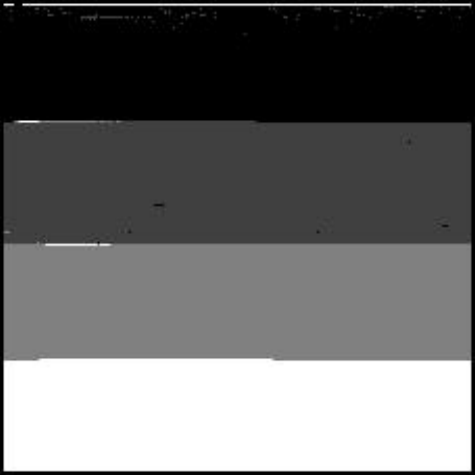}
         \caption{Horizontal \\ Gradient \\ \\} \label{image_gradient_horizontal_reconstructed}
     \end{subfigure}
     \hfill
     \begin{subfigure}[b]{0.08\textwidth}
         \centering
         \includegraphics[width=1.6cm]{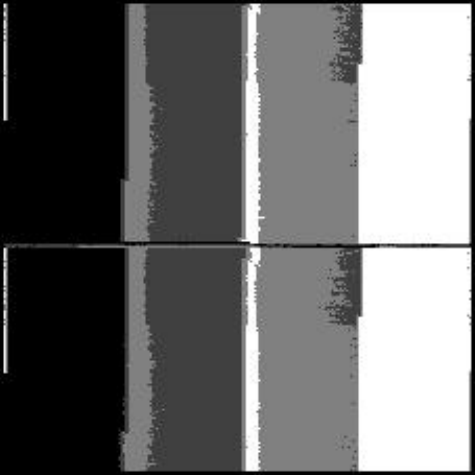}
         \caption{Vertical \\ Gradient \\ \\} \label{image_gradient_vertical_reconstructed}
     \end{subfigure}
     \hfill
        \caption{\small Reconstructed images of the test image patterns for pixel value leakage.}
        \label{reconstructed_images_checkered}
\end{figure*}

The results of this baseline validation are shown in Figure~\ref{reconstructed_images_checkered}. By comparing the reconstructed images (Figure~\ref{reconstructed_images_checkered}) to the originals (Figure~\ref{original_input_images_checkered}), we can make several key observations. The low-frequency patterns, such as the large-scale horizontal and vertical splits (Figures~\ref{half_black_white_horizontal_reconstructed} and~\ref{half_black_white_vertical_reconstructed}) and the 2x2 checkerboard (Figure~\ref{2x2_checkered_one_reconstructed}), are recovered with high fidelity, confirming that the fundamental attack is effective. However, the reconstruction quality clearly degrades as the spatial frequency of the patterns increases. The finer checkerboards, particularly the 8x8 (Figure~\ref{8x8_checkered_one_reconstructed}) and 16x16 (Figure~\ref{16x16_checkered_one_reconstructed}) patterns
have significant reconstruction artifacts, likely due to timing jitter or systematic noise in the measurement setup, which limited the sensor's ability to resolve high-frequency spatial patterns.
Furthermore, the results for the grayscale gradients reveal limitations in the sensor's intensity resolution. While the overall orientation of the gradients is correctly captured, the smooth transitions are lost and replaced with discrete, quantized bands of gray. This suggests that while our sensor can distinguish between significantly different power levels (e.g., black vs. white), it is less precise when differentiating the subtle, continuous changes of a smooth gradient. Additionally, a consistent horizontal line artifact is visible across all reconstructed images, likely stemming from systematic noise in our measurement setup.

These results establish a crucial baseline: our attack can successfully recover the general structure and low-frequency components of an image, but it is less effective at capturing fine-grained textures. This provides the confidence to proceed with applying the methodology to more complex ImageNet images.

\end{document}